\documentclass[preprint]{aastex61}
\usepackage{natbib}

\newcommand{\al}{{\it align2d}\ }
\newcommand{\microns}{$\mu$m}

\begin{document}
\title{The JCMT Gould Belt Survey: SCUBA-2 Data-Reduction Methods and Gaussian Source Recovery Analysis}
\author{Helen Kirk}
\affil{NRC Herzberg Astronomy and Astrophysics Research Centre, 5071 West Saanich Rd, Victoria, BC, V9E 2E7, Canada}
\affil{Department of Physics and Astronomy, University of Victoria, 3800 Finnerty Road, Victoria, BC, Canada V8P 5C2}
\author{Jennifer Hatchell}
\affil{Physics and Astronomy, University of Exeter, Stocker Road, Exeter EX4 4QL, UK}
\author{Doug Johnstone}
\affil{NRC Herzberg Astronomy and Astrophysics Research Centre, 5071 West Saanich Rd, Victoria, BC, V9E 2E7, Canada}
\affil{Department of Physics and Astronomy, University of Victoria, 3800 Finnerty Road, Victoria, BC, Canada V8P 5C2}
\author{David Berry}
\affil{East Asian Observatory, 660 N. A`oh\={o}k\={u} Place, University Park, Hilo, Hawaii 96720, USA}
\author{Tim Jenness}
\affil{Joint Astronomy Centre, 660 N. A`oh\={o}k\={u} Place, University Park, Hilo, Hawaii 96720, USA}
\affil{LSST Project Office, 933 N. Cherry Ave, Tucson, AZ 85719, USA}
\author{Jane Buckle}
\affil{Astrophysics Group, Cavendish Laboratory, J J Thomson Avenue, Cambridge, CB3 0HE, UK} 
\affil{Kavli Institute for Cosmology, Institute of Astronomy, University of Cambridge, Madingley Road, Cambridge, CB3 0HA, UK}
\author{Steve Mairs}
\affil{East Asian Observatory, 660 N. A`oh\={o}k\={u} Place, University Park, Hilo, Hawaii 96720, USA}
\author{Erik Rosolowsky}
\affil{Department of Physics, University of Alberta, Edmonton, AB T6G 2E1, Canada}
\author{James Di~Francesco}
\affil{NRC Herzberg Astronomy and Astrophysics Research Centre, 5071 West Saanich Rd, Victoria, BC, V9E 2E7, Canada}
\affil{Department of Physics and Astronomy, University of Victoria, 3800 Finnerty Road, Victoria, BC, Canada V8P 5C2}
\author{Sarah Sadavoy}
\affil{Harvard-Smithsonian Center for Astrophysics, 60 Garden Street, Cambridge, MA 02138, USA}
\author{Malcolm Currie}
\affil{Joint Astronomy Centre, 660 N. A`oh\={o}k\={u} Place, University Park, Hilo, Hawaii 96720, USA}
\affil{RAL Space, Rutherford Appleton Laboratory, Harwell Oxford, Didcot, Oxfordshire, OX11 0QX, UK}
\author{Hannah Broekhoven-Fiene}
\affil{Department of Physics and Astronomy, University of Victoria, 3800 Finnerty Road, Victoria, BC, Canada V8P 5C2}
\author{Joseph C. Mottram}
\affil{Leiden Observatory, Leiden University, PO Box 9513, 2300 RA Leiden, The Netherlands}
\affil{Max-Planck Institute for Astronomy, K{\"o}nigstuhl 17, 69117 Heidelberg, Germany}
\author{Kate Pattle}
\affil{Jeremiah Horrocks Institute, University of Central Lancashire, Preston, Lancashire, PR1 2HE, UK}
\affil{Institute of Astronomy and Department of Physics, National Tsing Hua University, Hsinchu 30013, Taiwan}
\author{Brenda Matthews}
\affil{NRC Herzberg Astronomy and Astrophysics Research Centre, 5071 West Saanich Rd, Victoria, BC, V9E 2E7, Canada}
\affil{Department of Physics and Astronomy, University of Victoria, 3800 Finnerty Road, Victoria, BC, Canada V8P 5C2}
\author{Lewis B.~G. Knee}
\affil{NRC Herzberg Astronomy and Astrophysics Research Centre, 5071 West Saanich Rd, Victoria, BC, V9E 2E7, Canada}
\author{Gerald Moriarty-Schieven}
\affil{NRC Herzberg Astronomy and Astrophysics Research Centre, 5071 West Saanich Rd, Victoria, BC, V9E 2E7, Canada}
\author{Ana Duarte-Cabral}
\affil{School of Physics and Astronomy, Cardiff University, The Parade, Cardiff, CF24 3AA, UK}
\author{Sam Tisi}
\affil{Department of Physics and Astronomy, University of Waterloo, Waterloo, Ontario, Canada N2L 3G1}
\author{Derek Ward-Thompson}
\affil{Jeremiah Horrocks Institute, University of Central Lancashire, Preston, Lancashire, PR1 2HE, UK}

\begin{abstract}
The JCMT Gould Belt Survey was one of the first Legacy Surveys with the James Clerk Maxwell
Telescope in Hawaii, mapping 47 square degrees of nearby ($< 500$~pc) molecular clouds in 
both dust continuum emission at 850~\microns\ and 450~\microns, as well as a more-limited area 
in lines of various CO isotopologues.  
While molecular clouds and the material that forms stars have structures
on many size scales, their larger-scale structures are difficult to observe 
reliably in the submillimetre regime using ground-based facilities.  
In this paper, we quantify the extent
to which three subsequent data-reduction methods employed by the JCMT GBS accurately
recover emission structures of various size scales, in particular, dense cores
which are the focus of many GBS science goals.
With our current best data-reduction procedure, we expect to recover
$100$\% of structures with Gaussian $\sigma$ sizes of $\le $30\arcsec\ and intensity 
peaks of at least five times 
the local noise for isolated peaks of emission.  
The measured sizes and peak fluxes of these compact structures
are reliable (within 15\% of the input values), but source recovery and
reliability both decrease significantly for larger emission structures
and for fainter peaks.
Additional factors such as source crowding have not been tested in our analysis.
The most recent JCMT GBS data release includes pointing corrections, and we demonstrate
that these tend to decrease the sizes and increase the peak intensities 
of compact sources in our
dataset, mostly at a low level (several percent), but occasionally with notable improvement.
\end{abstract}

\section{Introduction}
The James Clerk Maxwell Telescope (JCMT) Gould Belt Survey \citep[GBS;][]{WardThomp07} 
is one of the  initial set of
JCMT Legacy Surveys, and has the goal of mapping and characterizing dense star-forming
cores and their environments across all molecular clouds within $\sim$500~pc.
The JCMT GBS included extensive maps of the dust continuum emission at 850~\microns\ and
450~\microns\ of all nearby molecular clouds observable from Maunakea using 
SCUBA-2 \citep[Submillimetre Common User Bolometer Array-2;][]{Holland13}, 
as well as more-limited spectral-line 
observations of various CO isotopologues using HARP 
\citep[Heterodyne Array Receiver Program;][]{Buckle09}.
For this paper, we focus on the SCUBA-2 portion of the survey.

\begin{figure}[htbp]
\includegraphics[width=5.9in]{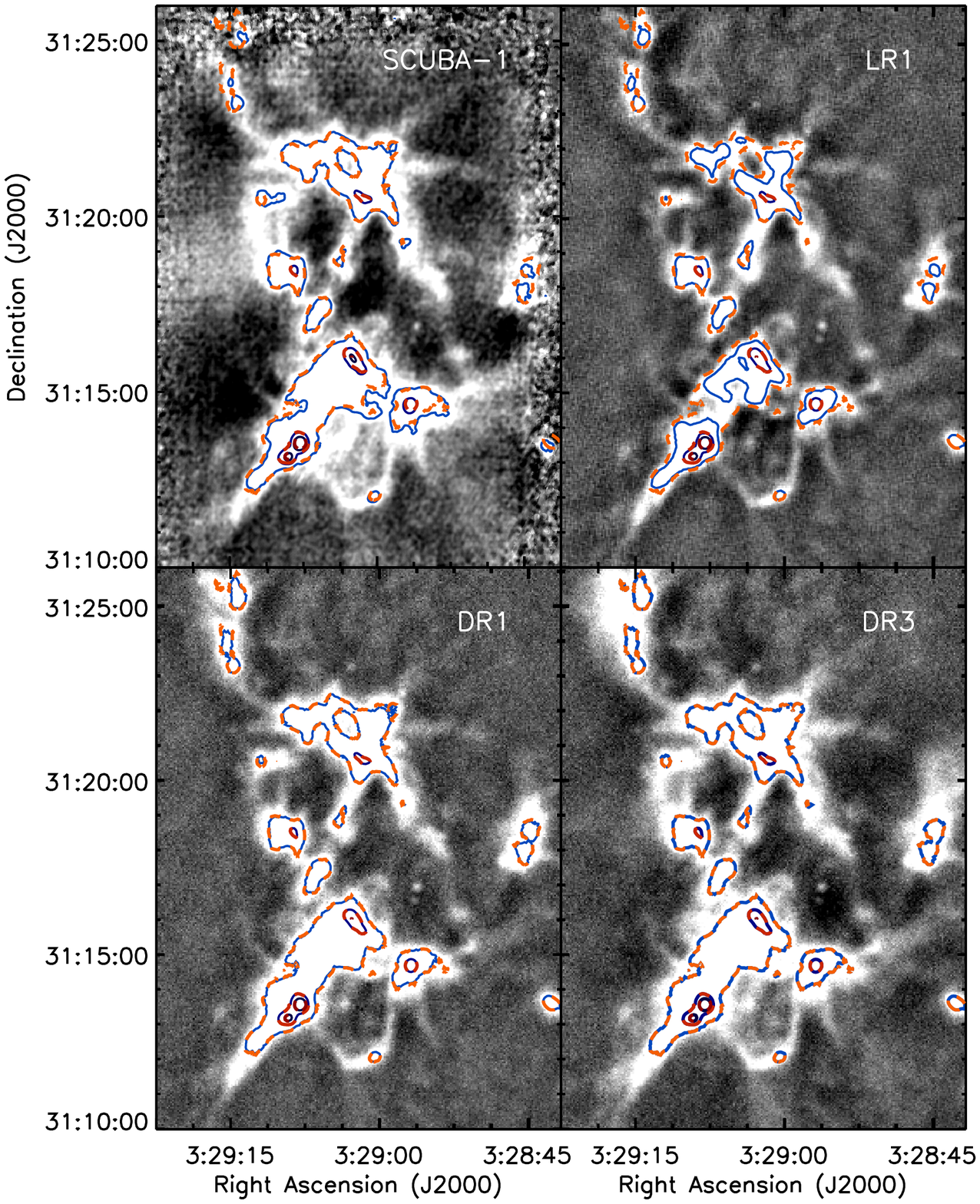}
\vspace{-0.75in}
\caption{A comparison of emission observed in NGC~1333.  The top left panel shows
	data from SCUBA, published in \citet{Sandell01}, while the remaining three panels
	show SCUBA-2 observations, converted into Jy~bm$^{-1}$ flux units assuming a 
	14\farcs6 beam as in \citet{Dempsey13}.  The SCUBA-2 reductions shown are
	the JCMT Legacy Release 1 (LR1; top right), JCMT GBS Data Release 1 (DR1; bottom left)
	and JCMT GBS Data Release 3 (DR3; bottom right), all discussed further in
	this paper. 
	In all panels, faint emission is emphasized in the grey scale which ranges from
	-0.05~Jy~bm$^{-1}$ and 0.1~Jy~bm$^{-1}$.  Both the solid blue and dashed orange 
	contours indicate emission at 0.2~Jy~bm$^{-1}$, 1~Jy~bm$^{-1}$, 
	and 3~Jy~bm$^{-1}$.  The solid blue contours trace the greyscale image shown
	in that panel, while the dashed orange contours show the SCUBA-2 DR3 map 
	for reference.
	}
\label{fig_scuba_scuba2}
\end{figure}

The SCUBA-2 instrument is an efficient and sensitive mapper of thermal emission from
cold and compact dusty structures such as dense cores, the birthplace of future stars.  One of
the science goals of the JCMT GBS is to identify and characterize these dense cores,
which includes estimating their sizes and total fluxes (masses).  These are
challenging observations to make from the ground, as the Earth's atmosphere
is bright and variable at submillimetre wavelengths.  
As such, all ground-based observations in the submillimetre
regime use some form of filtering.  Often this filtering is done in the form of 
`chopping', where fluxes are measured in some differential form 
\citep[see, e.g.][and references therein]{Haig04}.
SCUBA-2, however, combines a fast scanning pattern during observing with an 
iterative filtering technique during the data reduction process, which 
has the similar consequence of
removing both contributions from the atmosphere and extended source emission
\citep[e.g.,][]{Holland13,Chapin13}.
Regardless of the method, the largest scales of emission cannot be recovered
from ground-based submillimetre observations, as it is not possible to 
disentangle such signal from that of the atmosphere.  
Nonetheless, it is desirable
for star-formation science to obtain accurate measurements of emission structures on as
large a scale as possible.  
New instrumentation, observing techniques, and 
data-reduction tools allow for better recovery of larger-scale emission structures than
was feasible in the past.  As an example, Figure~\ref{fig_scuba_scuba2} shows
the emission observed in the NGC~1333 star-forming region in the Perseus molecular
cloud as seen with the original SCUBA detector \citep{Sandell01} compared with the
same map obtained with SCUBA-2, as part of the GBS survey, and reduced using
several different techniques.  The SCUBA-2 map was first presented in \citet{Chen16}
using the GBS Internal Release 1 reduction method, but is shown in Figure~\ref{fig_scuba_scuba2}
using several more recent SCUBA-2 data reduction methods, all of which are discussed 
further throughout this paper. 
While bright and compact emission structures appear the same in all panels, 
the GBS DR3 map clearly recovers the most faint and extended structure, while
suffering the least from artificial large-scale features,
such as that seen at the centre left
of the SCUBA image.  In this paper, we focus on the reliability of the GBS SCUBA-2
maps, and do not present any quantitative comparisons with SCUBA data.

While not the focus of our present work, we note that space-based submillimetre 
facilities such as the {\it Herschel Space Telescope} avoid the challenge of observing through 
the atmosphere, and therefore offer the ability to obtain observations with much less filtering.
At the same time, space-based submillimetre facilities have much lower angular resolutions, due to 
the difficulty in placing large dishes in space.  Previous work by JCMT GBS members provide 
a comparison of star-forming structures observed using {\it Herschel} and SCUBA-2 
\citep{Sadavoy13,Pattle15,WardThomp16,Chen16}, although all of these analyses used earlier 
SCUBA-2 data-reduction methods than the methods analyzed here.  The loss of larger-scale
emission structures inferred by comparing SCUBA-2 and {\it Herschel} observations will 
therefore be somewhat less severe when the current data products are used instead.

To achieve the various science goals of the GBS, it is important to have
a thorough understanding of the completeness and reliability of the sources detected.
Uncertainties in source detection and characterization can arise both from the observations
and map reconstruction efforts, as well as from the tools used to identify and
characterize the emission sources.
In the analysis presented here, we aim to investigate thoroughly 
the first of these issues,
i.e., quantifying how well a source of known brightness and size is recovered
in a JCMT GBS map, when an idealized source-detection algorithm is used
in an ideal (non-crowded) environment.

The JCMT GBS has released several versions of ever-improving data products to the
survey team for analysis: Internal Release 1 (IR1), Data Release 1 (DR1)\footnote{This
is called the `GBS Legacy Release 1' in \citet{Mairs15}.}, Data Release 2 (DR2),
and Data Release 3 (DR3), while the JCMT has also released maps of
all data 850~$\mu$m data obtained between 2011 February 1 and 2013 August 1
through their `JCMT Legacy Release~1'
(LR1; Graves et al, in prep; 
see also \texttt{http://www.eaobservatory.org/jcmt/science/archive/lr1/}).  
Table~\ref{tab_published} summarizes all currently
published GBS maps.  The last three GBS data releases, intended to be made 
fully public,
are the focus of this paper.  
We also provide an approximate comparison of the GBS data 
products to the JCMT's LR1 maps, which are qualitatively similar to the 
intermediate `automask' GBS data products discussed in the text.   

Within the GBS data releases, 
DR2 improves on DR1 through the use of improved data-reduction techniques that enhance 
the ability to recover faithfully 
large-scale emission structures.  Many of these improvements were outlined in
\citet{Mairs15}, but it was beyond the scope of that work to replicate fully the
data-reduction process used for DR1 and DR2 and quantify how well structure
in the maps is recovered.  Additionally, several small modifications to the data-reduction
procedure were made after the testing performed in \citet{Mairs15}.  
The majority of this paper focuses on a careful comparison between the recovery of
structure using the exact JCMT GBS DR1 and DR2 methodologies.
Unlike DR1 and DR2, DR3 does not involve a completely new re-reduction
of all JCMT GBS observations with improved recipes.  Instead, DR3 focuses on 
estimating the pointing offset errors present in the observations, and adjusting the final 
DR2 maps to correct for them.  

\begin{deluxetable}{clll}
\tablecolumns{8}
\tablewidth{0pc}
\tabletypesize{\scriptsize}
\tablecaption{GBS Published Maps\tablenotemark{a}\label{tab_published}}
\tablehead{
\colhead{Region} &
\colhead{Data Version} &
\colhead{Reference} &
\colhead{DOI\tablenotemark{b}} 
}
\startdata
CrA &		DR1 & 	Bresnahan et al (in prep) & 	pending \\
Auriga &	DR1 & 	\citet{BroekhovenFiene18} & 	https://doi.org/10.11570/17.0008 \\
IC5146 &	DR1 & 	\citet{Johnstone17} & 	https://doi.org/10.11570/17.0001 \\
Lupus & 	DR1 & 	\citet{Mowat17} & 	https://doi.org/10.11570/17.0002 \\
Cepheus & 	DR1 & 	\citet{Pattle17} & 	https://doi.org/10.11570/16.0002 \\
Orion~A &	DR1 automask & 	\citet{Lane16} & https://doi.org/10.11570/16.0008 \\
Taurus~L1495 & 	IR1 & 	\citet{WardThomp16} & 	https://doi.org/10.11570/16.0002 \\
Orion~A\tablenotemark{c} & DR1 & \citet{Mairs16} & https://doi.org/10.11570/16.0007 \\
Perseus & 	IR1 & 	\citet{Chen16} & 	https://doi.org/10.11570/16.0004 \\
Serpens~W40 & 	DR1 & 	\citet{Rumble16} & 	https://doi.org/10.11570/16.0006 \\
Orion~B & 	DR1 & 	\citet{Kirk16} & 	https://doi.org/10.11570/16.0003 \\
Ophiuchus & 	IR1 & 	\citet{Pattle15} & 	https://doi.org/10.11570/15.0001 \\
Serpens~MWC297 & IR1 & 	\citet{Rumble15} & 	https://doi.org/10.11570/15.0002 \\
\enddata
\tablenotetext{a}{This table includes only published GBS papers where the submillimetre
	map was publicly released alongside the paper.}
\tablenotetext{b}{Digital Object Identifier is a permanent webpage where a static version
	of the GBS data is stored for public distribution.}
\tablenotetext{c}{While analysis was performed only in the southern portion of the map,
	the entire map is provided at the DOI.}

\end{deluxetable}

Quantifying the quality and fidelity of our JCMT GBS maps is a crucial step for the 
over-arching science goals of the survey.  
For example, one goal is to measure the distribution of core masses and compare this
distribution
with the initial (stellar) mass function \citep{WardThomp07}.  
Without detailed knowledge of source recoverability 
and whether or not 
there is any bias in real versus observable flux, the obtained core mass function 
could be misinterpreted.
A wide range of artificial Gaussians were used in our testing, ranging from sources
that should be difficult to detect (e.g., peak brightnesses similar to the image noise
level) to those that should be easy to recover accurately (e.g., compact sources with
peaks at 50 times the image noise level).  We emphasize that especially for the former
case, the recovery results we present here represent an unachievable
ideal case for realistic analysis: knowing precisely {\it where} to look for the injected
peaks, as well as precisely {\it what} to look for (known peak brightness and width) allows
us to recover sources that would never be identifiable in a real observation.
A full quantification 
of completeness would require including non-Gaussian sources (e.g., also filamentary 
morphologies, and elongated cores with non-Gaussian radial profiles), 
testing the effects of source crowding, testing several of
the commonly used source-finding algorithms and determining the influence of
false positive detections, and not tuning the source-finding algorithm to 
look for emission in known locations.
Such an analysis is beyond the scope of this paper, although some aspects have
been examined by previous studies 
\citep[e.g.,][]{Rosolowsky08,Pineda09,Kainulainen09,Kauffmann10,Shetty10,Rosolowsky10,Reid10,Ward12,Menshchikov13}.

The paper is structured as follows.
In Section~2, we discuss the JCMT GBS observations and the general data-reduction procedure.
In Section~3, we describe our method for testing source recoverability and fidelity in source
recovery in the DR1 and DR2 maps, and the results are discussed in 
Section~\ref{sec_completeness_results}.
These tests provide essential metrics for future analyses of GBS data where
the
role of bias and the recoverability of real structure 
in the observations will need to be understood.
In Section~\ref{sec_IR4}, we introduce two independent methods for measuring the 
telescope-pointing errors in each observation, and demonstrate that the final DR3 maps should
have little residual relative pointing error.  This analysis 
provides us with confidence that the properties of
emission structures measured in DR3 should not be substantially more blurred out than
expected from the native telescope resolution.

\section{Observations}
\label{sec_obs}

SCUBA-2 observations were obtained between 2011 October 18 and 2015 January 26.
Observations were made in Grade 1 ($\tau_\mathrm{225GHz} < 0.05$) and 
Grade 2 ($0.05 < \tau_\mathrm{225GHz} < 0.08$) 
weather conditions.  Grade~1 weather provides good measurements at both 850~\microns\ and
450~\microns, while Grade~2 weather is suitable for 850~\microns\ and provides poorer
measurements at 450~\microns.
Each field was observed four to six times depending on the local weather conditions,
to obtain approximately constant noise levels across the survey at 850~\microns.
Observations at 450~\microns\ are more sensitive to the atmospheric conditions,
and hence show a significantly larger variation in noise properties.

\begin{figure}[htb]
\plottwo{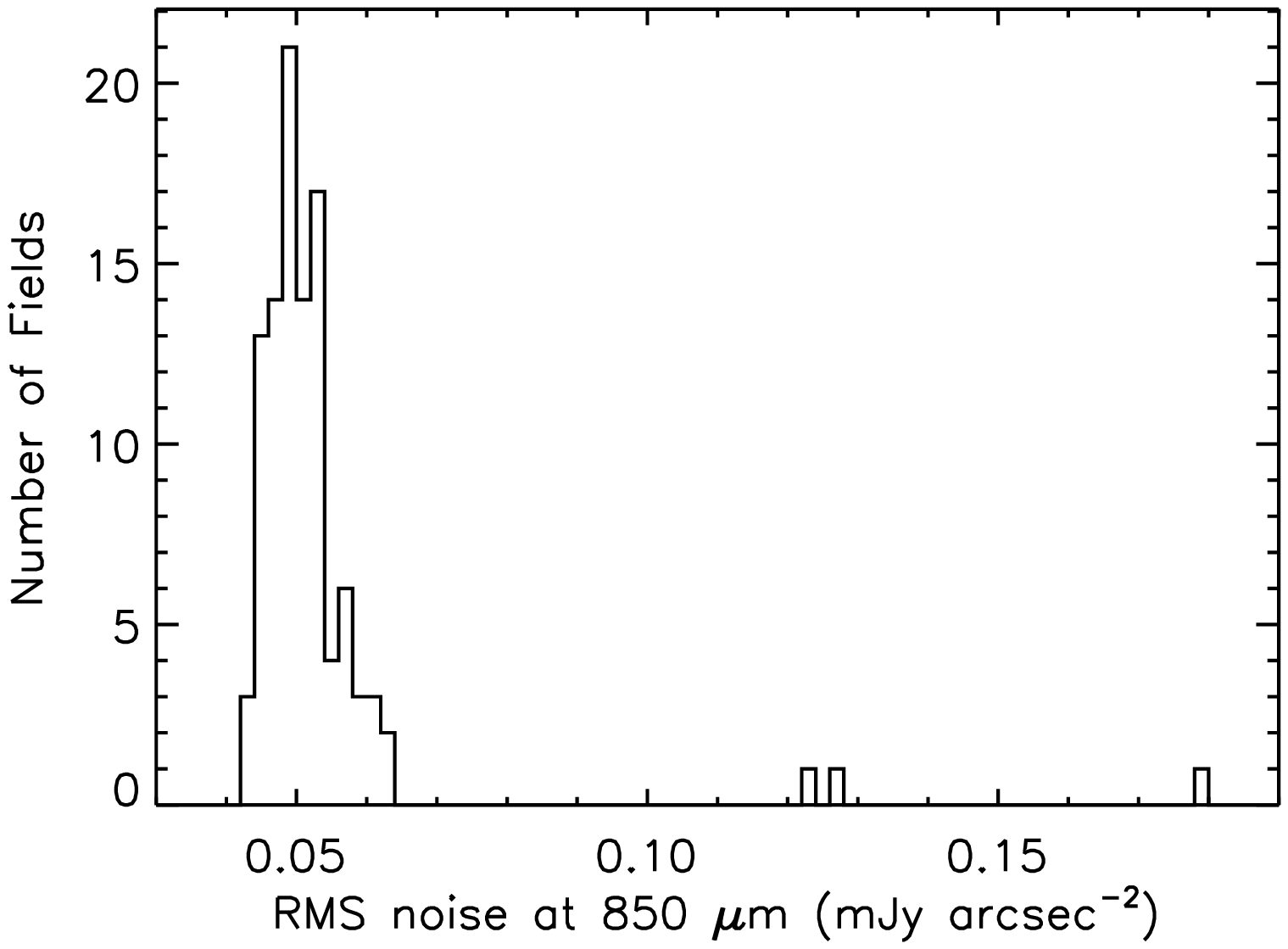}{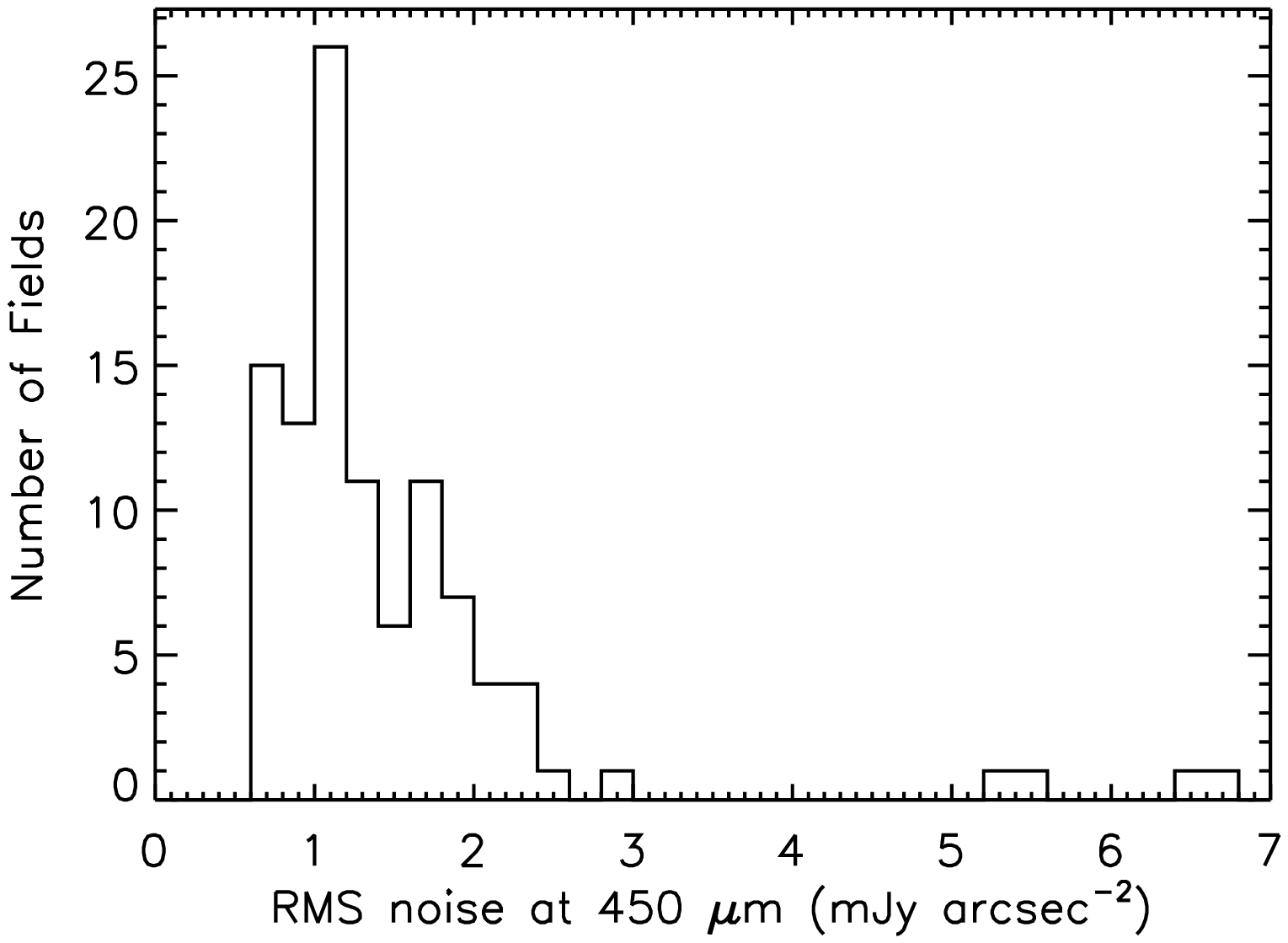}
\caption{The distribution of rms values for each field mapped by the GBS.  The left hand
	panel shows the rms values at 850~$\mu$m, while the right hand panel shows the
	rms values at 450~$\mu$m.  1~mJy~arcsec$^{-2}$ corresponds to 
	approximately 242~mJy~beam$^{-1}$ at 850~$\mu$m and 109~mJy~beam$^{-1}$ at 
	450~$\mu$m.
	Note that the three highest rms values in each
	panel correspond to science verification fields which were not observed
	to their full depth.  The final high-noise outlier at 450~$\mu$m corresponds
	to one of the fields in Lupus, which was observed during marginal weather
	at a low elevation, conditions that adversely affect 450~$\mu$m data to a much
	greater extent than 850~$\mu$m.}
\label{fig_rms_histogram}
\end{figure}

Table~\ref{tab_pong_rms} summarizes the approximate noise level in each field
of the survey, while Figure~\ref{fig_rms_histogram} shows the distribution of
noise levels.  We ran the Starlink {\sc Picard} \citep{Gibb13} recipe {\it mapstats} 
on each individual observation 
to calculate the noise in the central portion (i.e., inner
circle of radius 90\arcsec) of the observed area.  We 
then estimated the effective 
noise for each field in the final mosaic by accounting for the fact 
that the observations are combined
using the mean values weighted by the inverse square of the noise at that location.
For most of the paper, we focus on the 850~\microns\ data,
where the noise levels are more uniform.

The standard observing mode used for the GBS data was the PONG 1800 mode \citep{Kackley10},
which produces fully sampled 30\arcmin\ diameter regions.  The GBS obtained a total of 581
observations under this mode, as well as a handful of additional observations under the 
PONG 900 and PONG 3600 modes during SCUBA-2 science verification (SV).  
We focus our analysis here entirely
on the PONG 1800 observations.  Earlier testing by the GBS data-reduction team showed that
the other mapping modes have different sensitivities to large-scale structures.

We reduced the maps using the iterative routine known as {\it makemap}, which is distributed as part 
of the {\sc smurf} package \citep{Chapin13,smurf} in Starlink \citep{Currie14}.  
We used a gridding size
of 3\arcsec\ pixels at 850~\microns\ and 2\arcsec\ pixels at 450~\microns, and halted 
iterations when the map pixels changed on average by $<0.1$\% of the estimated map rms.
In both DR1 and DR2, we reduced each observation twice, following a similar overall
procedure.  In the first reduction, known as the automask reduction, pixels containing 
real astronomical signal were estimated using various signal-to-noise ratio (SNR) 
criteria applied
to the raw data time stream.  We then mosaicked together all maps of the same region, and 
determined more comprehensive areas of likely real astronomical signal.  These areas were
then supplied as a mask for the second round of individual reductions known as the
external-mask reduction.  The final mosaic was created using the output of the second
round of reductions.  

The final (external mask) mosaic tends to contain much more 
large-scale emission structure than is in the first (automask) mosaic.  
The reason for this difference
is that the map-making algorithm needs to distinguish between larger modes of variation
in the raw time stream data, which arise from scanning across true astronomical signals,
versus those induced by variations in the sky or instrumental effects, which it does through
the use of a mask.  By being able to combine four to six initially reduced maps together
to determine where real astronomical signal is likely, it is possible to identify
accurately emission over
a much larger area of sky than is evident from the raw data in a single observation.
The differences between our DR1 and DR2 procedures focused on methods of improving the
sensitivity to larger-scale structure in the initial automask reduction (e.g., reducing
large-scale filtering), as well as
creating more generous, but still accurate, masks for the external-mask reduction
(e.g., lowering the mask SNR criteria).
We note that in defining the mask, there are two competing challenges, as also
discussed in \citet{Mairs15}.
Masks that are smaller than the true extent of the source emission will prevent a 
full recovery of that emission, leading to artificially smaller and fainter sources.
At the same time, masks that include regions without real source emission are liable
to introduce false large-scale structure which may artificially increase the total
size and brightness of real sources.
Appendix~\ref{app_dr_params} outlines the full reduction procedure and {\it makemap}
parameters applied for both DR1 and DR2.

Although not identical, the data-reduction procedure for the JCMT's Legacy Release~1
(LR1) dataset is similar to the GBS DR1 automask procedure: only one round of reduction
is run, and strong spatial filtering is applied to suppress real and artificial 
large-scale structures.

In DR3, we use the DR2 reductions for each observation, and then search for possible
offsets between observations of the same field due to telescope-pointing errors.  If
positional offsets are found, we apply the appropriate shift to the observation before
creating the final mosaicked image.  This procedure is discussed in more detail in
Section~5.

One final data-reduction parameter which we do not refine beyond the standard
recommended procedure is the appropriate flux conversion factor applied to each 
observation.  As discussed in \citet{Dempsey13}, the standard observatory-derived
FCF values appear stable over time, with a scatter of less than 5\% at 850~$\mu$m
and about 10\% at 450~$\mu$m in relative calibration, while the absolute calibration 
factors are approximately 8\% and 12\% at 850~$\mu$m and 450~$\mu$m, respectively.  
The JCMT Transient Survey demonstrates that it is
possible to improve the relative calibration at 850~$\mu$m to 2\%-3\% \citep{Mairs17a},
however, the Transient Survey procedure requires multiple bright point sources per observation, 
which many GBS fields do not possess.  We therefore simply note that the GBS source
flux estimates should be accurate to 8\% at 850~$\mu$m and 12\% at 450~$\mu$m
using the default calibrations, as confirmed in \citet{Mairs17a}.
A small fraction of sources may also have variable emission, although most of the 
variable candidates identified in the Transient Survey show variations in flux of less than
a few percent over the course of the typically short (days or months) time span between
typical GBS observations of the same field \citep{Mairs17b,Johnstone18}.
Only one source of the $\sim$150 monitored by the Transient Survey shows 
variability of more than 10\% over short time scales 
\citep[EC~53 in Serpens][]{Yoo17,Johnstone18}.

For completeness, we note that where available, all GBS data releases additionally
include `CO-subtracted' maps.  The $^{12}$CO(3-2) emission line lies within the
850~$\mu$m bandpass, and therefore can contribute flux to the emission measured
\citep[e.g.,][]{Drabek12}.
This `CO contamination' is typically $<10$\% of the total flux measured, although it
can be significantly higher (up to 80\%) in rare cases where there is an outflow in
a lower density environment.
Where appropriate measurements of the $^{12}$CO(3-2) integrated intensity were available 
to the GBS, we ran an additional round of reductions for each of DR1, DR2, and DR3
with the CO emission properly subtracted from the 850~$\mu$m map.  A brief summary
of our CO subtraction procedure is given in Appendix~A.4.

\section{Source Recovery Measurements}
Here, we discuss our procedure for measuring our accuracy in recovering
emission structures.  

\subsection{Test Setup}
As discussed in Section~\ref{sec_obs}, Starlink's {\it makemap} is the standard software for
reducing SCUBA-2 mapping observations.  
Using {\it makemap}, the user can insert artificial sources
directly into an observation's raw-data time stream, providing an easy mechanism to
measure how well idealized model emission structures are recovered under different
data-reduction settings.  
Our approach was guided by the aim to test systematically
the best-case scenario of isolated point sources that are 
not confused by a local background.
We emphasize that many of the dense cores identified in the GBS will have some
degree of crowding and / or hierarchical structures, which will reduce the 
reliability of the recovered emission.
We used the GBS 850~$\mu$m 
observations of the OphScoN6 field as the
basis for our testing.  It is the GBS field that contains the least amount of real
signal, i.e., the observation mostly closely resembling a pure-noise field.
OphScoN6 was observed seven times rather than the standard six times for observations
obtained in Grade~2 weather, so we excluded one of the observations (20130702\_00031) 
to make the dataset more similar to a standard GBS field.
This excluded observation was taken under marginal weather conditions with higher noise levels
than is typical for most GBS observations.
The noise at 850~$\mu$m in the mosaic of the six OphScoN6 maps 
is 0.049~mJy~arcsec$^{-2}$, which is similar to those of other GBS fields 
(cf. Table~\ref{tab_pong_rms} and Figure~\ref{fig_rms_histogram}).

\begin{figure}[htb]
\includegraphics[width=3in]{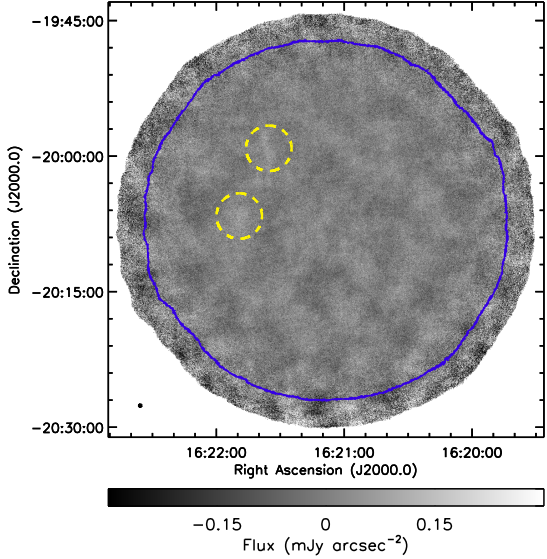}
\caption{The OphScoN6 field as observed by the JCMT GBS at 850~$\mu$m.  The image shown
	is the external-mask reduction for DR2.  Both the DR1 and DR2 reductions (automask
	and external mask) appear similar in this region due to the lack
	of structure detected.
	[We note that the DR2 reductions show faint large-scale
	mottling which is not present in the DR1 reductions, as DR1 included additional
	large-scale filtering outside of masked regions.  See Appendix~A for more details.]   
	The yellow dashed circles show the approximate location of two 
	possible faint emission
	structures in the map, while the blue contour shows a separation of 3\arcmin\
	from the edge of the mosaic.  The small black dot at the bottom left
	indicates the SCUBA-2 beam.}
\label{fig_oph_obs}
\end{figure}

Figure~\ref{fig_oph_obs} shows the DR2 automask reduction of the OphScoN6 data used here.
A careful visual examination of the map shows that there are two faint  
zones of potentially real
emission to the east of the field but, with the low peak signal level, 
neither are definite detections.
Nonetheless, we take care in our completeness testing to avoid potential biases
due to low-level emission in these regions.

We generate artificial, radially-symmetric Gaussian sources with a range of 
peak intensities and widths 
to add to each raw observation, to test how well they are
recovered in the final reduced mosaic.  
We constrained all fake sources to lie in angular separation
at least 3 Gaussian $\sigma$ away from the 
outer 3\arcmin\ of the map (where the local noise is significantly higher), and 
also away from the zone of potential emission in the east of the mosaic, defined 
as two circles of 2.5\arcmin\ radius, with the centres set by eye.  Both of these 
excluded map areas are shown in Figure~\ref{fig_oph_obs}.

For any given set of Gaussian parameters (i.e., amplitude and width), we 
randomly placed 500 sources, eliminating those which landed in the edge or possible 
emission zones noted above, or those located 
less than 6 $\sigma$ away from a previously placed source.  In the
case of the narrowest ($\sigma =$10\arcsec) Gaussians we tested,
this process resulted in more than 100 inserted sources
per map.  For the widest ($\sigma =$150\arcsec) Gaussians that we tested, 
however, only one or two sources could
be placed in a map while still satisfying all of the above criteria.  We therefore created
multiple maps with artificial sources added for the widest Gaussians to improve our
statistics.  We note, however, that our statistics are still poorer for the widest Gaussian
cases.  It is too computationally intensive to run hundreds of reductions for each wide 
Gaussian to match the number of sources able to be inserted in a single narrow Gaussian test
image\footnote{For reference, the reduction of each SCUBA-2 raw observation requires 
approximately 1.5 hours running on a dedicated 100~GB RAM, 12 core CPU machine, while each 
test Gaussian input field uses a mosaic of six raw observations, and requires four reductions 
(DR1 and DR2, automask and external mask).}.  
Table~\ref{tab_NGauss} summarizes the Gaussian parameters used for testing,
and lists the total number of artificial sources used for each combination of width
and peak.
In total, we inserted 3196 artificial Gaussian sources into the maps.  We explored
63 different Gaussian widths and amplitudes, with a total of 306 test fields, to 
boost our statistics.

\begin{figure}[htb]
\plotone{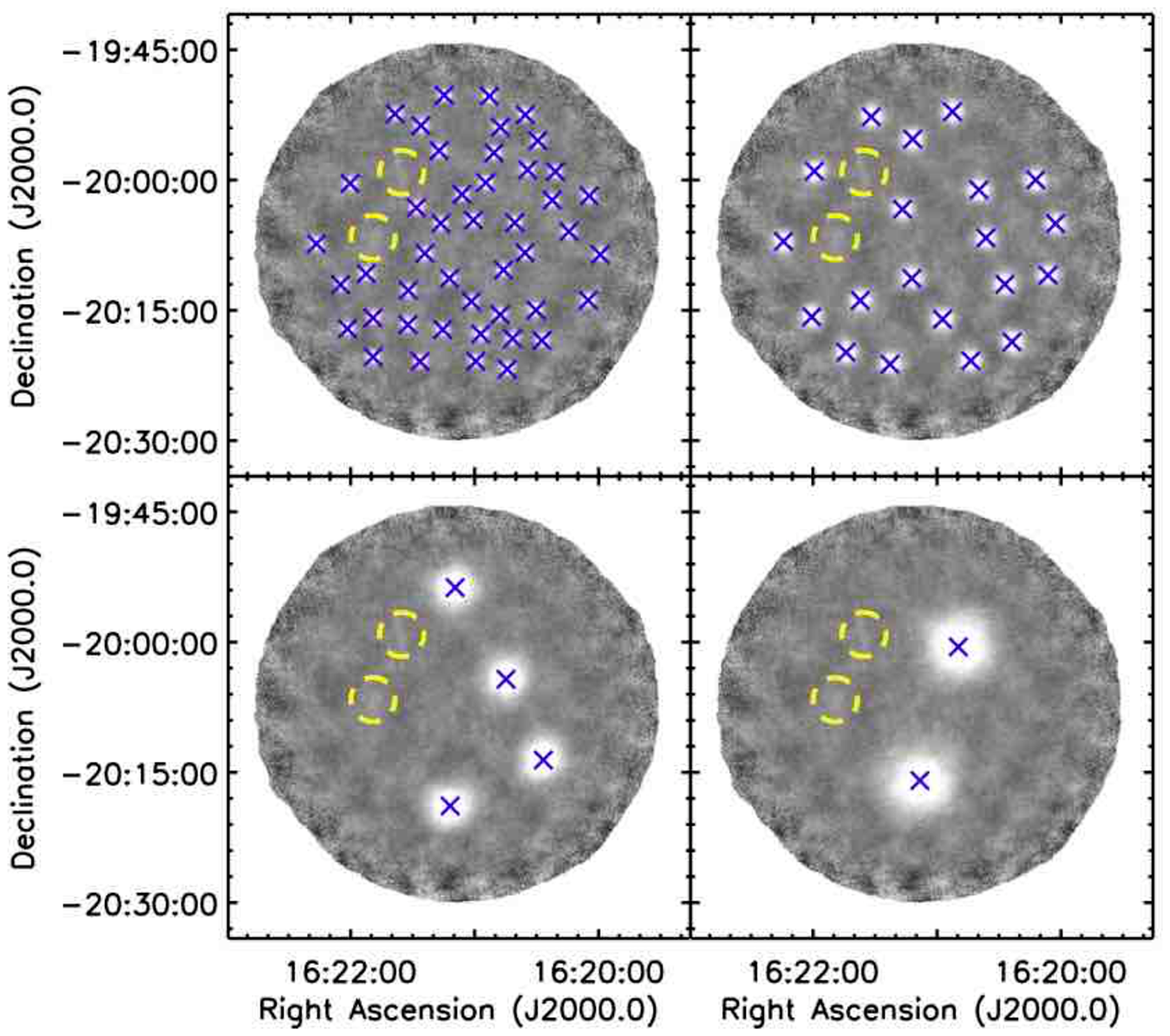}
\caption{An illustration of the artificial Gaussian test set-up.  The background greyscale image
	shows the original OphScoN6 DR2 external-mask mosaic with the artificial random
	Gaussians added {\it directly to the image}.
	As in Figure~\ref{fig_oph_obs}, the grayscale ranges from -0.3~mJy~arcsec$^{-2}$ 
	to 0.3~mJy~arcsec$^{-2}$.
	These images represent the idealized case where the data-reduction process 
	perfectly returns all structure in the area.
	The blue crosses denote the centres
	of each random Gaussian, while the dashed yellow circles show regions of the mosaic 
	where the Gaussians were not allowed to be placed due to low-level potential
	emission structures present in the mosaic.
	All of the Gaussians shown have amplitudes of 10 times the mosaic rms noise.
	From top left to bottom right, the Gaussians have widths ($\sigma$) of 
	[30, 50, 100, 150] arcseconds, respectively.
	}
\label{fig_test_image}
\end{figure}

Figure~\ref{fig_test_image} shows an example of the test setup, with the artificial
Gaussian sources added directly to the original mosaic.  Since these Gaussians have
not passed through our data-reduction pipeline, deviations from perfect Gaussians are
entirely attributable to the background noise in the mosaic.

\begin{deluxetable}{c|lllllll}
\tablecolumns{8}
\tablewidth{0pc}
\tabletypesize{\scriptsize}
\tablecaption{Number of Gaussian Peaks Analyzed\label{tab_NGauss}}
\tablehead{
\colhead{Amplitude} &
\multicolumn{7}{c}{Sigma (arcsec)}\\
\colhead{($N_\mathrm{rms}$)} &
\colhead{10} &
\colhead{30} &
\colhead{50} &
\colhead{75} &
\colhead{100} &
\colhead{125} &
\colhead{150} 
}
\startdata
1  & 163 & 50 & 57 & 36 & 24 & 14 & 10 \\
2  & 170 & 45 & 61 & 37 & 21 & 20 & 11 \\
3  & 151 & 52 & 63 & 37 & 23 & 16 & 9  \\
5  & 171 & 48 & 55 & 42 & 22 & 19 & 9  \\
7  & 147 & 53 & 53 & 37 & 24 & 20 & 9  \\
10 & 181 & 45 & 57 & 37 & 21 & 16 & 12 \\
15 & 159 & 45 & 58 & 34 & 20 & 19 & 11 \\
20 & 166 & 48 & 57 & 35 & 20 & 15 & 11 \\
50 & 152 & 52 & 58 & 37 & 23 & 17 & 11 \\
\hline
N$_{repeat}$\tablenotemark{a} & 1 & 1 & 3 & 5 & 6 & 9 & 9\\
\enddata
\tablenotetext{a}{The number of reductions run for each input Gaussian sigma value, 
	done to increase the total number of artificial sources available for analysis.}

\end{deluxetable}

\subsection{Data Reduction}
\begin{figure}[htb]
\plotone{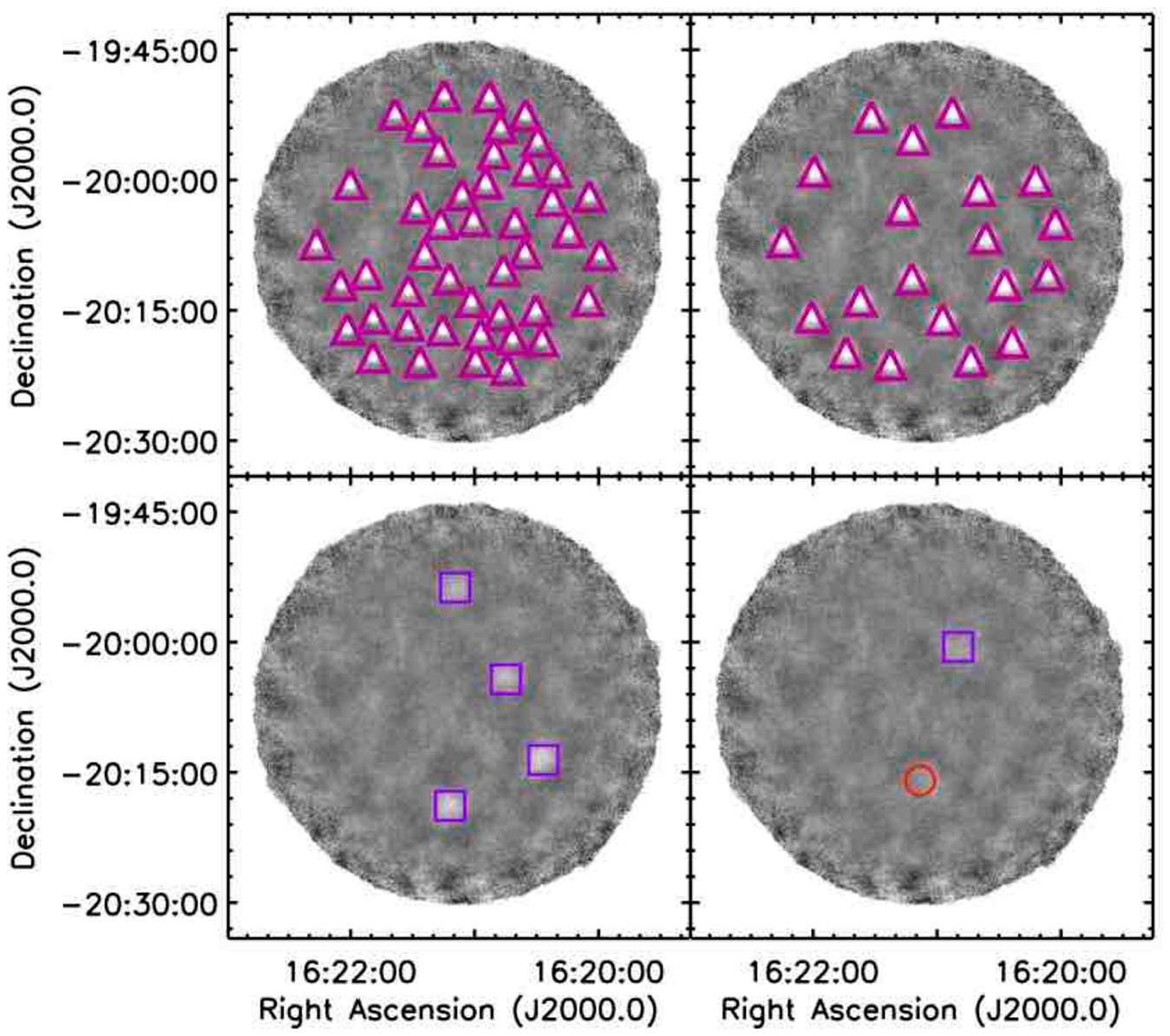}
\caption{Examples of final mosaics using the DR2 external-mask reduction method, 
	with the artificial Gaussians added into the 
	raw data prior to processing.  This figure shows the same artificial
	Gaussian fields as Figure~\ref{fig_test_image}, with the same grayscale and
	other plotting conventions.
	The thick maroon triangles show the sources which were recovered within an
	external mask and the thin purple squares show the sources which were recovered
	outside of an external mask (i.e., are too faint to satisfy masking criteria;
	see Section~4.2), 
	while the red circle shows a non-detection.}
\label{fig_sample_reducs}
\end{figure}

After creating each instance of artificial Gaussians, we run our standard GBS 
data-reduction procedure with the artificial Gaussians added directly into the raw-data 
time stream for each of the six observations of OphScoN6 using the `fakemap' parameter
in {\it makemap}.  The standard reduction procedure
is outlined in Section~2.  We emphasize that the external mask is created 
separately for each set of artificial Gaussians, based on the individual automask reductions.
We follow these steps for both the DR1 and DR2 reduction procedures.
For each set of added artificial Gaussians, we therefore have four maps to examine:
DR1 and DR2, automask and external-mask reductions.  Figure~\ref{fig_sample_reducs}
shows the DR2 external-mask reductions for the four 
test cases from Figure~\ref{fig_test_image}.
Comparison of these two figures reveals a clear difference in the quality of
source recovery for smaller and larger sources, which will be analyzed quantitatively
in Section~4.

\subsection{Source Recovery}
We next use an automated method to determine how well the artificial Gaussians are recovered
in each of the maps.  In normal scientific analyses, uncertainties in where real emission is
located and its true structure can complicate emission recovery.  
Here, we take advantage of knowing precisely 
where the emission is located and what the brightness profile 
should look like to reduce the uncertainties
associated with source recovery.  For each known artificial Gaussian peak position, we use 
{\it mpfit} \citep{Markwardt09} to search the surrounding 3.0~$\sigma$ radius for a 
given input Gaussian size.  This search window is large enough
to encompass the model Gaussian peak to 0.003 times the peak brightness, which corresponds
to about one tenth of the image rms for the brightest model Gaussians.
To eliminate spurious noise features being identified, we 
discard any fits that did not converge, had large fitting uncertainties\footnote{
Specifically, we discarded fits where the ratio of the peak flux or width and its
associated fitting error was less than three, i.e., any fits where the peak flux or 
width was uncertain by at least 100\% within the standard 3-sigma uncertainty range.
We also excluded fits where the uncertainty in the location of the peak exceeded
50\% of the input Gaussian width.},
were dominated by an artificial background term\footnote{Small fitted background
terms may be reasonable if the source lies near the peak or valley of a noise feature
in the mosaic.  We excluded fits where the absolute background exceeded half of the
input peak flux or one third of the fitted peak flux.}, 
or had properties too different from the input values\footnote{This criterion
required true source recoveries to have a peak location within 1.25~$\sigma$ 
of the true centre -- a radius of 1.25~$\sigma$ 
corresponds roughly to the full width at half maximum, or FWHM.  We
also required the fits 
to be approximately round (axial ratios less than 1.5), 
to have a peak no more than 2.5 times the real value, and to have 
a width no more than twice 
the real value.  Finally, we excluded fits which were offset from their input
locations by more than the input Gaussian width divided by the square root of the peak
signal-to-noise of the input Gaussian; brighter Gaussians should have more accurately
determined centres.}.  
We did {\it not} eliminate sources which were much fainter or smaller
than the input Gaussians, as we expect the data-reduction process to create smaller and
fainter sources than we started with, as shown by \citet{Mairs15} and we wish to
quantify this effect.  Finally, to 
ensure that noise spikes or underlying larger-scale structure from the original data were 
not contaminating our results, we performed a similar Gaussian fit on the original mosaics 
(i.e., maps with no artificial sources added).  We then removed from our list 
of recovered artificial sources any fits which had consistent fit parameters to the 
original mosaic fit (within 1.5 times the fit uncertainty in all fit parameters).   
We emphasize that our
entire Gaussian-fitting procedure gives the best case possible for source recovery.  Many
of the faintest sources that we can find in our maps would not be identifiable using
a standard source-detection algorithm that was not targeted to known positions and
Gaussian properties.

We note that all of our source recovery tests discussed here and in the following
sections focus on the 850~$\mu$m observations.  We expect that the 450~$\mu$m data would 
follow qualitatively similar trends, but would not behave identically.  Early testing by the
data reduction team showed that in general, large-scale structure is better recovered
when larger pixel sizes are adopted.  The GBS uses smaller pixels for the 
450~$\mu$m maps (2\arcsec) than the 850~$\mu$m maps (3\arcsec) to account for 
the smaller beamsize of the former.  Therefore, we expect that for structures of
equal size and the same peak brightnesses signal to noise ratios, recovery will
be poorer in the 450~$\mu$m map\footnote{Source recovery testing at 450~$\mu$m is
also complicated by the fact that we apply the 850~$\mu$m-based mask for the 
450~$\mu$m data reduction.}.

\section{Analysis: Artificial Source Recovery}
\label{sec_completeness_results}

In our analysis below, 
we examine the final reduced mosaics to determine the effectiveness of each reduction
in recovering the artificial Gaussians introduced into the raw-data time stream. 
The quantitative metrics that we examine are the fraction of Gaussians recovered,
as well as the recovered peak flux, total flux, and size compared with the input values,
as well the recovered axial ratio and offsets in the recovered peak position.
We also note that our artificial source recovery also gives us a tool not available for 
normal observations.  By comparing the reduced maps with and without the artificial sources
added to the raw-data time stream, we can measure precisely how much flux each artificial
source contributes to the final reduced map.  The analysis of the difference maps
is presented in Appendix~B.

\subsection{Recovery Rate}
\begin{figure}[htb]
\begin{tabular}{cc}
\includegraphics[width=3in]{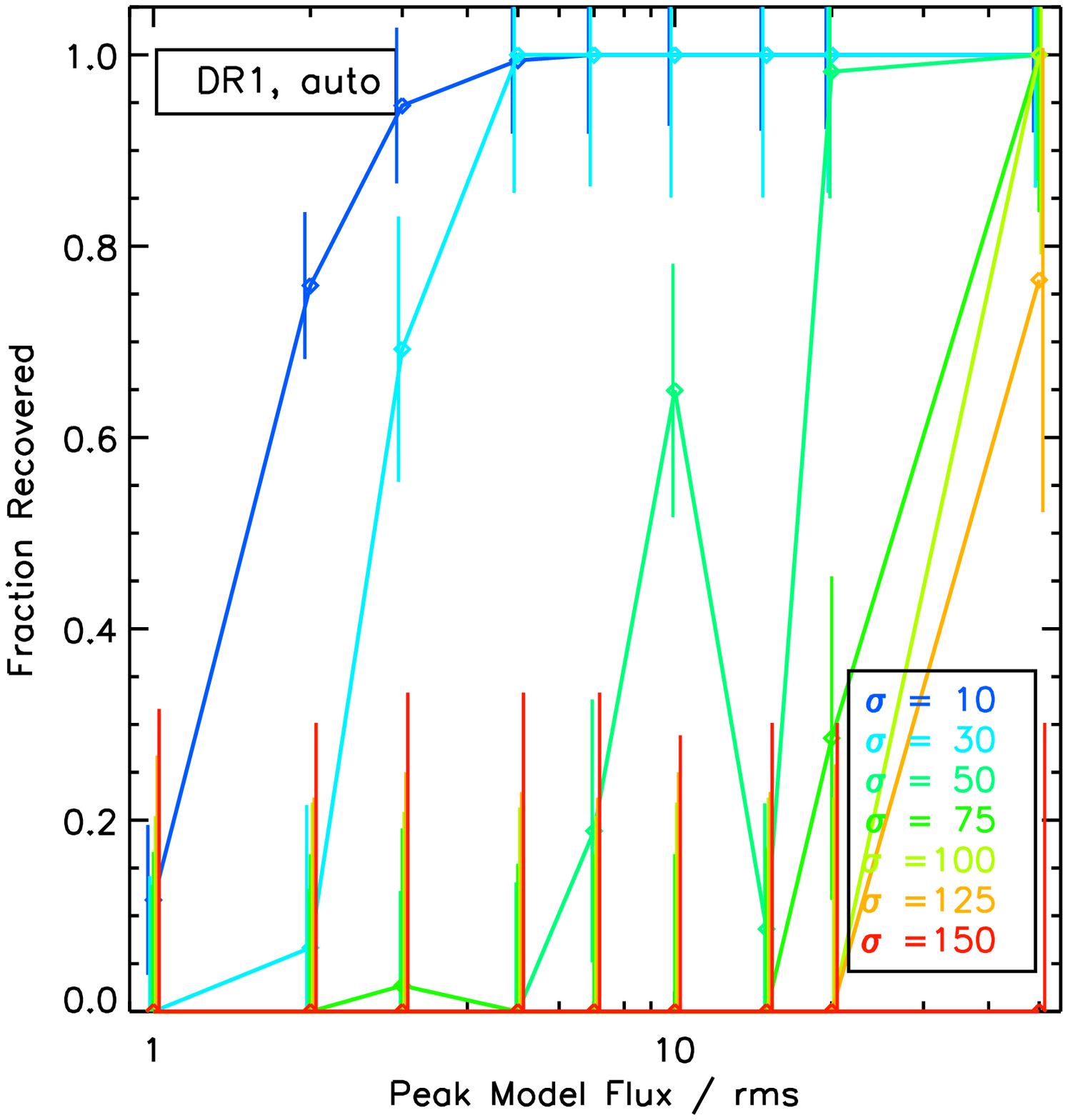} &
\includegraphics[width=3in]{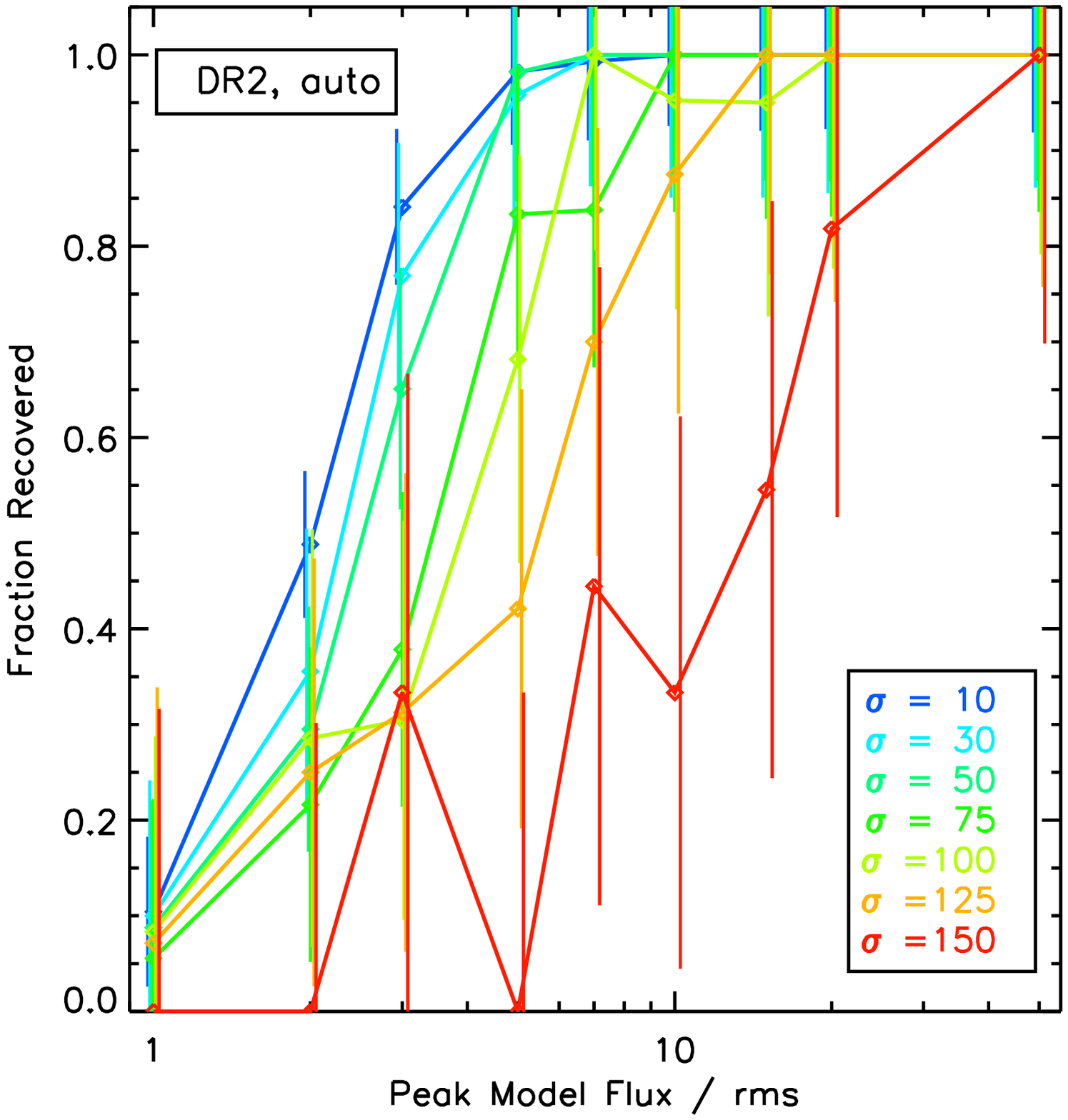} \\
\includegraphics[width=3in]{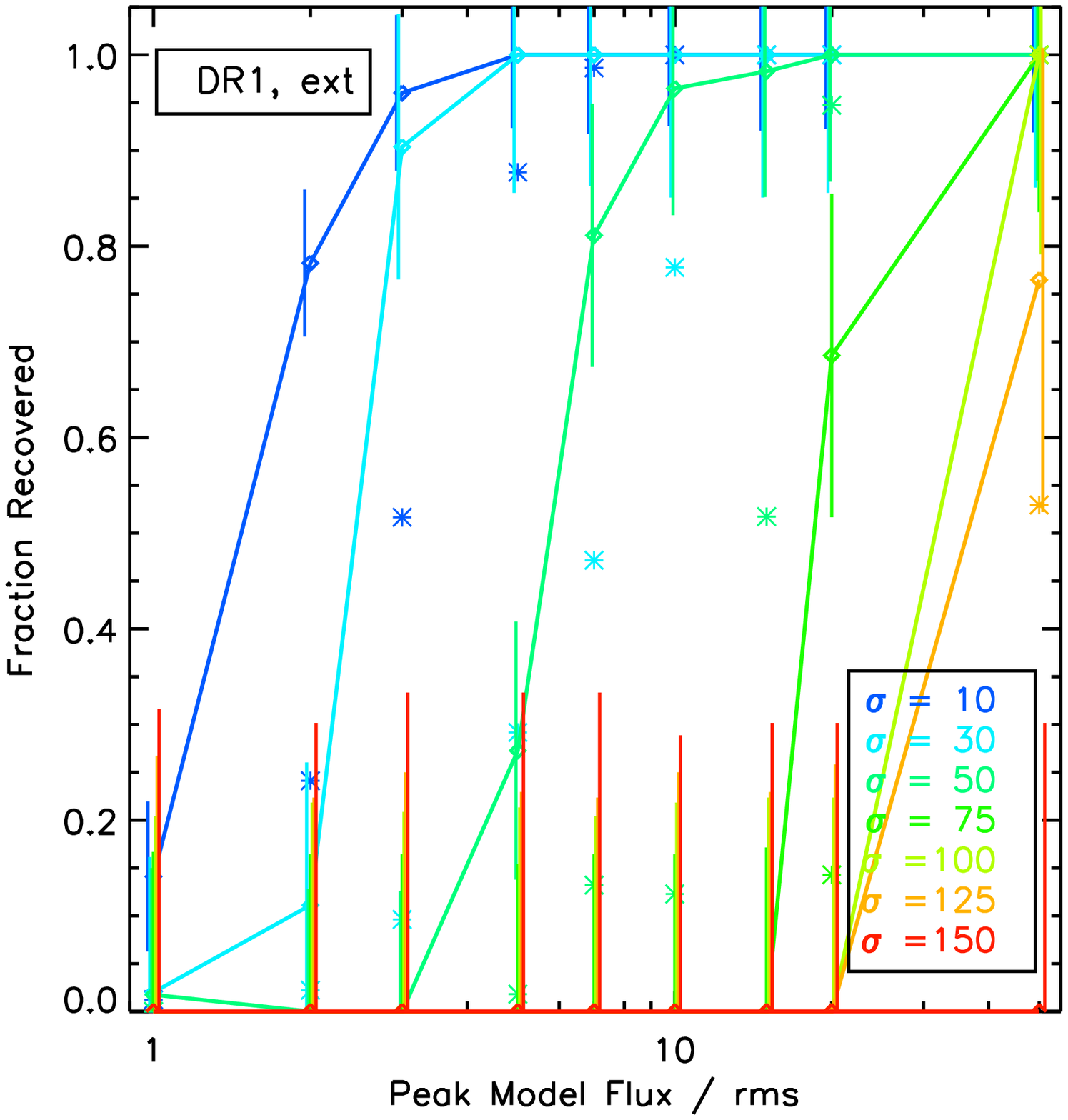} &
\includegraphics[width=3in]{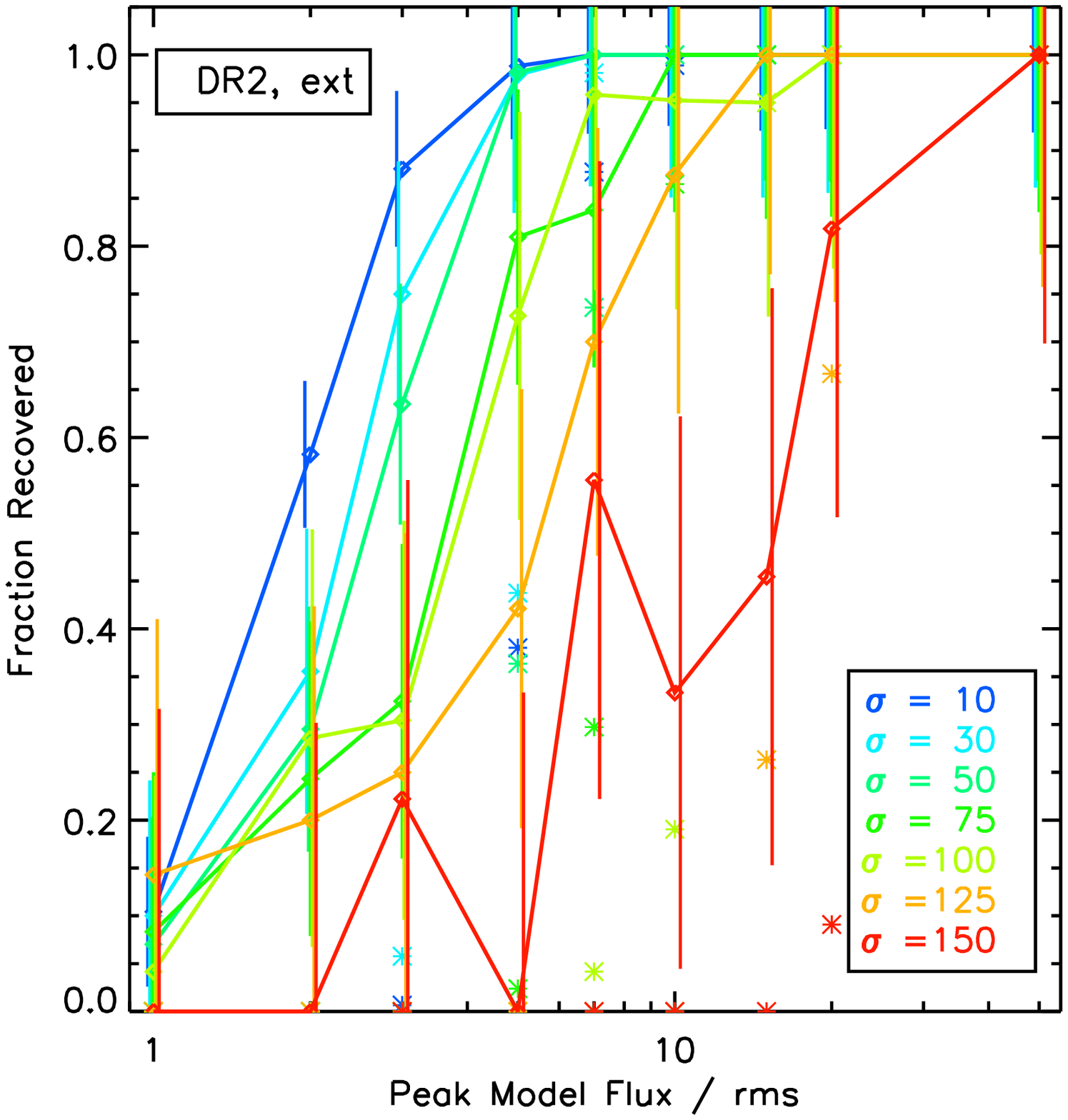} \\
\end{tabular}
\caption{The fraction of artificial sources recovered in each reduction method.  Top row:
	automask reductions for DR1 (left) and DR2 (right).  Bottom row: external-mask
	reductions for DR1 (left) and DR2 (right).  In each panel, the fraction of sources
	recovered is shown versus the input source amplitude, in units of the mosaic rms
	(0.049~mJy~arcsec$^{-2}$).  Each colour shows sources with a 
	different input width (Gaussian $\sigma$ values are shown in the legend at the bottom right, 
	in arcseconds).  The error bars denote the Poisson error for each input Gaussian
	test case (the square root of the number of input Gaussians).  The diamonds denote
	all recovered sources, while the asterisks denote recovered sources which 
	lie within the external mask (bottom panels only).}
\label{fig_completeness}
\end{figure}

The first metric that we analyze is the recoverability of the artificial Gaussians in the 
final maps.  
We emphasize that this recovery rate is an upper limit to the detection rate that would
be possible to measure in real observations, where source properties are not known and 
complications such as source crowding exist.
Figure~\ref{fig_completeness} shows the fraction of sources recovered
versus the peak flux for Gaussians of various widths using different data-reduction
methods.  
Bright and compact sources are always recovered, regardless of the
reduction method.  Compact sources become
poorly recovered only at extremely low intensity levels, i.e., peak amplitudes of one or
two times the mosaic rms.  On the other hand, extended sources are 
more difficult to recover.  
We also examine the subset of recovered sources that lie within the mask, whose properties
are expected to be better recovered (in the external-mask reduction), 
as demonstrated in \citet{Mairs15}.  
Some of the recovered sources are only marginally brighter than the local
noise level, and in many of these cases, are sufficiently faint that they did
not satisfy the masking criteria we adopted.  Therefore, we find that the total
source recovery rate is poorer for sources which lie within the mask, although it
follows the same general trend as the full set of recovered sources 
(i.e., a higher recovery rate for brighter and more
compact input Gaussians). 
The recovery rate of sources which lie within a mask is generally a better
representation of the detection rate of sources which could be confidently 
identified in real observations, however, an important exception is that moderately bright
but very compact sources may have too few pixels to satisfy the DR2 masking criteria,
even though they are clearly detectable.
A comparison of the DR1 (left column) and DR2 (right column) reductions in
Figure~\ref{fig_completeness} shows that the
latter is much better at retaining larger sources in
the automask and external-mask reductions.  For large ($\sigma \ge 100$\arcsec) and
bright (input peaks $\ge 10 \times$ rms) input Gaussians, we typically recover at
least twice as many of the Gaussians in DR2 maps as we can in DR1 maps.

In Figure~\ref{fig_completeness}, we also see slight improvement in the 
fraction of recovered sources between the automask and external-mask reductions.  
We generally expect the external-mask reductions to improve the reliability 
of recovered source properties (as examined
in the following sections), rather than the recovery fraction itself.  Indeed, for
a source to be included in the external mask, by definition it must be visible already in
the automask reduction.  Therefore, the similarity in source recovery fractions between 
automask and external-mask reductions is expected.  We attribute the marginal difference
in recovered sources in the external-mask reduction to added sources near
our detection limit and do not consider the difference in source recoveries to
be significant.  These faint or extended Gaussians are 
barely distinguishable from the background-map noise, even with our generous 
recovery criteria.  
As also noted in \citet{Mairs15}, 
the recovery rate for sources in the GBS DR1 map should therefore be similar to
the JCMT LR1 maps, which are similar to the GBS DR1 automask reductions.

Our source recovery rates compare favourably with the JCMT Galactic Plane Survey (JPS),
a JCMT Legacy Survey which focussed on mapping 850~$\mu$m emission of large areas
of the Galactic Plane using the larger PONG3600 mapping mode.  
\citet{Eden17} ran a series of completeness tests injecting artificial
Gaussians of FWHM = 21\arcsec\ ($\sigma\sim9$\arcsec) with a range of peak brightnesses,
using the CUPID source detection algorithm {\it FellWalker} to measure their
observable properties.
They report a 90\% to 95\% detection rate for sources with peak fluxes
of 5 or more times the noise level, and did not test the detection rate for
larger sources.  For comparison, we recover 100\% of the $\sigma = 10$\arcsec\
sources with peak fluxes of 5 or more times the noise in both external mask reductions. 

Table~\ref{tab_complete} summarizes the percentage of sources recovered in each
of the external-mask reductions, as a function of data-reduction method and input 
artificial Gaussian parameters.  In Table~\ref{tab_complete}, the left hand portion
of the table provides statistics for all recovered sources, while the right hand
portion provides statistics for the subset of recovered sources within an external
mask.  We again emphasize that these values represent upper limits to the {\it observable}
detection rate, where a blind search is run on sources with varying levels of crowding.

\begin{deluxetable}{cl|lllllllll|lllllllll}
\tablecolumns{20}
\tablewidth{0pc}
\tabletypesize{\scriptsize}
\tablecaption{Source Recovery for External Mask Reductions\label{tab_complete}}
\tablehead{
\colhead{} &
\colhead{} &
\multicolumn{18}{c}{Percentage of Sources Recovered (\%)}\\
\hline
\colhead{DR Method} &
\colhead{$\sigma$\tablenotemark{a}} &
\multicolumn{9}{c}{Peak-all ($N_\mathrm{rms}$)\tablenotemark{b}} &
\multicolumn{9}{c}{Peak-mask ($N_\mathrm{rms}$)\tablenotemark{c}}  \\
\colhead{}&
\colhead{(\arcsec)} &
\colhead{1} & 
\colhead{2} & 
\colhead{3} & 
\colhead{5} & 
\colhead{7} & 
\colhead{10} & 
\colhead{15} & 
\colhead{20} &
\colhead{50} &  
\colhead{1} &
\colhead{2} &
\colhead{3} &
\colhead{5} &
\colhead{7} &
\colhead{10} &
\colhead{15} &
\colhead{20} & 
\colhead{50} 
}

\startdata
 DR1 &  10 &   14  &   78  &   96  &  100  &  100  &  100  &  100  &  100  &  100  &    1  &   24  &   51  &   87  &   98  &  100  &  100  &  100  &  100 \\
 DR1 &  30 &    2  &   11  &   90  &  100  &  100  &  100  &  100  &  100  &  100  &    0  &    2  &    9  &   29  &   47  &   77  &  100  &  100  &  100 \\
 DR1 &  50 &    1  &    0  &    0  &   27  &   81  &   96  &   98  &  100  &  100  &    0  &    0  &    0  &    1  &   13  &   12  &   51  &   94  &  100 \\
 DR1 &  75 &    0  &    0  &    0  &    0  &    0  &    0  &    0  &   68  &  100  &    0  &    0  &    0  &    0  &    0  &    0  &    0  &   14  &  100 \\
 DR1 & 100 &    0  &    0  &    0  &    0  &    0  &    0  &    0  &    0  &  100  &    0  &    0  &    0  &    0  &    0  &    0  &    0  &    0  &  100 \\
 DR1 & 125 &    0  &    0  &    0  &    0  &    0  &    0  &    0  &    0  &   76  &    0  &    0  &    0  &    0  &    0  &    0  &    0  &    0  &   52 \\
 DR1 & 150 &    0  &    0  &    0  &    0  &    0  &    0  &    0  &    0  &    0  &    0  &    0  &    0  &    0  &    0  &    0  &    0  &    0  &    0 \\
\hline
 DR2 &  10 &   10  &   58  &   88  &   98  &  100  &  100  &  100  &  100  &  100  &    0  &    0  &    0  &   38  &   87  &   98  &  100  &  100  &  100 \\
 DR2 &  30 &   10  &   35  &   75  &   97  &  100  &  100  &  100  &  100  &  100  &    0  &    0  &    5  &   43  &   98  &  100  &  100  &  100  &  100 \\
 DR2 &  50 &    7  &   29  &   63  &   98  &  100  &  100  &  100  &  100  &  100  &    0  &    0  &    0  &   36  &   73  &  100  &  100  &  100  &  100 \\
 DR2 &  75 &    8  &   24  &   32  &   80  &   83  &  100  &  100  &  100  &  100  &    0  &    0  &    0  &    2  &   29  &   86  &  100  &  100  &  100 \\
 DR2 & 100 &    4  &   28  &   30  &   72  &   95  &   95  &   95  &  100  &  100  &    0  &    0  &    0  &    0  &    4  &   19  &   95  &  100  &  100 \\
 DR2 & 125 &   14  &   20  &   25  &   42  &   70  &   87  &  100  &  100  &  100  &    0  &    0  &    0  &    0  &    0  &    0  &   26  &   66  &  100 \\
 DR2 & 150 &    0  &    0  &   22  &    0  &   55  &   33  &   45  &   81  &  100  &    0  &    0  &    0  &    0  &    0  &    0  &    0  &    9  &  100 \\
\enddata
\tablenotetext{a}{The Gaussian width, sigma, of the inserted artificial Gaussians.}
\tablenotetext{b}{The peak flux of the inserted artificial Gaussians, given in
	units of the rms noise of the map.  The source recovery fractions listed in these
	columns give all of the recoveries within the map.}
\tablenotetext{c}{The peak flux of the inserted artificial Gaussians, given in
	units of the rms noise of the map.  The source recovery fractions listed in these
	columns give only the recoveries that lie within the external mask.}
\end{deluxetable}

\subsection{Recovered Properties: Peak Flux}
For the artificial Gaussians which were recovered, we now examine how well their
measured properties match the input properties.  We measure the
mean and standard deviation of the recovered Gaussian fit values and compare them to
the input Gaussian values.  
Table~\ref{tab_recov_props} summarizes the recovered Gaussian properties for each
of the external-mask reductions.  For each reduced map, we report on the mean
and standard deviation of the fraction of the measured Gaussian property with the
initial input value.  Table~\ref{tab_recov_props} includes the peak flux (discussed
here), as well as the Gaussian width $\sigma$ and the total flux (discussed in the
following sections).

\begin{figure}[htb]
\begin{tabular}{cc}
\includegraphics[width=3in]{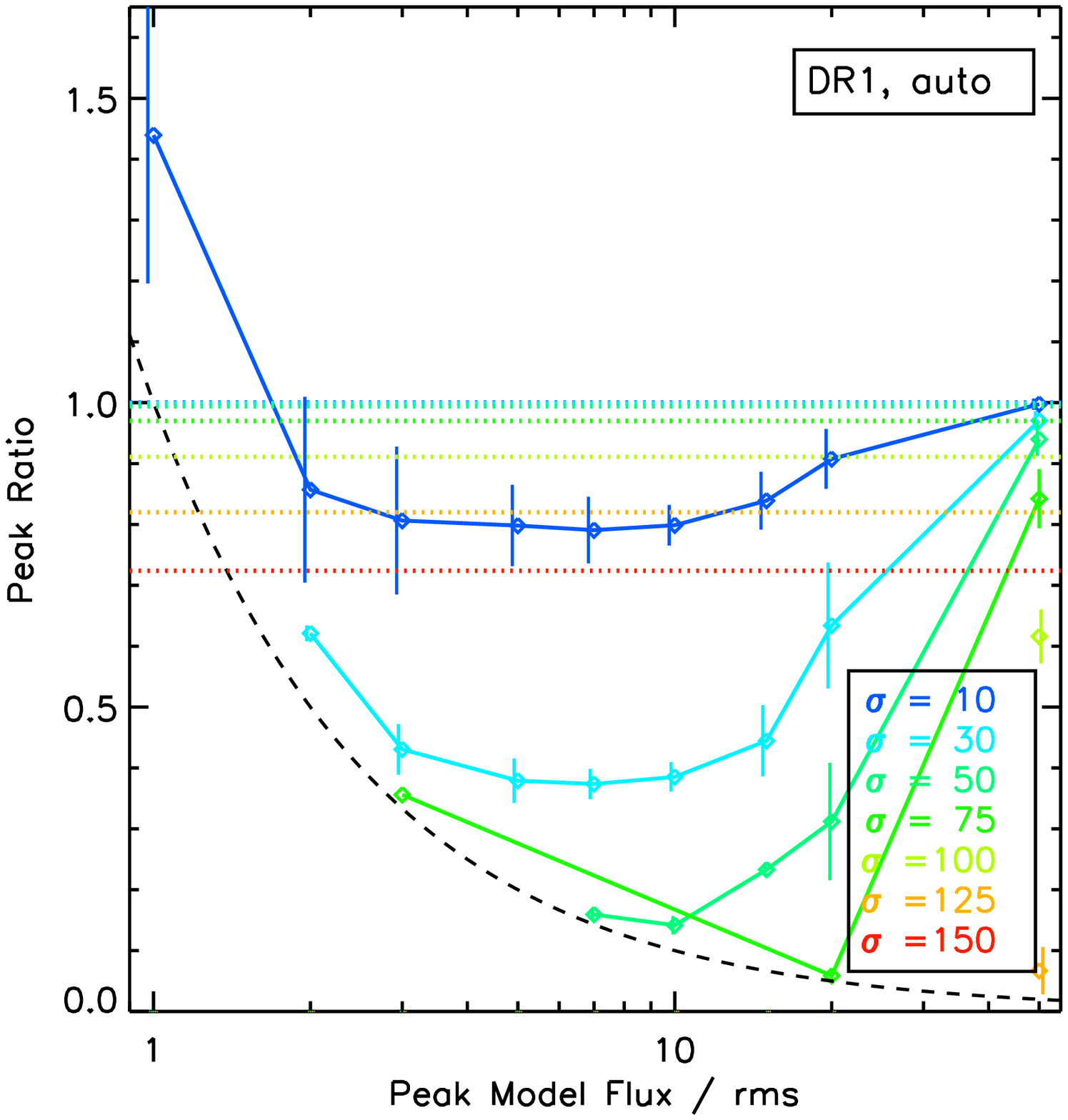} &
\includegraphics[width=3in]{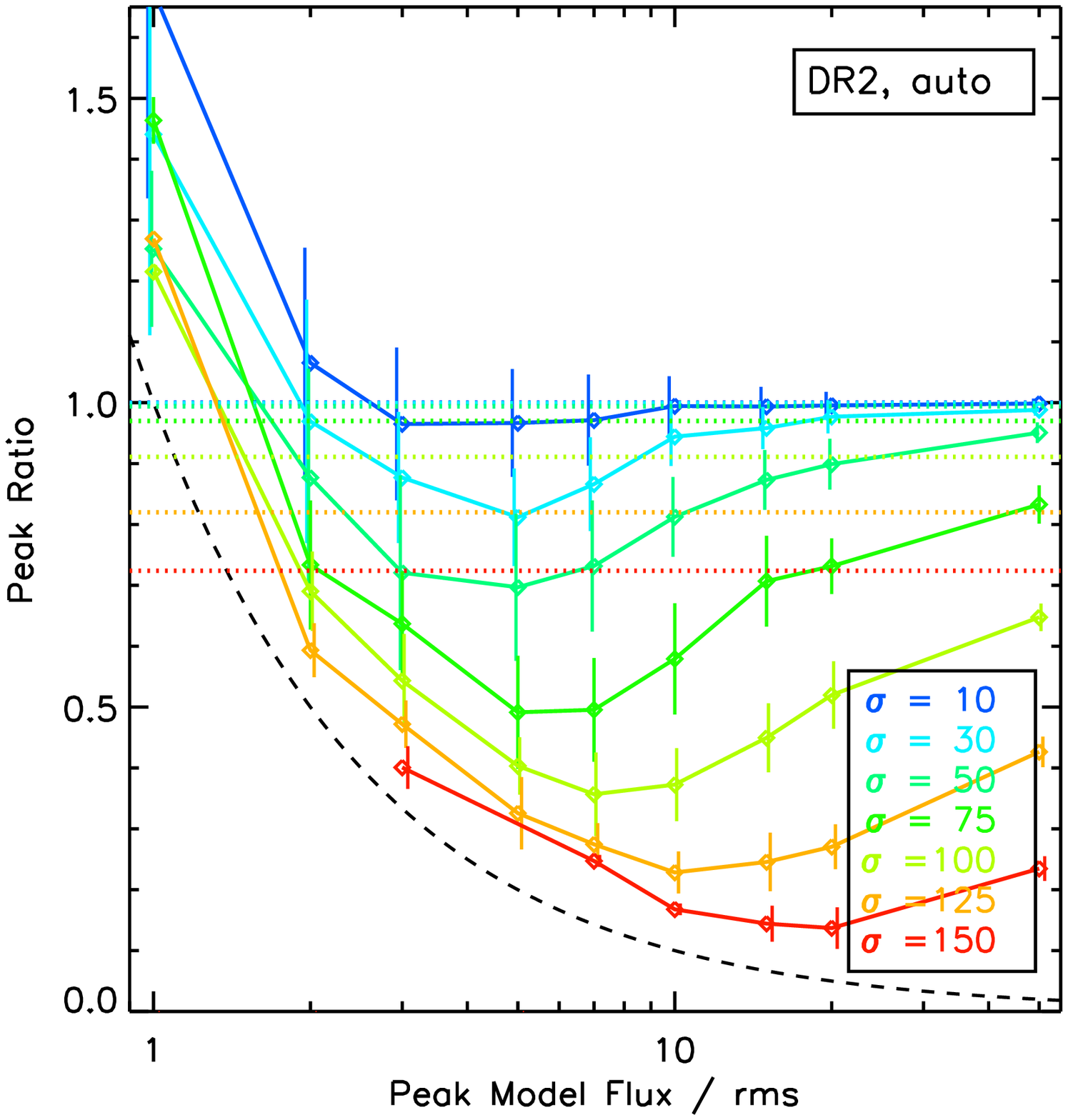} \\
\includegraphics[width=3in]{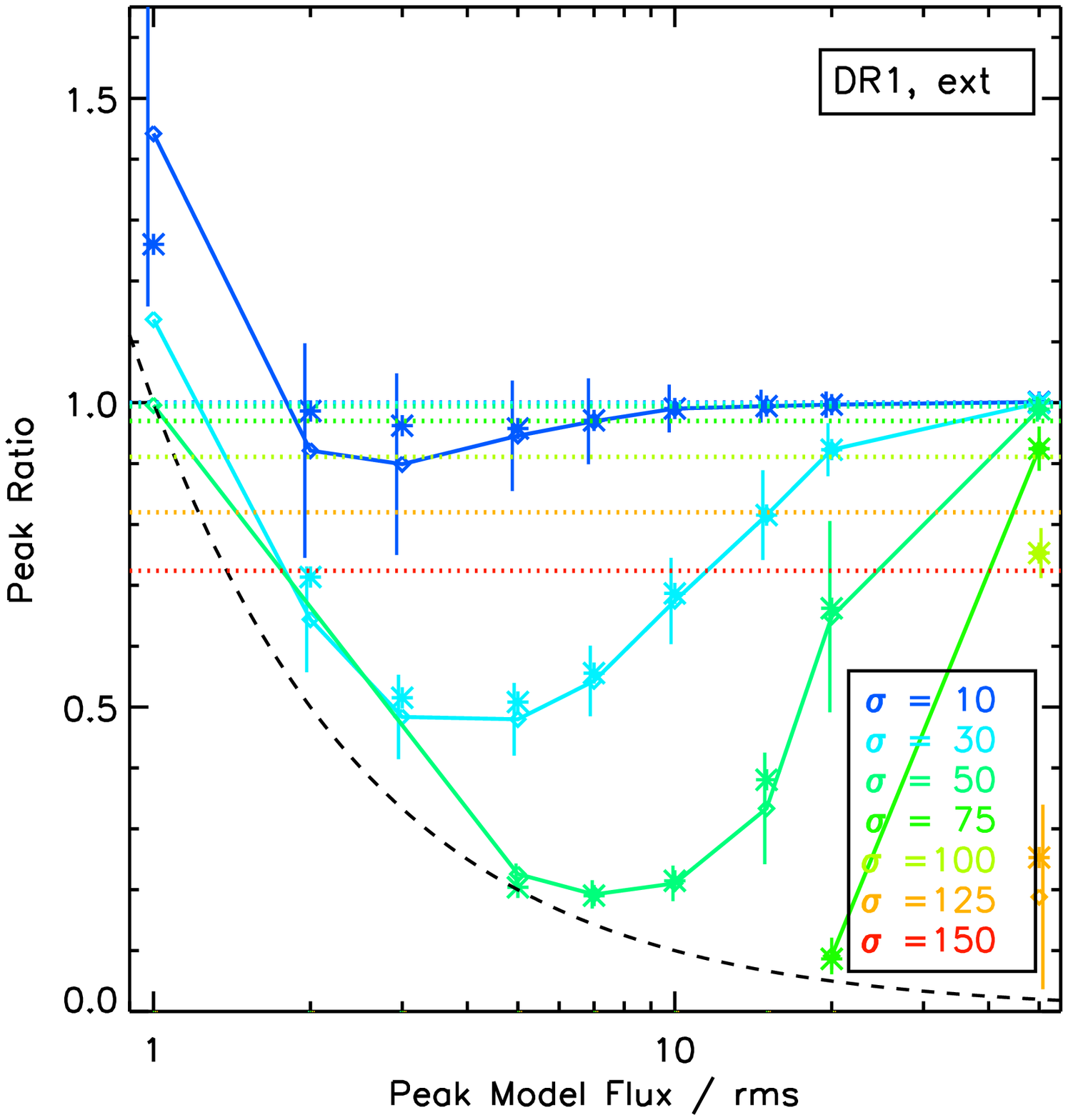} &
\includegraphics[width=3in]{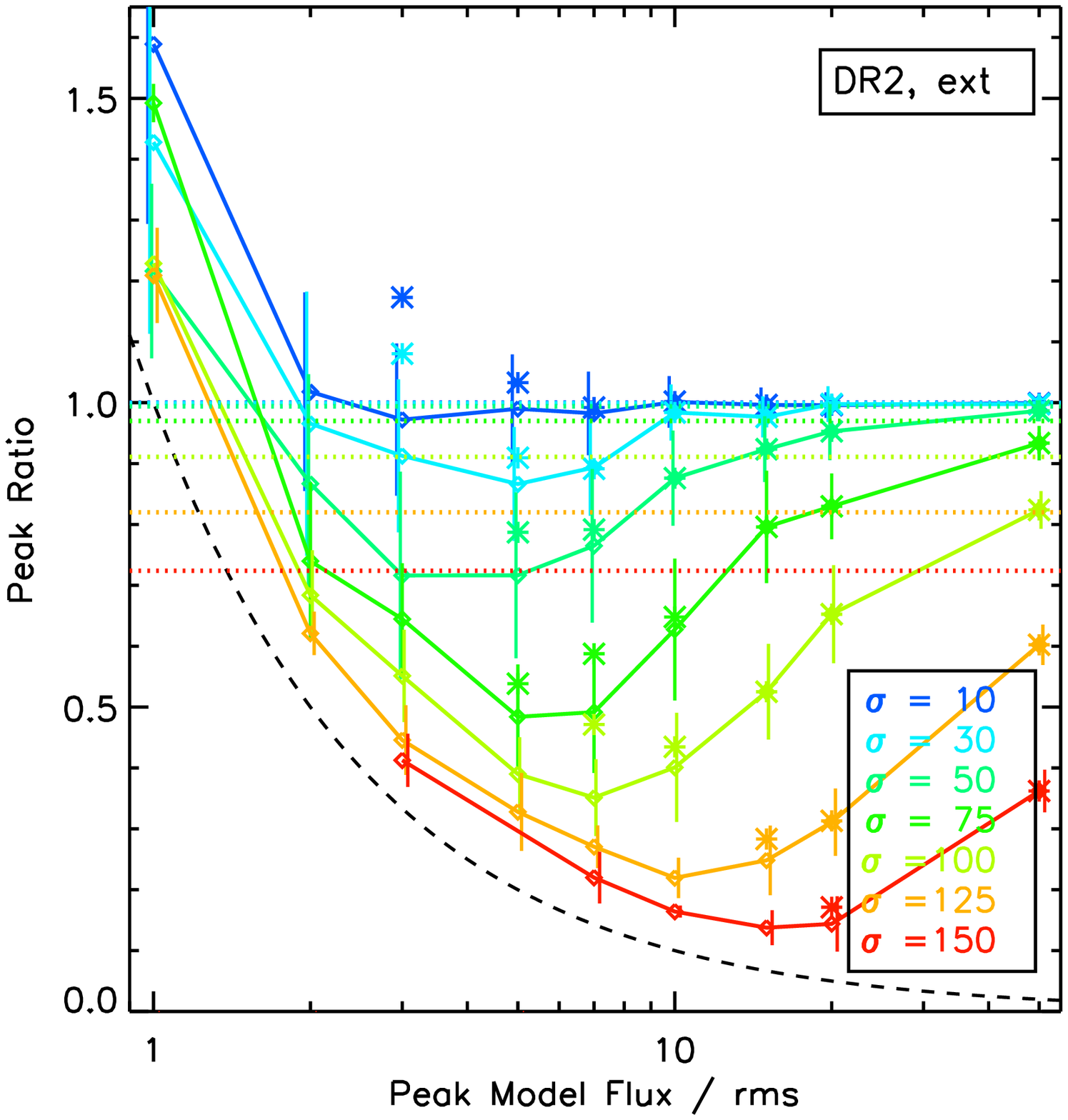} \\
\end{tabular}
\caption{Peak brightnesses measured for the recovered artificial Gaussian sources, as a fraction
	of their input values.  As in Figure~\ref{fig_completeness}, the four panels show the
	automask (top panels) and external-mask (bottom panels) reductions using
	the GBS DR1 procedure (left panels) and DR2 procedure (right panels).  The horizontal
	axis indicates the different input peak values tested, while the colours indicate the
	different Gaussian widths tested.  The error bars 
	indicate the standard deviation in
	values measured for each set of Gaussians.  The black dashed line indicates the
	expected peak flux values where the recovered peak has a value of the local rms noise.
	The dotted horizontal coloured lines represent model values for 600\arcsec\ 
	filtering.  Diamonds denote values for all recovered sources, while asterisks
	denote values for sources lying within the mask (for the external-mask reduction 
	only).  See the text for details.}
\label{fig_peaks}
\end{figure}

Figure~\ref{fig_peaks} shows how well the peak flux is recovered 
for all recovered sources.  For the faintest input Gaussians, the recovered
peak flux is typically larger than the input value, i.e., the peak-flux ratio is
above one.  Faint 
artificial sources with peak fluxes near the typical noise level in the map
(one or two times the rms) are easier to recover when these sources are coincident with
positive noise features in the map.  We therefore expect the recovered faintest peaks
to have peak fluxes biased towards higher values.
\citet{Eden17} report a similar behaviour for their completeness testing in the JPS maps.
The black dashed curve in Figure~\ref{fig_peaks} shows the approximate effect of this bias
by showing a measurement of a peak flux equal to the local rms.  While not identical,
the shape of this curve gives a reasonable approximation of the measured peak flux
ratios at low input peak flux values for DR2.

Figure~\ref{fig_peaks} also shows that the peak fluxes are better recovered for
compact sources than larger sources.  Larger sources have a great fraction of their
flux at larger size scales, and are thus expected to be more sensitive to filtering.
We constructed a simple model of the large-scale spatial 
filtering that occurs during data
reduction to see how well it predicts the observed source recovery behaviour.
Accordingly, we created a series of two-dimensional Gaussian models matching 
our artificial Gaussian sources.  We approximate the filtering as a 
single-scale boxcar smoothed version of the model being subtracted from the original.
We fit the resulting filtered
model with a two-dimensional Gaussian (including a constant zero-point term to
alleviate fitting challenges with slight negative bowling)
to calculate the fractional reduction in peak flux
and size.  

Previous tests of the initial data-reduction method employed by the GBS 
(IR1, not examined
here) suggested that source recovery was consistent with a simple single filtering
scale of about 1\arcmin.  The subsequent data-reduction methods examined here
(DR1 and DR2) were expected to recover more emission, i.e., be described by a larger
filtering scale.  Our test results confirm the larger scale of filtering, although we also
find that a single filter scale is insufficient to describe the recovered source properties
for the full range of artificial Gaussians tested.
Figure~\ref{fig_peaks} shows the predictions for recovered peak-flux ratio for
a filter scale of 600\arcsec\ (dotted horizontal lines).
This filter scale is equal to
the large-scale filtering formally applied during data reduction, via the 
{\tt flt.filt\_edge\_largescale=600} parameter.
Peak-flux ratios lying below the model line imply they have been subject to more filtering
than in the model, i.e., filtering on a smaller size scale.
The 600\arcsec\ filtering scale matches the smallest artificial Gaussian 
sources, of sizes of below about 75\arcsec\ 
for the external-mask reduction of DR2 -- i.e., the dashed filtering model curves
are a good match for the recovered peak flux ratios at the highest SNR values.
At the same time, the model clearly under-predicts the
amount of filtering for larger sources for that same reduction. 

Comparing the reductions, Figure~\ref{fig_peaks} clearly shows that DR2 recovers
more reliable peak flux values than DR1.  Also, although the difference is subtle, 
the external-mask 
reductions improve on the automask reductions, especially for the largest and
brightest of input Gaussians.  In cases where not all of the recovered sources lie
within the external mask, the subset of sources that are included in the mask tend
to have recovered peak fluxes which more closely correspond to the input value than
the full sample of sources do.  As an example, in DR2, a Gaussian with $\sigma$=100\arcsec\ 
and a peak flux of 10 times the noise has recovered peak fluxes of about 37\% of their
true value in the automask reduction, while this rises to roughly 40\% 
of their true value in the external-mask reduction for all sources, and 43\% for
those lying within the mask.
In contrast, no sources are reliably recovered in either the automask or
external-mask reduction of DR1 for these Gaussian properties.
As noted in \citet{Mairs15}, sources are only accurately recovered in the external-mask
reductions when the mask encompasses the true extent of the source.  
The superiority of the DR2 reductions over the DR1 reductions is therefore partly
attributable to the mask-making procedures, which better reflect the true source
extents in DR2 than in DR1.

\subsection{Recovered Properties: Sizes}
\begin{figure}[htb]
\begin{tabular}{cc}
\includegraphics[width=3in]{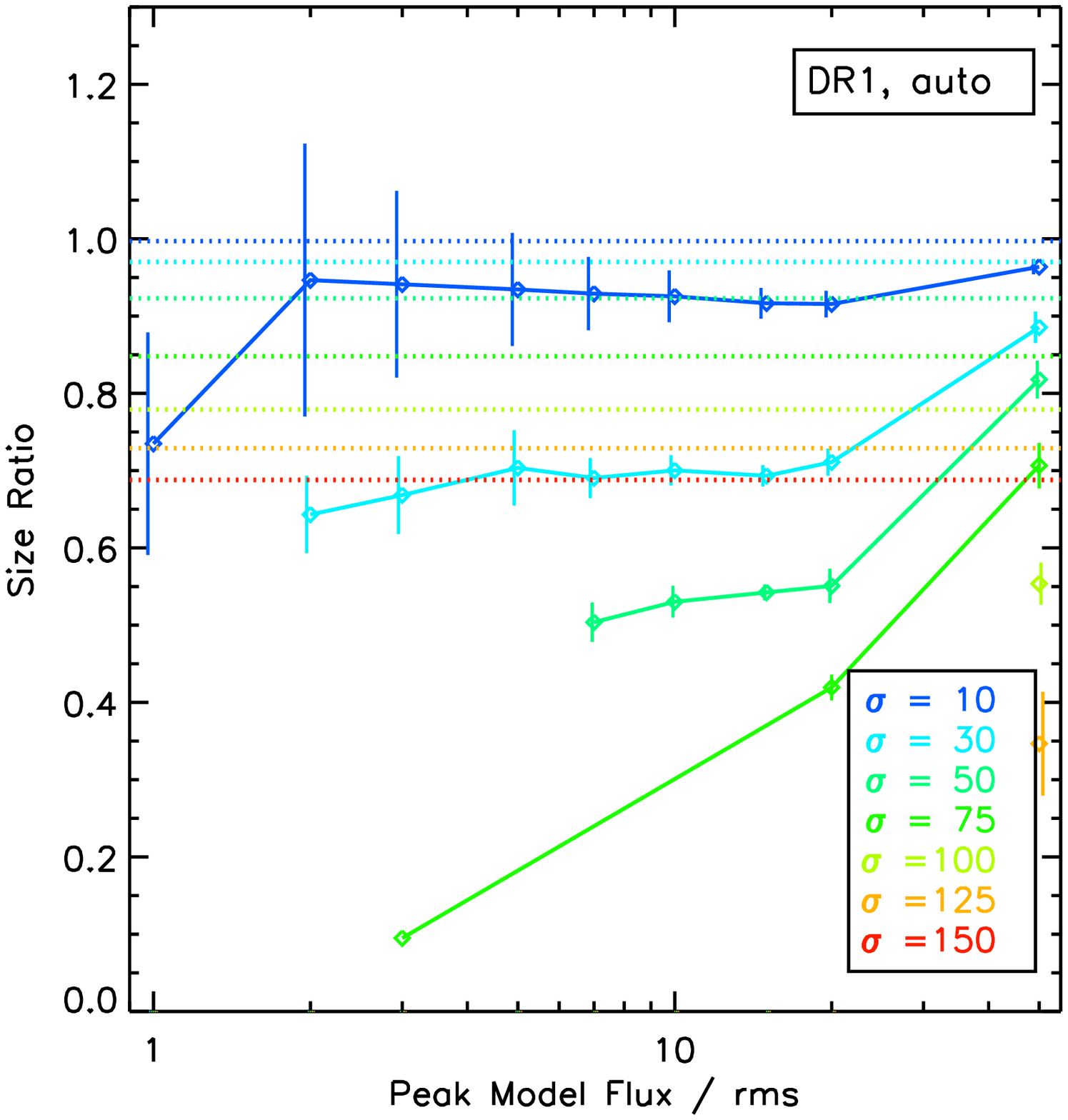} & 
\includegraphics[width=3in]{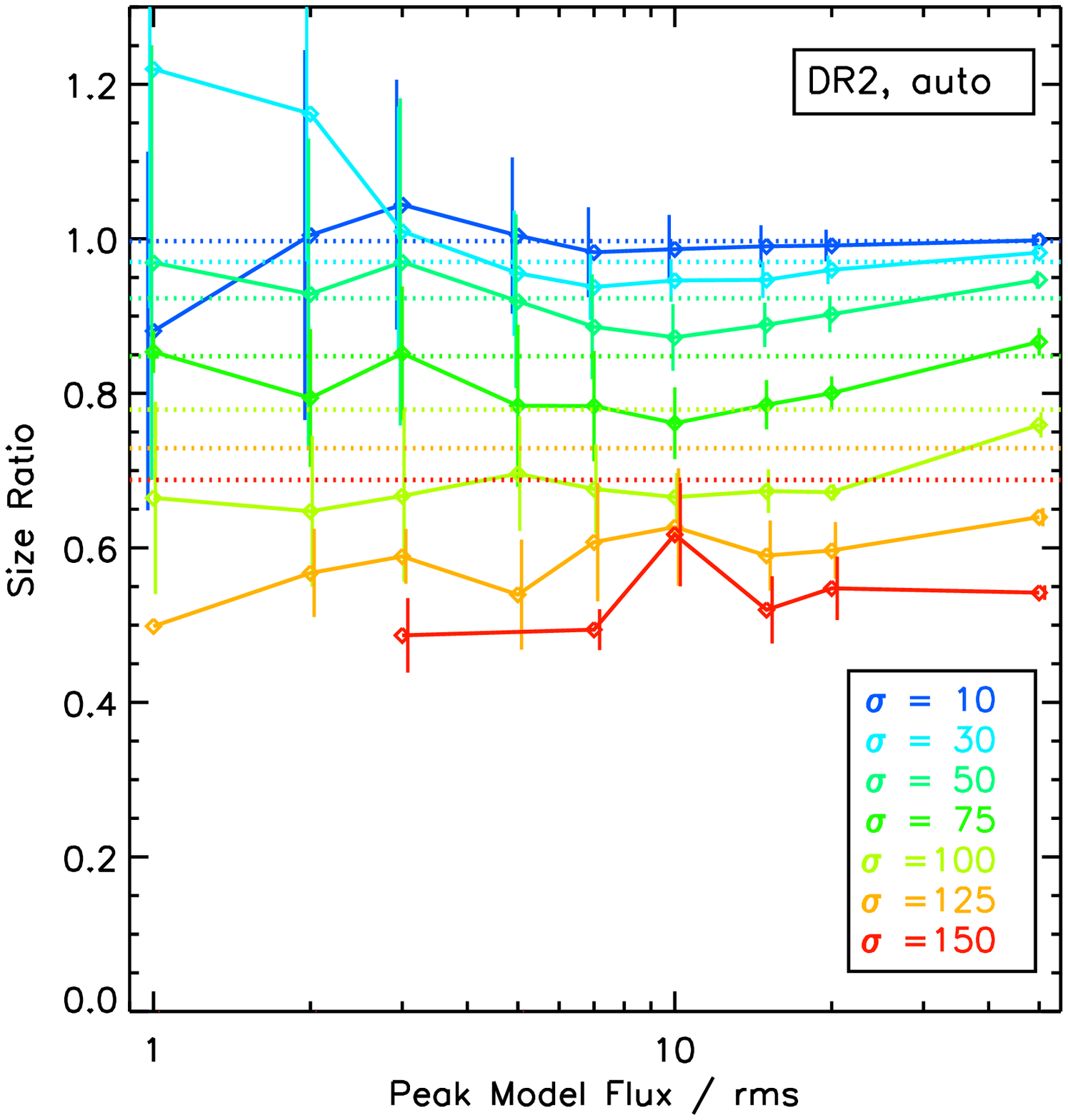} \\
\includegraphics[width=3in]{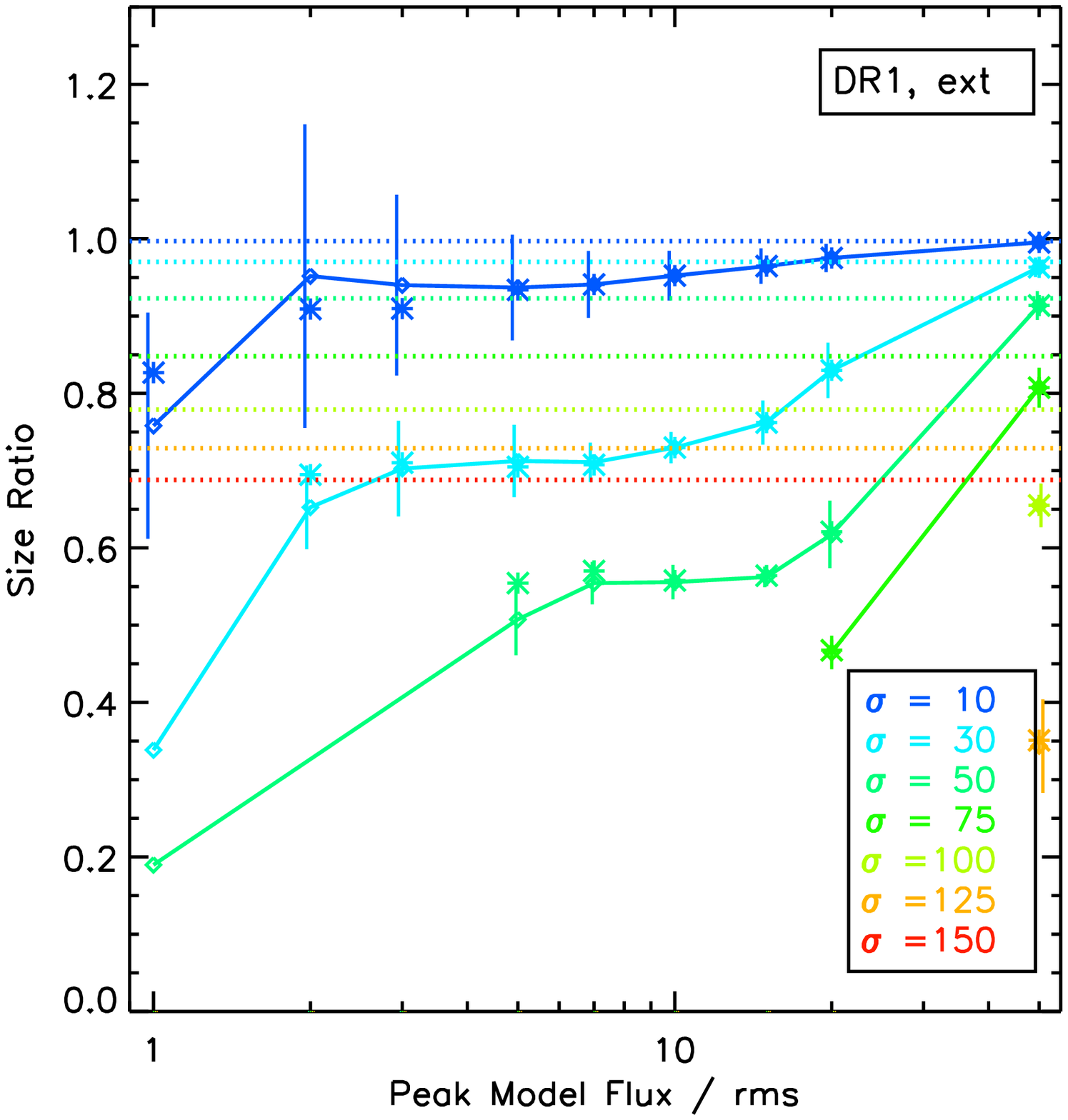} &
\includegraphics[width=3in]{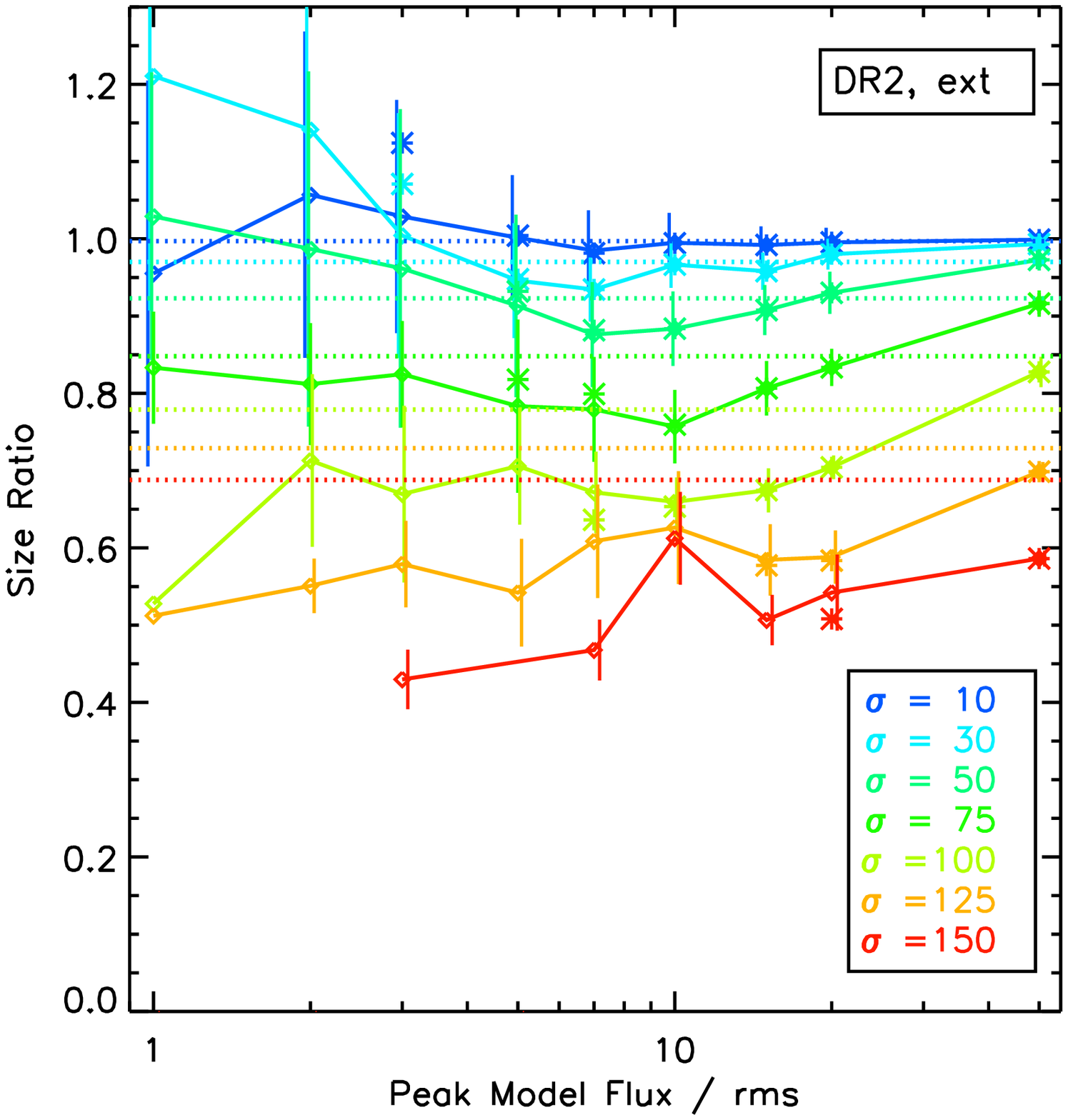} \\
\end{tabular}
\caption{Sizes measured for the recovered artificial Gaussian sources, as a fraction
	of the input values.  See Figure~\ref{fig_peaks} for the plotting conventions
	used.}
\label{fig_widths}
\end{figure}

Figure~\ref{fig_widths} shows the size ratios measured for
the artificial Gaussians.  We remind the reader that the input Gaussians were
all round, though we allowed fits for sources with axial ratios up to 1.5:1 (Section~3.3).
We present here a single size estimate based on the geometric mean of the two Gaussian
$\sigma$ values.  In general, we find similar results to those previously presented: (1)
compact and brighter Gaussians have their sizes recovered more accurately, (2) 
DR2 tends to show more reliable structures than DR1, and (3) the external-mask 
reduction is an improvement
over the automask reduction, particularly for sources which lie within a masked area.
Using a single size to describe the recovered sources is reasonable, as
they tend to be quite round, with mean axial ratios of no
more than $\sim$1.4:1 for any of the reductions.  For sources that are
bright or recovered within a mask (or both), the mean axial ratio is almost always
lower than 1.2:1, and the sources recovered in the DR2 reduction
furthermore tend to have lower axial ratios than those in the corresponding DR1 reduction.

As in Figure~\ref{fig_peaks}, we also consider the effects of filtering on the model
Gaussians.  The dashed lines in Figure~\ref{fig_widths} shows the ratio of the mean
measured filtered size to the input size for the grid of model Gaussians with which we
applied a 600\arcsec\ filter (see previous section for details).  As expected,
these models show that the more-extended model Gaussians have a greater reduction in size than
their more compact counterparts.  The filtering model predicts approximately the
size ratio for sources smaller than 50\arcsec\ (blue points and lines), 
but under-predicts
the amount of filtering for the largest sources (i.e., predicts size ratios which are
too large).

\subsection{Total Flux}
\begin{figure}
\begin{tabular}{cc}
\includegraphics[width=3in]{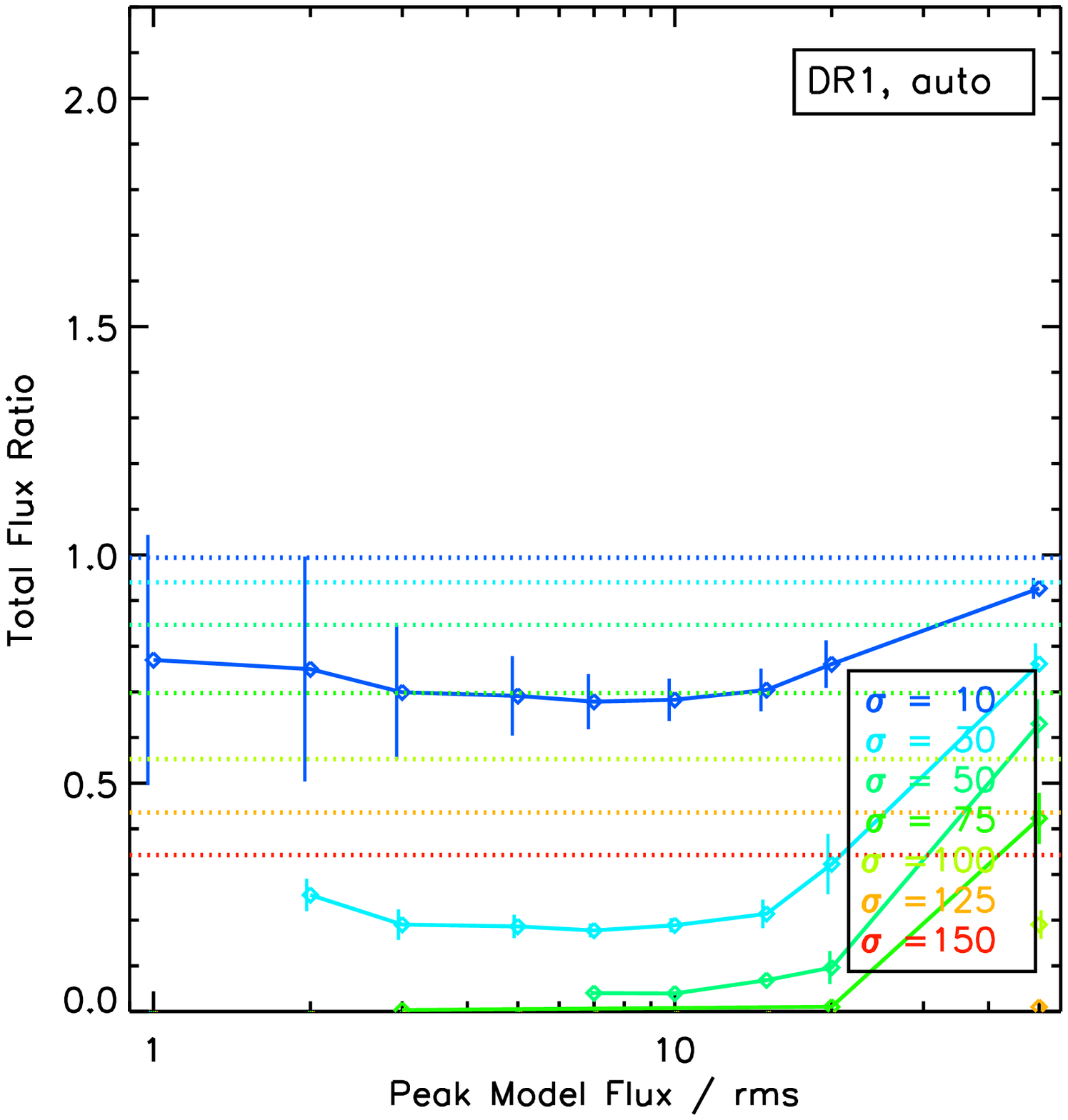} &
\includegraphics[width=3in]{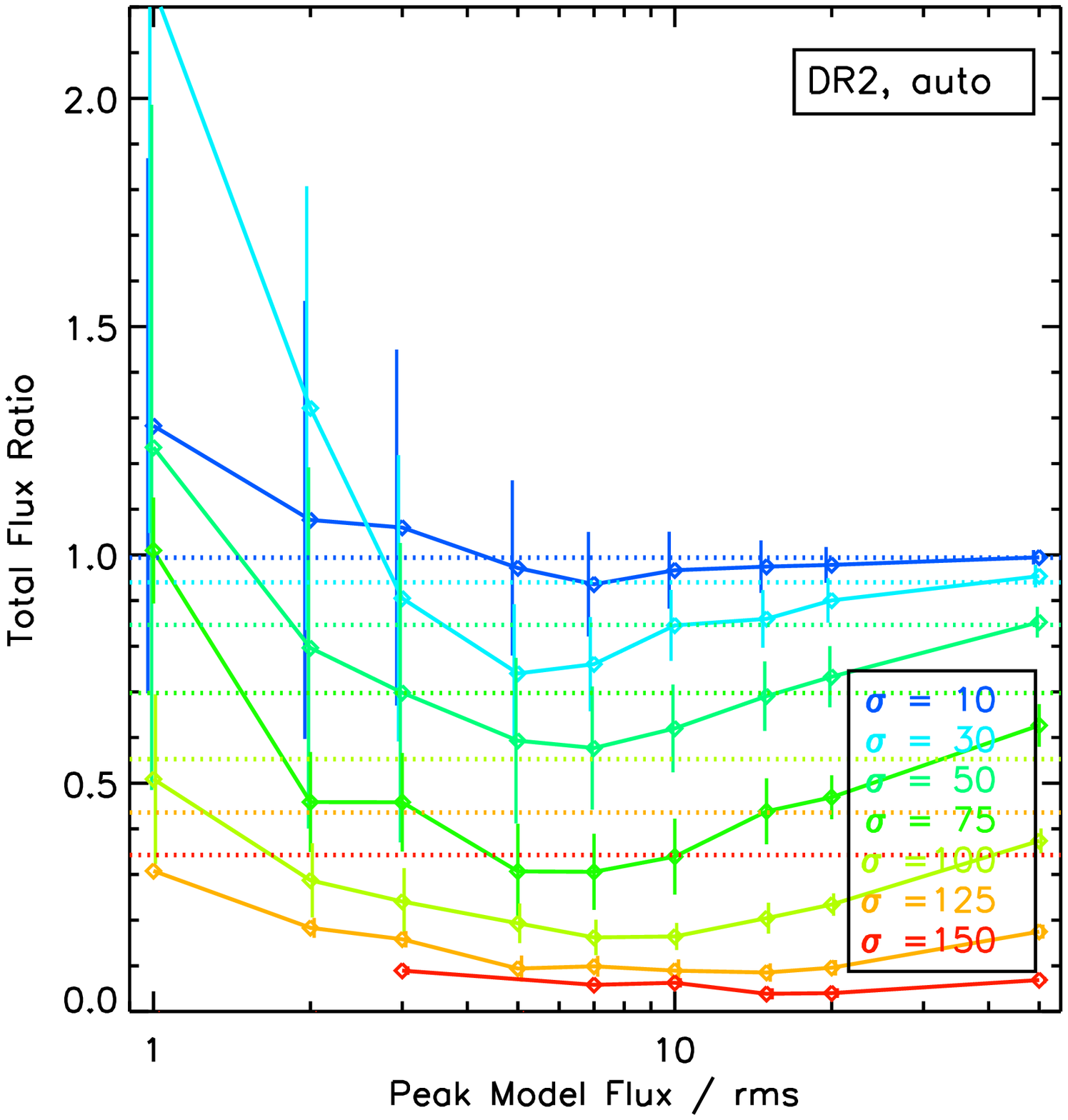} \\
\includegraphics[width=3in]{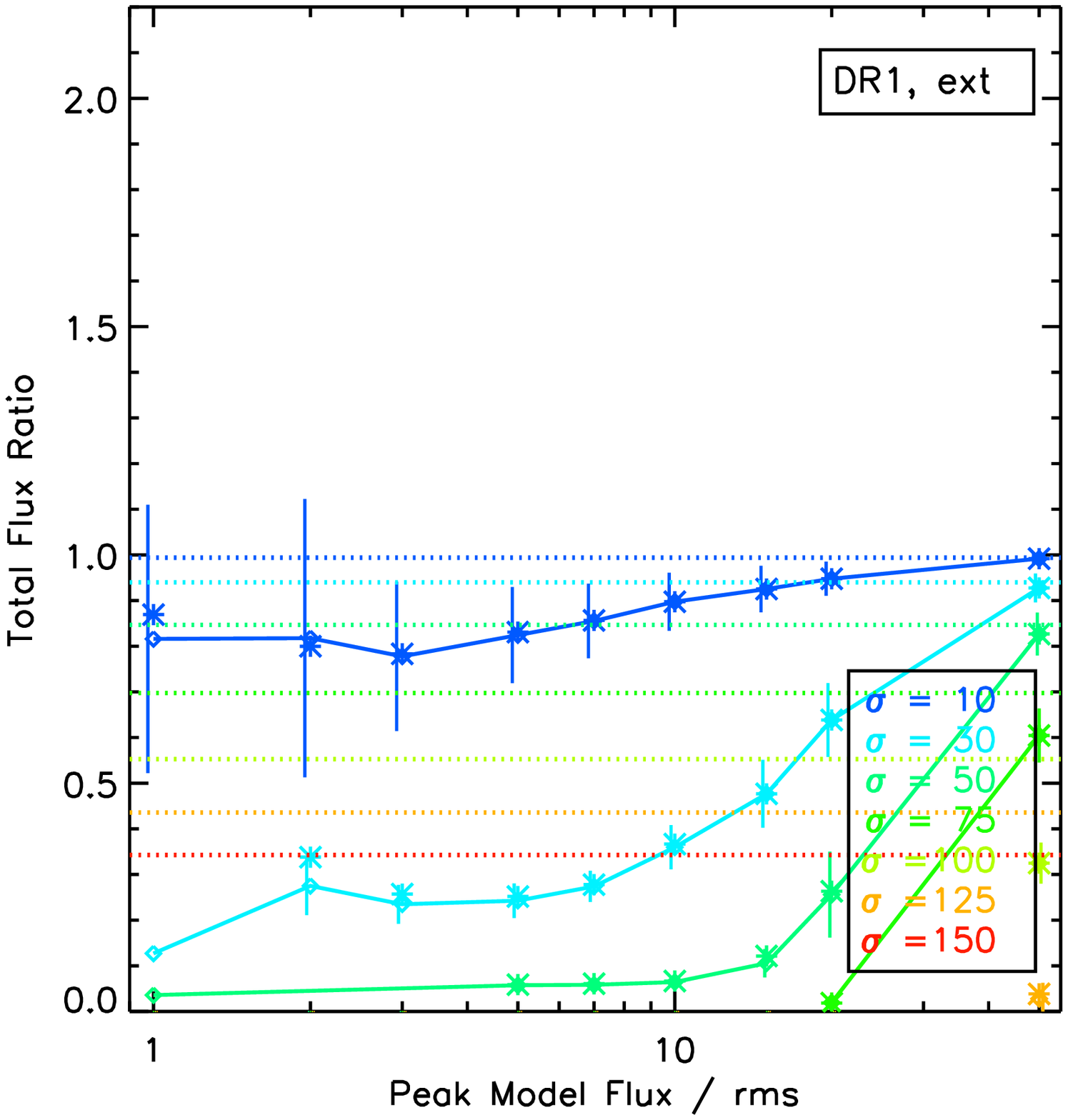} &
\includegraphics[width=3in]{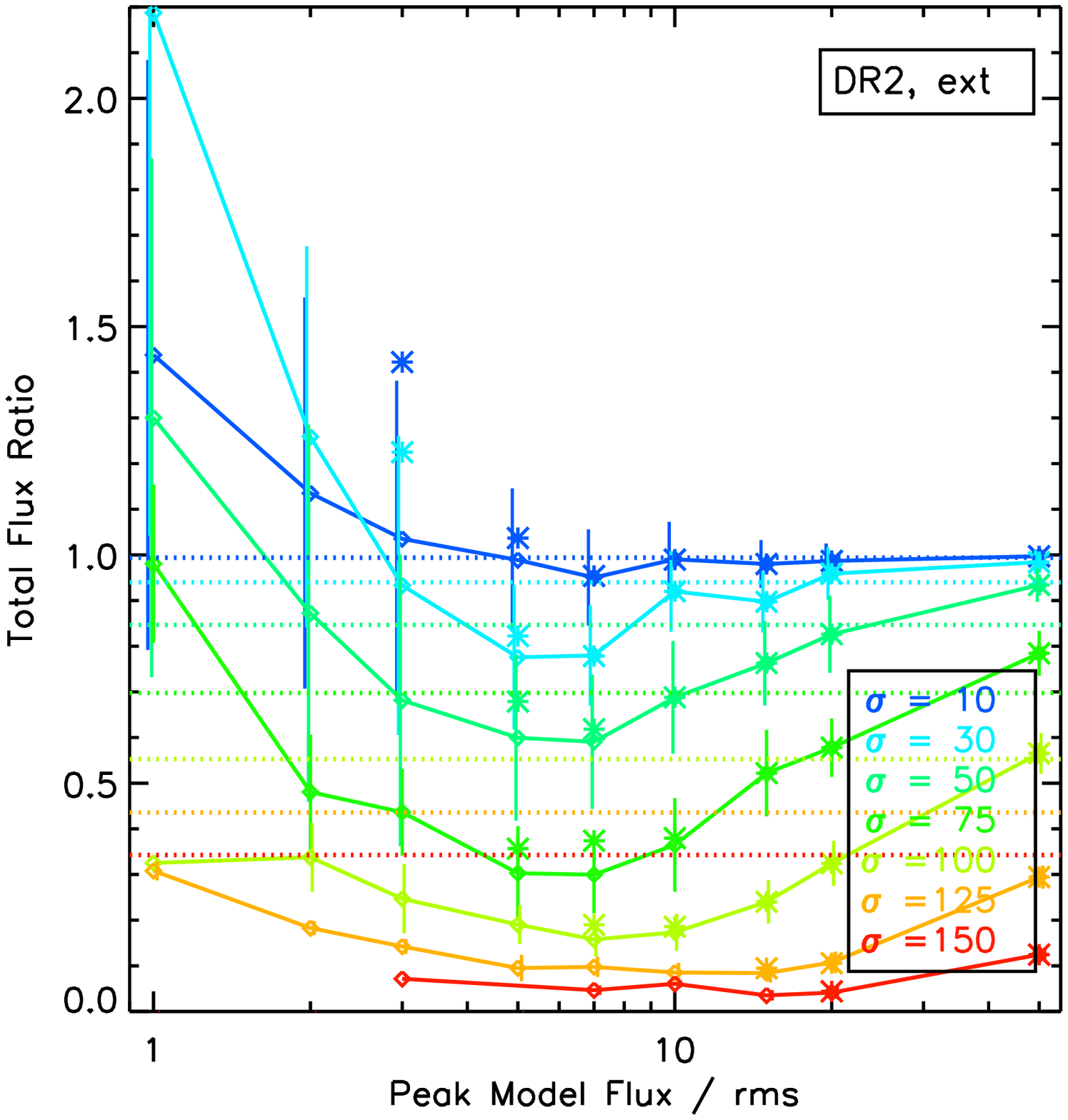} \\
\end{tabular}
\caption{Total fluxes recovered for the artificial Gaussian sources, as a fraction of the
	input values.  See Figure~\ref{fig_peaks} for the plotting conventions used.}
\label{fig_totflux}
\end{figure}

Here, we present results for the total flux recovered.  For individual recovered cores,
the trends discussed in the previous two sections (for peak flux and size) are at work.
Figure~\ref{fig_totflux} shows the total flux recovered.  The values shown in this plot
and Table~\ref{tab_recov_props} should be considered when analyzing dense core mass functions
in GBS data.  For compact sources ($\sigma \le 30$\arcsec) which are brighter than 3 times 
the local noise, the total fluxes are recovered to better than 25\% in DR2.

\subsection{Location}
We also examined the positional offset of the centre of the recovered Gaussian compared
with their true input location (not shown).  Since the central location is expected to
become less certain as the input Gaussian becomes larger, we measured the ratio of the
positional offset to the input Gaussian width, $\sigma$.  Using this measure, we find that the
offset ratios are typically small, with mean values of 0.3, i.e., offsets of no more than
30\% of $\sigma$, with significantly lower values obtained for model peaks of 10 or more
times the rms.  For the DR2 external-mask reductions, for cases where the
model peak is 10 or more times the rms, the mean offset ratio is
$\le$0.05 for all input $\sigma$.  For $\sigma = 30$\arcsec,
this implies a typical positional accuracy of better than 1\farcs5.  We therefore
find that the data-reduction and source-recovery processes do not 
typically induce significant shifts to the true source positions.

\subsection{Summary}
Our artificial-source recovery tests confirm that the GBS maps more reliably reproduce
true sky emission using the newer DR2 method than the earlier DR1 method, and that using
the two-step reduction 
process of automask reductions, mask creation, and external-mask reductions
also provides improvements over a single automask reduction, particularly for bright 
extended sources.  
Sources with $\sigma \gtrsim$ 100\arcsec\ are generally 
not recovered well, with
peak fluxes and sizes often recovered at values of less than half of their true values,
especially for fainter sources.  Compact sources, however, are well recovered.
For sources with $\sigma \le 30$\arcsec, peak fluxes and sizes are nearly always recovered
at better than 90\% of their true value for the DR2 external-mask reduction.  These compact
scales are of the greatest interest to the GBS, as they represent the typical scales of 
dense cores.  
We note that often analyses of dense cores discuss their sizes in terms of FWHM values
instead of Gaussian $\sigma$ widths; a core with $\sigma = 30$\arcsec\ has a corresponding 
FWHM of 71\arcsec.  Within the Gould Belt, where clouds are between $\sim$ 100~pc and 500~pc,
71\arcsec\ corresponds to a physical size of 0.03~pc to 0.17~pc.

For measurements such as the core mass function, where only compact structures
are being analyzed, we expect that only slight
corrections to the measured fluxes and sizes 
will be needed for cores recovered with peak fluxes between 3 and 
10 times the noise
in the map when using the DR2 external-mask reduction\footnote{At least in the absence
of significant source crowding.}.  For analyses using the DR1 
external-mask reduction, more caution is needed if a significant number of the cores have sizes
closer to $\sigma = $30\arcsec, although those with sizes closer to $\sigma = $10\arcsec\ are 
still very well
recovered.  
Users of the JCMT LR1 catalogue 
(which is produced by the JCMT directly, rather than the GBS) 
should take note that sources
detected in that catalogue will have significantly underestimated peak fluxes, total
fluxes, and sizes\footnote{The goal of the JCMT LR1 catalogue is to identify
where peaks of emission exist, and not to provide an accurate estimate of the total
flux present.}.  The GBS DR1 automask-reduction results provide an approximate guide
to the level of underestimation in each of these source properties.

In Table~\ref{tab_recov_props}, we summarize the peak-flux ratios and size ratio
data shown in Figure~\ref{fig_peaks} and \ref{fig_widths}, so that accurate completeness 
can be estimated for future core-population studies.
We emphasize that even for analyses of relatively compact sources using the
DR2 external-mask reduction, extra attention should be paid to three factors.
First, the population of sources near the completeness limit (peak fluxes of 3 to 5 times 
the noise) likely have contributions from even fainter sources (peak fluxes of 1 to 2 times
the noise) which have been boosted to higher fluxes through noise spikes, etc.  If the 
true underlying source population is expected to increase with decreasing peak flux, 
then this contribution of fainter sources could be significant.  Second, faint compact
sources could be either intrinsically faint and compact, or they could be brighter and 
larger sources that are not fully recovered.  Examination of the size distribution of
the brighter sources in the map should help determine what the expected properties of the
fainter sources are.  Third, for analyses where the source detection rate is important
(e.g., applying corrections to an observed core mass function), the source recovery
rates presented in Section~4.1 should not be blindly applied, as they do not include
factors such as crowding or the limitations of core-finding algorithms running without
prior knowledge on a map, both of which are expected to decrease the real observational 
detection rate.  Furthermore, while the results presented here are uncontaminated by
false-positive detections, such complications will need to be carefully considered
when running source identification algorithms on real observations.

\section{Further Refinements - Telescope Pointing Offsets}
\label{sec_IR4}
For our final data release (DR3), we correct for telescope-pointing errors,
using the same reduction strategy for individual observations as in DR2.
We emphasize that the completeness tests in Section~4 inject the artificial Gaussian
sources at the same pixel position on every stacked map, so they are always perfectly 
aligned.  Hence the results from DR2 discussed in Section~4 also apply to DR3, in 
both cases reflecting the properties of the sources in the coadded map.  For 
astronomical sources in DR2 (as well as DR1), the measured properties 
will be artificially broadened and weakened slightly by the map misalignments.
We quantify and correct for these map alignments in DR3, as discussed in this section.

Recently, the JCMT Transient Survey \citep{Herczeg17} investigated methods
to calibrate SCUBA-2 data at high precision to increase their sensitivity to small
variations in flux within protostellar cores.  One facet included in their calibration is
telescope-pointing errors, which can often be in the range of 2\arcsec\ to 6\arcsec. 
The Transient Survey has been able to decrease this error to $< 1$\arcsec\ 
for their final maps \citep{Mairs17a}.
Indeed, pointing errors of several arcseconds could be large enough to
influence the sizes and peak fluxes of the dense cores we identify in the GBS, especially
at 450~$\mu$m.  Therefore, we investigated two independent methods to improve the positional
accuracy of our observations.  Directly adopting the exact method used by the 
Transient team
is not possible for the GBS.  For example, 
the Transient method requires multiple
bright, compact sources in their fields to estimate relative positions, 
whereas the GBS requires a method
which will supply good absolute positions for fields that may not contain many bright compact
sources.

\subsection{Absolute Positions}
To obtain good absolute positional accuracy,
we implement first a modification of the Transient Survey method.  
The Transient team uses
its first observation of each region as the template from which to measure all
subsequent image offsets \citep{Mairs17a}.  If the first observation has a large
associated pointing error, however, all subsequent observations will be corrected
to the wrong position\footnote{The Transient Survey is primarily concerned with
relative offsets, and does not contain adjacent observing areas for mosaicking,
so this issue is not a problem for them.}.
This approach could lead to additional deficiencies for the GBS,
however, since
mosaics could then have blurred structures in areas of overlap between adjacent maps.
Instead, we assume that on average, pointing errors for a given
field are small.  While individual observations may have errors, the mosaic of all
observations (four to six per field) should be relatively more accurate.
We therefore adopted the GBS DR2 mosaics as our reference template by which we align
individual observations.  In our final pointing-corrected mosaics, we do not see any
evidence of source blurring in field-overlap areas, suggesting that this approach
was reasonable.

\subsubsection{Method 1: Gaussian Fits}
The first alignment method that we tested follows a similar procedure to that adopted
by the Transient team \citep{Mairs17a}.  There, \citet{Mairs17a} fit bright and compact
emission in each 850~$\mu$m observation with Gaussians, using the Starlink command
{\it gaussfit} \citep[part of the {\sc CUPID} package;][]{Berry07,Stutzki90}.
The relative offsets between Gaussian peaks in each observation
of the same field were then used to estimate the overall pointing offset in that observation.
Note that since the 850~$\mu$m and 450~$\mu$m observations are obtained simultaneously,
pointing offsets derived using the 850~$\mu$m data should also be applicable at 450~$\mu$m,
where the SNR is usually lower.

We followed a similar basic approach to \citet{Mairs17a}.  We relaxed some criteria,
such as the minimum peak brightness, however, to apply the method 
to a greater fraction
of the GBS data.  In detail, we first cropped each 850~$\mu$m image to a radius of 1200\arcsec\
to reduce the influence of noisy edge pixels in our later analysis.  We then created a mosaic
of each region, and fit Gaussians to all of the peaks therein, 
discarding any that lay below
0.3~mJy~arcsec$^{-2}$, which is slightly less than ten times the noise for most areas of
the mosaic.  We also discarded any peaks from features 
with sizes larger than $\sigma = $40\arcsec\ in 
either axis, as
larger-scale structures are less likely to yield reliable central positions that are
stable from observation to observation.  This set of Gaussian fits serve as the reference
by which individual observations were then compared.

For each individual observation, we first smoothed the map by 6\arcsec\ to reduce
pixel-to-pixel noise \citep[using the same smoothing kernel as in][]{Mairs17a}.  Next,
we fitted Gaussians to all peaks in the {\it individual} observation that lay 
above 0.5~mJy~arcsec$^{-2}$, which
is slightly less than ten times the noise for most individual observations\footnote{For
comparison, \citet{Mairs17a} required peaks to be brighter than 200~mJy~bm$^{-1}$, or
0.83~mJy~arcsec$^{-2}$, assuming a 14\farcs6 effective beamsize, as in
\citet{Dempsey13}.}.
We then searched for peaks in the individual observation which were less than 10\arcsec\
offset from a peak in the mosaic, and also had similar peak fluxes (i.e., within a factor
of two)\footnote{\citet{Mairs17a} also required a positional coincidence of $<10$\arcsec\,
but not the additional peak flux criterion since their matches were restricted to high
SNR peaks.}.
Our best estimate of the pointing offset for the individual observation was made by taking
the median of all individual peak offset measures (separately in Right Ascension and
declination).  For observations with three or more individual peak offset measures,
we additionally removed any individual offset measures which differed by more than one
standard deviation from the median of the full sample before making our final measurement
of the bulk offset value\footnote{\citet{Mairs17a} adopt a slightly different approach here,
using the mean offset and removing any individual measures which differ by more than 4\arcsec\
from other measures.}.

Our implementation of Gaussian fitting to identify pointing offsets in observations
is thus conceptually similar to that used in \citet{Mairs17a}, but allows estimates to
be made in cases with many fewer and fainter peaks than are present in any of the fields
covered by the Transient Survey.  Our relaxed criteria could also allow spurious
offsets to be measured in some cases.  For example, 
without any additional constraints, some observations
may be aligned based on a Gaussian fit to only one or two faint peaks, and therefore are
strongly susceptible to a variety of sources of error.  Nonetheless, we generally found
visually satisfactory results using this method.  As discussed in the following section,
however, we chose to adopt a different method, which is applicable to a broader 
swath of the GBS observations and appears to be slightly more reliable.

\begin{figure}[htb]
\begin{tabular}{cc}
\includegraphics[width=3.2in]{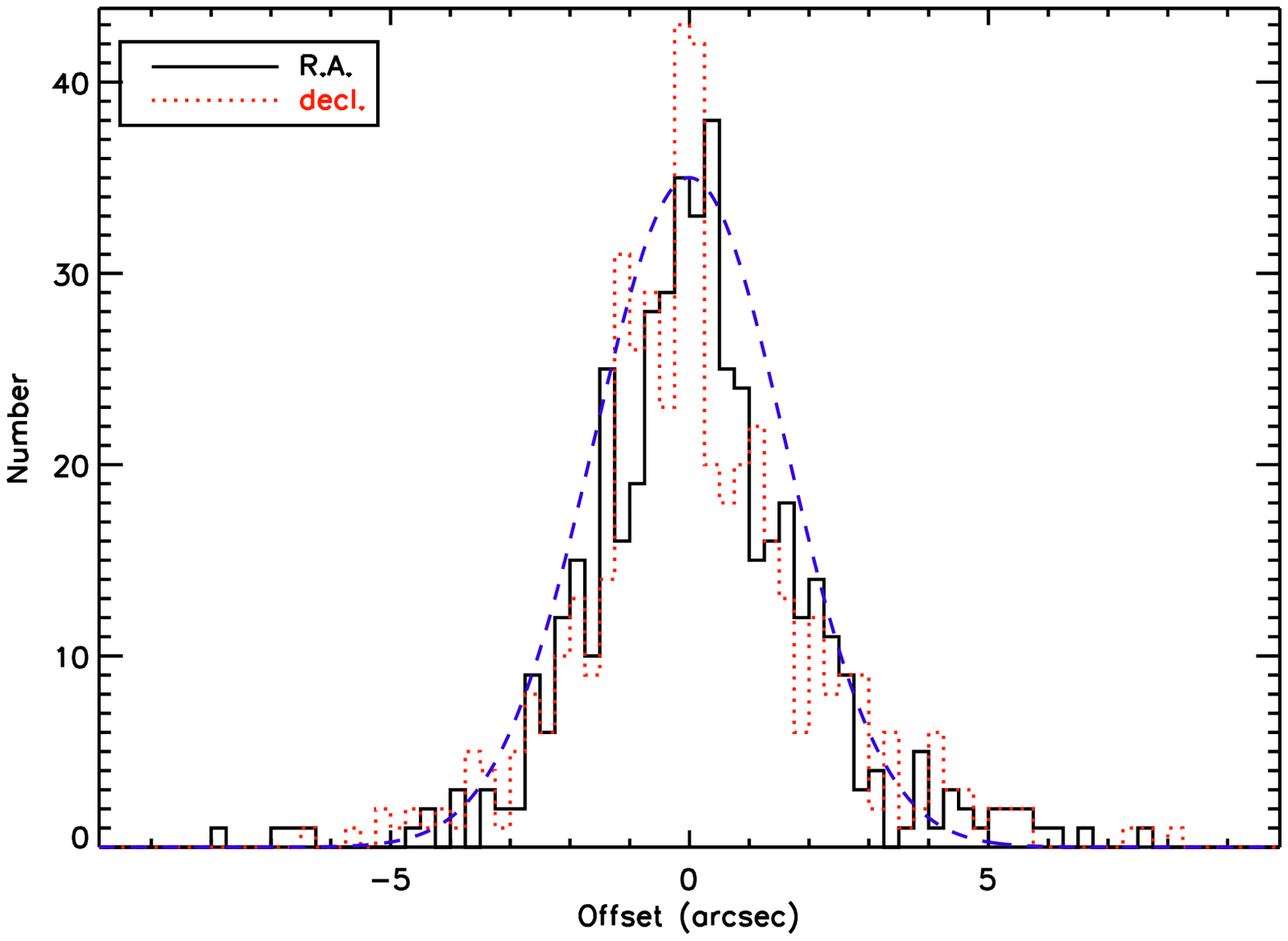} & 
\includegraphics[width=3.2in]{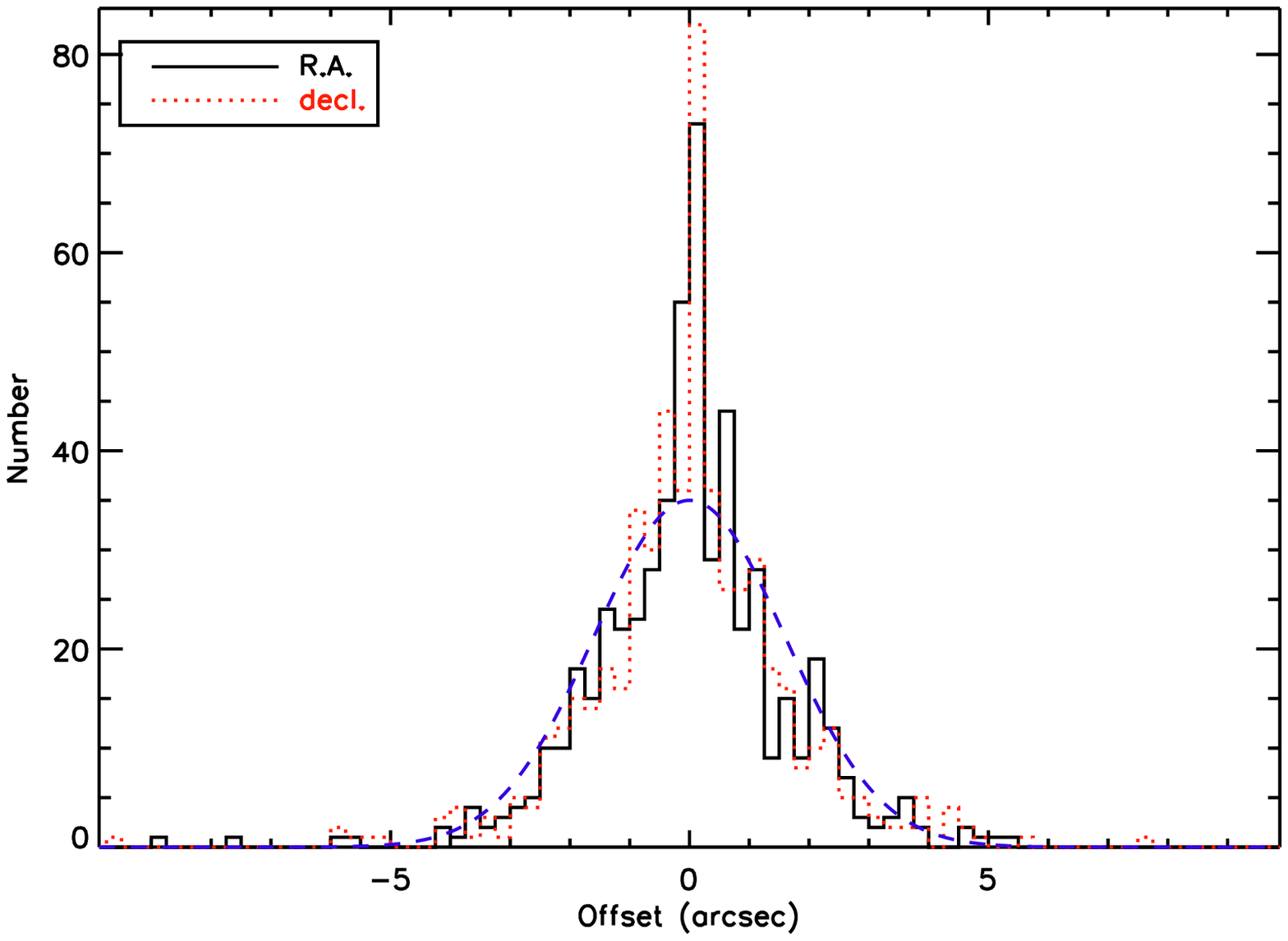} \\
\end{tabular}
\caption{Pointing offsets derived for all GBS observations, using the Gaussian-fit
        method (left) and \al method (right).  The solid black histogram shows
        offsets in Right Ascension, while the dotted red histogram shows offsets in
        declination.  Observations where the method was unable to be applied, e.g.,
        due to insufficient flux in the map,
        are excluded.  The blue dashed curve shows a Gaussian with $\sigma=$1\farcs6
        (left) and $\sigma=$1\farcs7 (right) for reference.
}
\label{fig_fit_offsets}
\end{figure}

Figure~\ref{fig_fit_offsets} (left panel) 
shows the pointing offsets estimated using the Gaussian-fitting technique.  
Of the 581 GBS observations, 115 did not fit our relaxed criteria
and no offsets could be measured.  Of the remaining 466 observations, 
the full range of offsets measured in Right Ascension and declination ran between
-7\farcs8 and 8\farcs2 (with similar minima and maxima for each of Right Ascension
and declination), with a standard deviation of 1\farcs9 and 2\farcs0 in Right
Ascension and declination, respectively.  The distribution of offsets is centred
on 0, with mean offsets of $< $0\farcs2 in both Right Ascension and declination.
A notable fraction of the observations showed significant offsets:
214 (46\%) had total offsets $\ge 2$\arcsec, 111 (24\%) had total 
offsets $\ge 3$\arcsec, and 33 (7\%) had total offsets $\ge 5$\arcsec.

\subsubsection{Method 2: Align2d}
The second method that we tested involved using Starlink's \al command \citep[part of the
{\sc KAPPA} package;][]{kappa}, which compares
all pixels with significant emission in both the observation and reference mosaic to determine
an optimal offset.  We assumed that the two maps differed only by a simple constant offset, and
did not include more-complex terms such as rotation or shear (as was also assumed
for the previous method).
We first slightly smoothed the observation (by two pixels using KAPPA's {\it gausmooth}
command), as we found this improved the reliability of the offsets measured 
compared to offsets measured using unsmoothed observations.
We also tested a range of thresholds for pixels to use in the  \al calculation,
and found that the recommended setting of \texttt{corlimit}=0.7\footnote{The 
\texttt{corlimit} parameter can
be varied between 0 and 1, with larger values causing more pixels to be excluded from the
calculation.} worked best.  Limiting the calculation to fewer, more reliable pixels
resulted in \al failing to measure an offset in more cases, while those that were measured
tended to be consistent between \texttt{corlimit} values, with typical variations of less than
1 pixel (3\arcsec).

The right-hand panel of Figure~\ref{fig_fit_offsets}
shows the offsets measured by \al for all of our
GBS fields.  Of the 581 GBS observations, \al was unable to measure offsets
in only 29 of them, compared with 115 observations without measureable offsets
using the Gaussian-fit method.
None of the 29 observations had
good Gaussian fits (i.e., fits where the offset is larger than the estimated uncertainty),
and 22 of the 29 observations had no Gaussian fit, due to insufficient emission features
in their respective maps.  In a few
cases, however, brighter emission was present, but did not yield a
single consistent offset value.  In these cases, multiple (usually two) peaks were
identified by the Gaussian-fit method, but the offsets derived from each peak were
mutually inconsistent.  The sparse nature of the emission structures in these exceptional
cases prevents any conclusion to be made on the cause of the inconsistency in offsets.

For the few observations where our implementation of
\al failed to calculate an offset, we attempted to calculate offset
values that would be derived under a variety of different implementations of
\al, using different values of the \texttt{corlimit} parameter, or using an un-smoothed
observation.  Sometimes, these variations in \al did yield offset
values, however, neither the magnitude nor sign of the derived offsets were
consistent between the different methods, again suggesting that simple linear
offsets may not be appropriate for these particular observations.

Using the implementation of \al described above, 
the full range of pointing offsets runs between
-9\farcs7 to 7\farcs6 (considering Right Ascension and declination separately;
both span a similar range).
The standard deviation of the pointing offsets is 1\farcs7 for Right
Ascension and 1\farcs8 for declination considered separately.
As can be seen from
Figure~\ref{fig_fit_offsets} (right panel), 
despite most observations having small pointing offsets,
a non-negligible
number of fields have significant pointing errors.  We find that 194, or 35\%, have
total offsets of more than 2\arcsec, and 86, or 16\%, have total offsets of more
than 3\arcsec, corresponding to the pixel size for the 450~$\mu$m and 850~$\mu$m maps,
respectively.  Eighteen maps, or about 3.3\%, have total offsets in excess of 5\arcsec,
which is a significant fraction of the 9\farcs8 450~$\mu$m beam.

\begin{figure}[htb]
\begin{tabular}{ccc}
\includegraphics[height=2.2in]{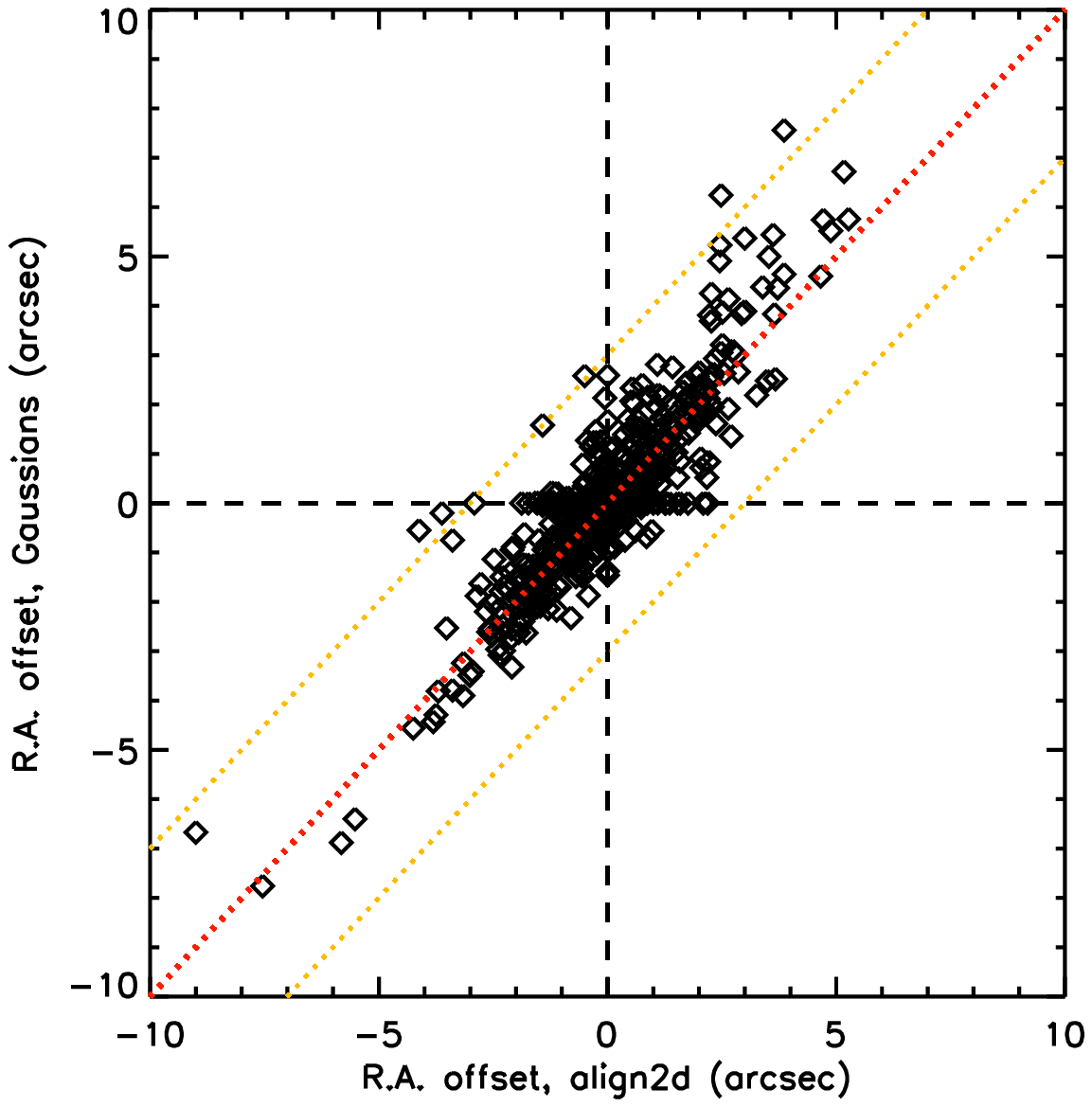} & 
\includegraphics[height=2.2in]{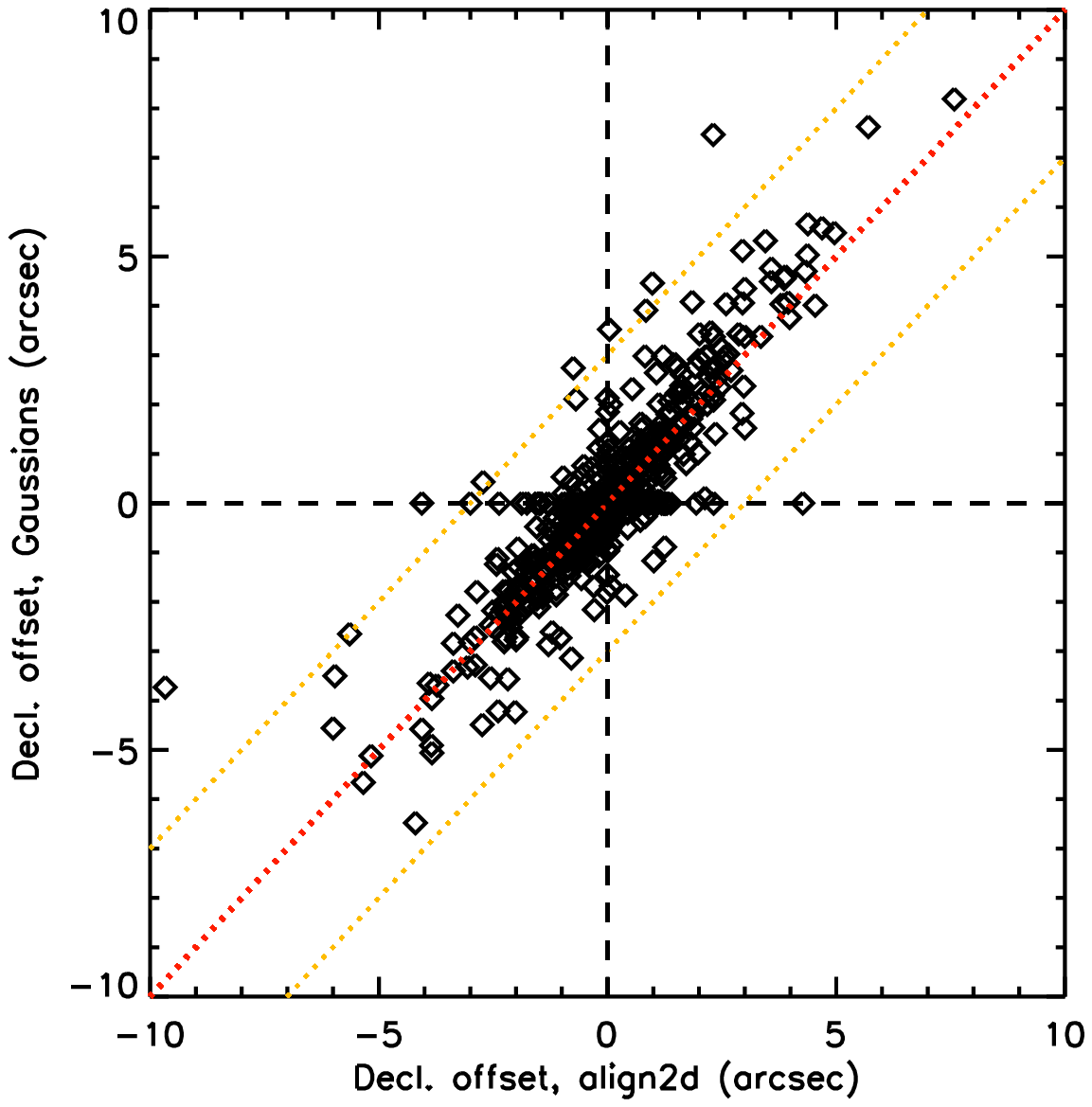} &
\includegraphics[height=2.2in]{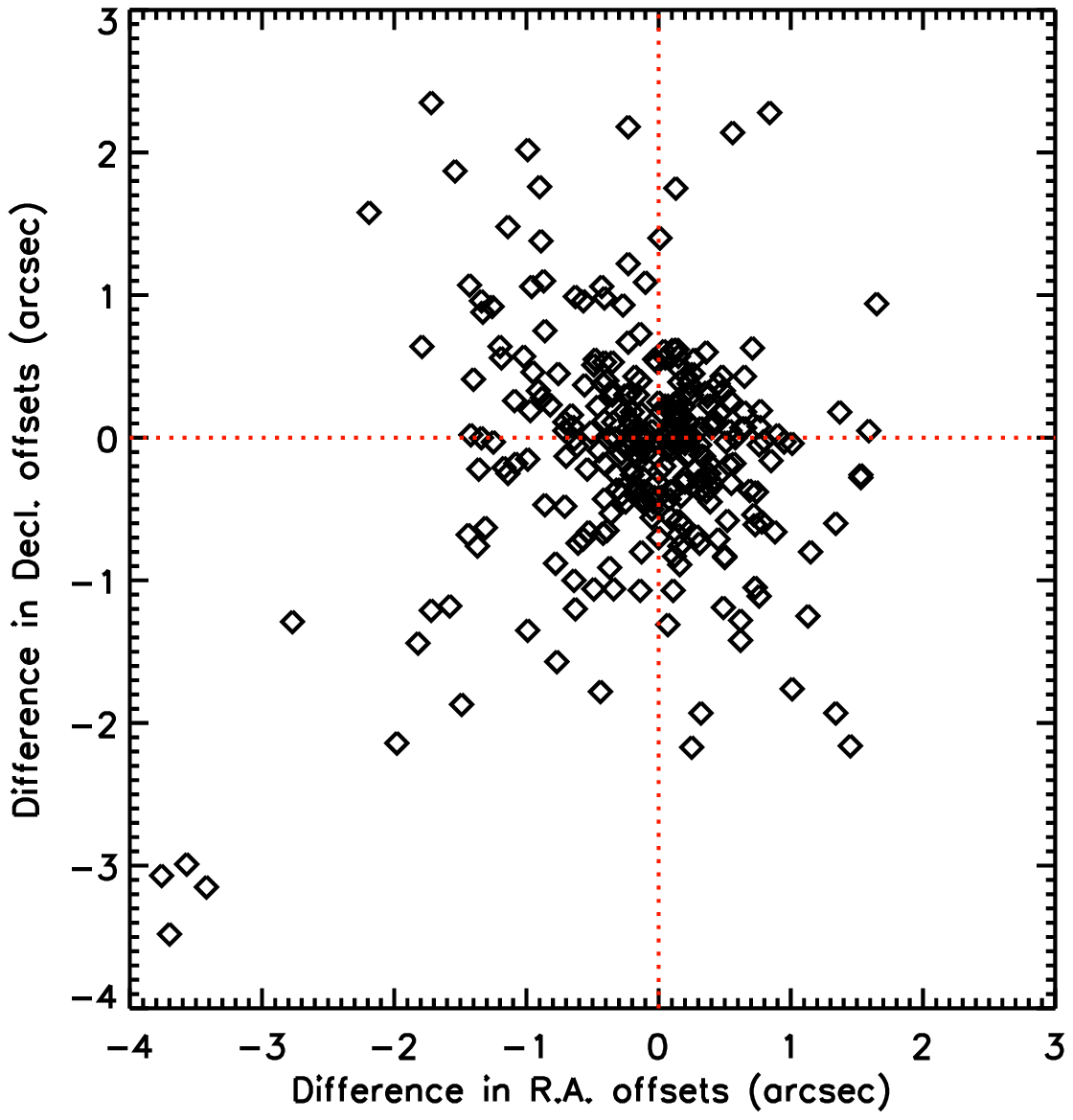} \\
\end{tabular}
\caption{Offsets derived for all GBS observations using \al and Gaussian fitting.
	The left and middle panels separately compare the offsets derived in
	Right Ascension (left) and declination (middle) for the two methods.
        Here, the red dotted line shows a
        one-to-one relationship, while the yellow dotted lines show discrepancies of 3\arcsec\
        (one pixel at 850~$\mu$m) between the two measures.  Offsets of 0 are assigned to
        the relatively few
	observations where the method was unable to determine a measurement.
	The right panel shows the difference in offsets (\al minus Gaussian) measured in
	Right Ascension and declination.  Here, only observations for which offsets
	were measured using both methods were included.}  
\label{fig_compare_offsets}
\end{figure}

In Figure~\ref{fig_compare_offsets}, we show a comparison of all of the offsets measured
using both the Gaussian-fit and \al methods.  Clearly, the vast majority of offsets are in
good agreement using either method.  We carefully visually examined the few observations
where \al and the Gaussian-fit method disagree by more than 3\arcsec\ (one pixel at
850~$\mu$m,
or about one third of the 450~$\mu$m beam), and found that the \al offset typically
appeared to be the more correct of the two measures.  None of the observations
with discrepant derived offsets contained many bright compact sources, where the
Gaussian-fit method is expected to perform its best.
We therefore adopt the align2d method for DR3.

\subsection{Impact on Mosaics}

\begin{figure}[htb]
\begin{tabular}{cc}
\includegraphics[width=3in]{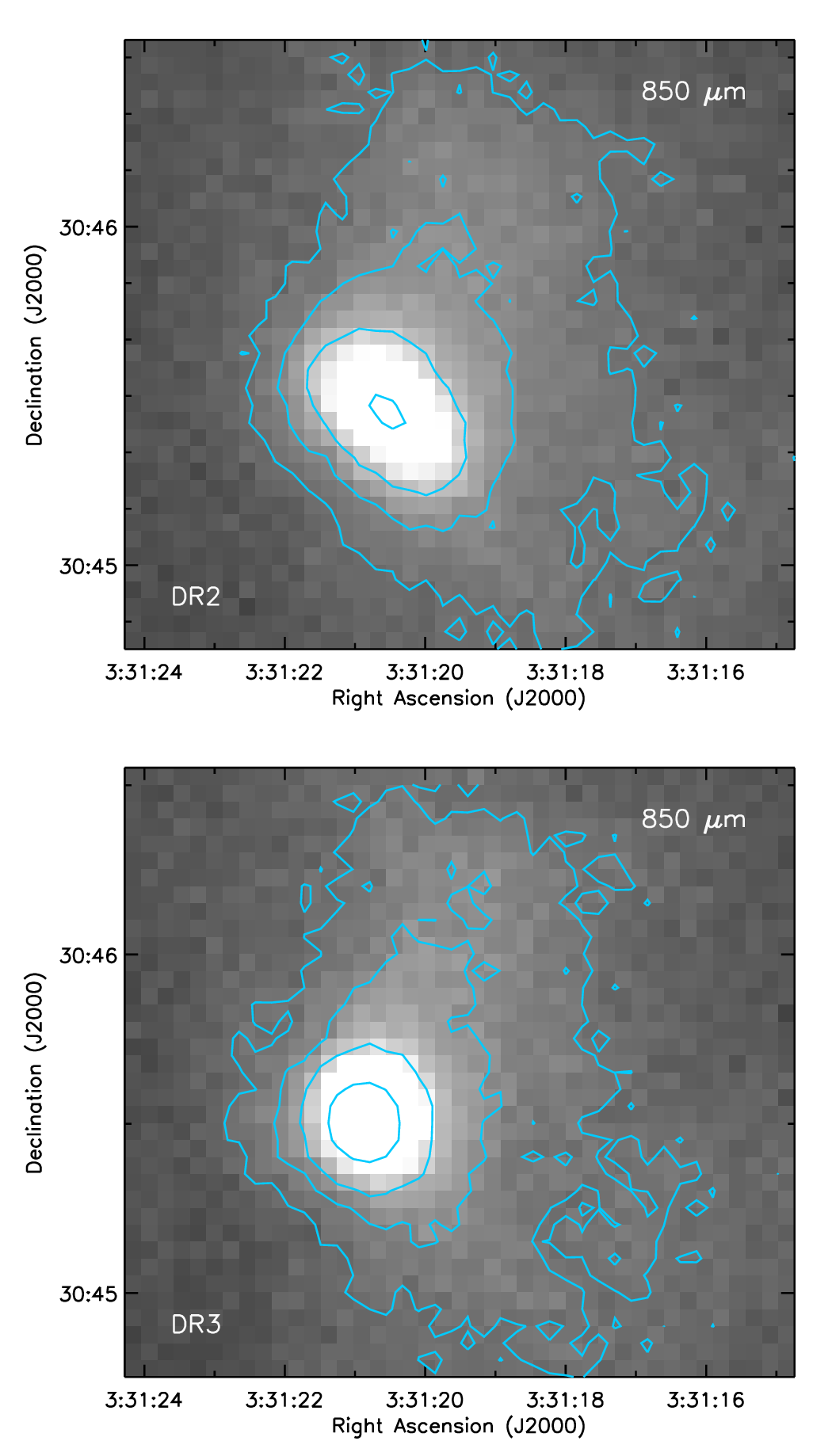} &
\includegraphics[width=3in]{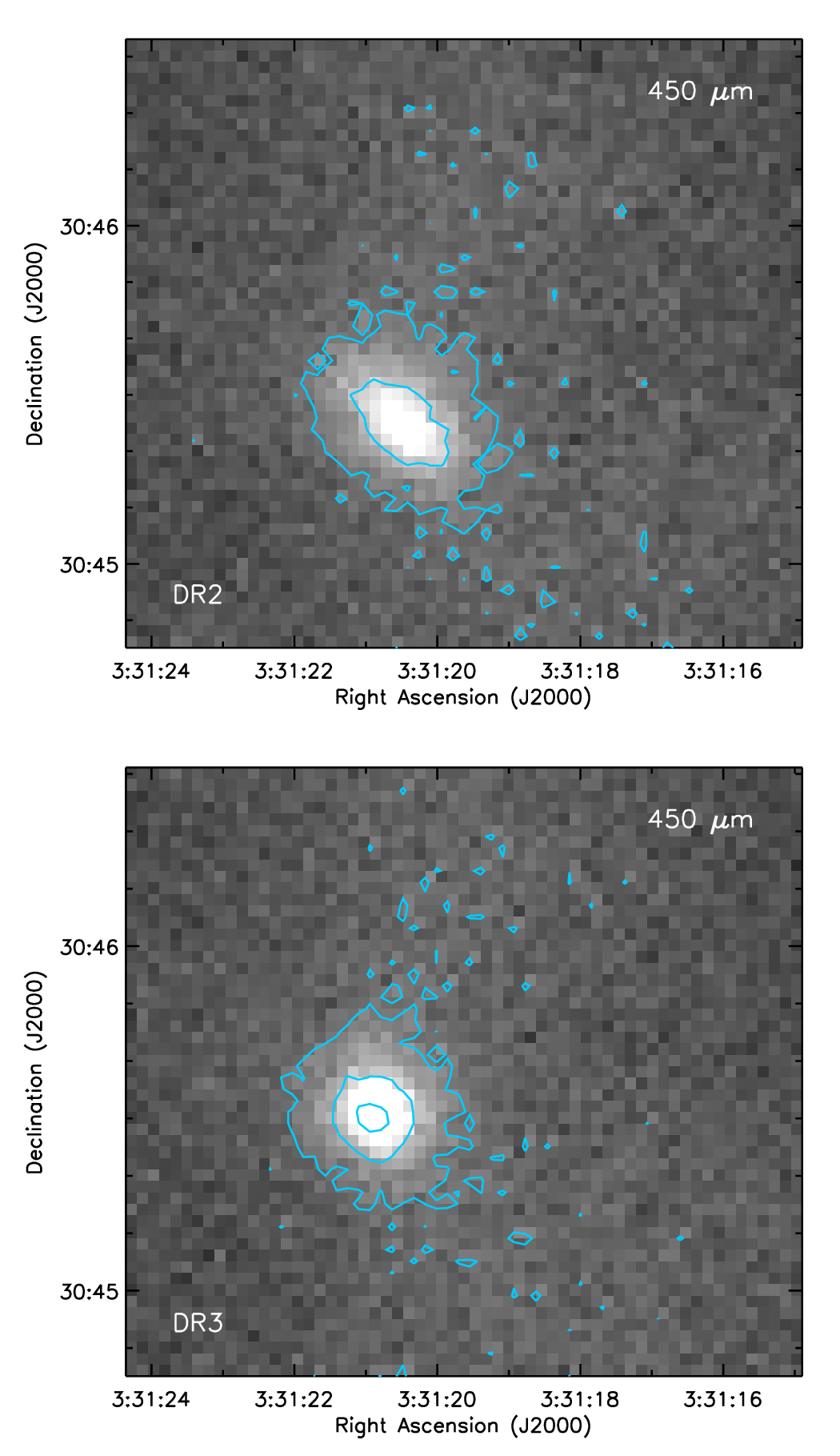} \\
\end{tabular}
\caption{A comparison of a bright source in the DR2 and DR3 mosaics in the B1-S field.  
	The top
        panels show the DR2 mosaics with no pointing corrections, while the
        bottom panels show the DR3 mosaics where pointing corrections have been
        included.  The left panels show the mosaics at 850~$\mu$m, where the greyscale
        ranges from $-$0.75~mJy~arcsec$^{-2}$ (black) to 1.5~mJy~arcsec$^{-2}$ (white).
	Contours are shown at [0.25,0.5,1,2.5]~mJy~arcsec$^{-2}$.
        The right panels show the mosaics at 450~$\mu$m, where the greyscale ranges
        from $-$7.5~mJy~arcsec$^{-2}$ (black) to 15~mJy~arcsec$^{-2}$ (white).
	Contours are shown at [3, 10, 25]~mJy~arcsec$^{-2}$.
        Correction for pointing offsets noticeably improves the point sources
        present in this particular field.
        }
\label{fig_b1s}
\end{figure}

In most fields, many, if not all, of the observations have positions that are corrected by
less than 3\arcsec, and thus the improvement in the DR3 mosaic over the DR2 mosaic is
subtle.  There are a handful of fields, however, where the pointing offsets are larger,
and the improvement in the final mosaic is more obvious.  The one (and only) dramatic
example of
this is the B1-S field within the PerseusWest mosaic, where pointing offsets for the
six observations comprising this field range from -9\farcs0 to 3\farcs9 in Right
Ascension and -9\farcs7 to 1\farcs0 in declination.
Figure~\ref{fig_b1s} shows a comparison of the brightest core in the B1-S field,
illustrating the extreme elongation and blurring of the core seen in the DR2 map, even
at 850~$\mu$m.
We emphasize that the B1-S field is an extreme outlier in terms of telescope-pointing errors
present in the original observations, but it does serve as a good exemplar of how
our pointing offset correction is effective.  In all other
fields, the improvement is subtle at best at 850~$\mu$m, and is still minor at 450~$\mu$m.
Because these offset corrections are small, it was not necessary to perform an
entire additional external-mask reduction with the masked areas shifted to account
for the offsets in the individual observations.  Even in the B1-S field, not shifting
the masks for each individual observation still leaves the majority of the compact source
emission (down to below 10\% of the local peak) lying within the mask for the reduction. 

\begin{figure}[htb]
\includegraphics[width=4in]{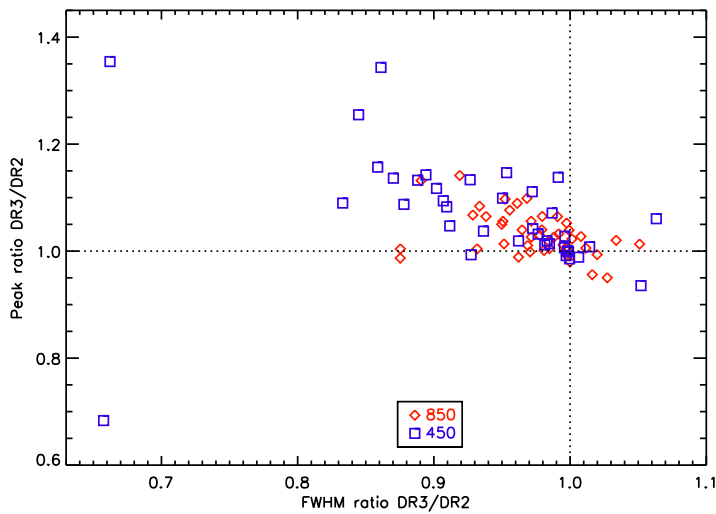}
\caption{A comparison of the FWHM and peak fluxes measured for bright compact sources
        in the DR2 and DR3 reductions.  The horizontal axis shows the ratio of FWHM
        values in DR3 versus DR2.  Ratios less than one indicate sources which became
        smaller in DR3.  The vertical axis shows the ratio of peak fluxes measured in DR3
        versus DR2.  Ratios greater than one indicate sources with brighter peak fluxes
        in DR3.  Red diamonds indicate ratios measured for compact sources at 850~\microns\
        while blue squares indicate ratios measured for compact sources at 450~\microns.
        In this plot, we only include sources with peak fluxes at least 50 times the
        local rms and sizes less than 25\arcsec\ to minimize errors due to uncertainties
        in the Gaussian fits.}
\label{fig_ir4_gauss}
\end{figure}

In Figure~\ref{fig_ir4_gauss}, we show the quantitative improvement of the DR3 maps over
the DR2 maps.  We ran CUPID's {\it gaussfit} on all of the mosaics
created using both DR2 and DR3, and then searched for positional matches between the
two catalogues, at each wavelength independently.  We restricted our analysis to
compact and bright sources
to minimize uncertainties in the Gaussian fit parameters.
We show only sources which had measured peak fluxes
of at least 50 times the local noise level and sizes of FWHM $<25$\arcsec.
In the figure, we see that most of the
fitted sources lie in the top left corner, where they would be expected to lie if
DR3 tended to reduce the amount of blurring present in the final mosaics.
At 850~\microns, the ratio of FWHM values for DR3 versus DR2 is $0.97 \pm 0.04$ and
the ratio of peak fluxes is $1.03 \pm 0.04$ (mean and standard deviation quoted for both).
At 450~\microns, those same ratios are $0.93 \pm 0.09$ and $1.07 \pm 0.11$ respectively.
As expected, the improvement in DR3 images tends to be larger at 450~\microns, although
there is significant scatter in all relationships.  We expect that some of the Gaussian
fits may be confused with the presence of diffuse extended structure around the 
compact-sources fit, and that a careful source-by-source fitting would reduce the 
scatter in the
ratios listed above.  Underlining this fact, we note that the 450~$\mu$m source 
with a peak-flux ratio less than 0.7 lies in the integral shaped filament within Orion~A,
in a region known for bright complex emission structures on a variety of scales.  
Excluding this one source, the FWHM ratio becomes $0.94 \pm 0.07$ and the peak-flux ratio
becomes $1.08 \pm 0.09$. 

\section{Conclusions}
\label{sec_conc}

In this paper, we present the data-reduction methodology employed by the JCMT 
Gould Belt Survey, through all three major data releases, DR1 through DR3.  
All of the DR3 data products, including final mosaics using the external mask
reductions, the mask files and CO subtracted 850~$\mu$m maps, are publicly 
available in conjunction with this paper.  
[An address for a DOI (permanent webpage) will be made available with the published version
of this paper.]
In Section~4, we measured the reliability of emission structures recovered in DR1 and
DR2.  There, we demonstrate that our two-step reduction process allows us to 
measure true source properties better than prior methods, 
and that DR2 provides significant improvements
over DR1, while both are expected to provide substantially better recovery
of extended structures than the JCMT LR1, as already shown in \citet{Mairs15}\footnote{We
note that the JCMT LR1 was designed to identify the locations of emission peaks but not
recover the total emission present.}

GBS science tends to concentrate on the more-compact emission structures
(cores and filaments)
where source recovery is best.  
For the DR2 method, in our idealized tests that assume isolated emission and a source
detection method tuned to the known source locations, 
we recover $>95$\% of artificial structures 
with peaks at 3 times the rms, for sizes of 30\arcsec\ and smaller, and 100\% of the
structures with peaks at 5 times the rms.  
These recovered structures also have reliable
properties measured, with typical peak flux and size measurements both lying within
15\% of the true values for sources with peak fluxes at least 3 times the rms,
while the total flux measurements lie within 25\% of the true values.
In all cases, the observed values can be corrected for the deficit in peak
flux, total flux, and size measured to a higher degree of accuracy than the
listed percentages.
Source recovery statistics and the reliability of measured parameters (peak flux and size)
for the full series of artificial Gaussian test inputs are provided for reference in 
Tables~\ref{tab_complete} and \ref{tab_recov_props}.  These numbers should be 
considered as best-case values if
measuring and interpreting source-population properties such as the dense-core
mass function.  Additional effects, such as
the presence of non-Gaussian sources, biases from source detection algorithms,
and biases due to source crowding have not been considered here, and are all 
expected to decrease the fraction of sources recovered and the reliability 
of their properties.
We strongly encourage readers to take care in considering these additional effects for
any analyses where our recovery and reliability statistics are being applied.

For the final GBS data release (DR3), we estimate the pointing offset present in each 
observation, by taking advantage of the fact that the survey observed each location
on the sky between four and six times.  We test two different methods for calculating
the offset present between repeated observations of the same field, 
and find that the KAPPA program \al
tends to produce the most-reliable results.  The pointing offsets estimated are  
typically small.  About 16\% of the fields
have total offsets of at least 3\arcsec, which
corresponds to one pixel in the 850~$\mu$m maps and 1.5 pixels in the 450~$\mu$m maps,
while 3.3\% have total offsets of at least 5\arcsec.  Most mosaics show little 
discernable difference before and after the pointing offset correction, however,
the B1-S field in the PerseusWest mosaic in particular is noticeably improved.
The full data-reduction procedure is given in Appendix~A (for DR1 and DR2), and 
Section~5 (for DR3) to allow other groups to reproduce our methods.  
We remind the reader that for DR3, we applied positional shifts to observations
reduced under the DR2 methodology, so all of the reduction parameters implemented
in {\it makemap} are identical to DR2.

\appendix
\section{Data Reduction Parameters}
\label{app_dr_params}

Here, we summarize the full procedure and parameters used to create maps in DR1 and DR2.
Settings are supplied to the {\it makemap} algorithm through a `dimmconfig' file\footnote{  
More information about the SCUBA-2 data reduction procedure can be found at 
{\tt http://starlink.eao.hawaii.edu/devdocs/sc21.htx/sc21.html}.}.

\subsection{DR1}
The DR1 automask dimmconfig file contains the following settings.

\begin{verbatim}
^$STARLINK_DIR/share/smurf/dimmconfig_bright_extended.lis
numiter=-300
flt.filt_edge_largescale=600
maptol=0.001
itermap=1
noi.box_size=-15 
flagfast=600 
flagslow=200 
flt.filt_edge_largescale_last=200
ast.skip=5 
flt.zero_snr=5
flt.zero_snrlo=3
noi.box_type=1 
flt.ring_box1=0.5
flt.filt_order=4
com.sig_limit=5
ast.zero_snr=5
ast.zero_snrlo=0
\end{verbatim}

The DR1 external-mask dimmconfig file is nearly identical, with only the final two 
parameters changed to the following assignments.
\begin{verbatim}
ast.zero_mask=1
ast.zero_snr=0
\end{verbatim}

In DR1, we created a mosaic of individually reduced observations using their mean, for both
the automask and external-mask mosaics.  Mask creation in DR1 was not completely identical
between regions, as individual region team leads experimented with different schemes.  The
most commonly adopted scheme was to include in the mask all pixels lying above a signal to 
noise threshold of 2 in the automask mosaic, and this scheme 
was the mask-creation method tested
in our analysis here.

\subsection{DR2}
The dimmconfig file for the DR2 automask reduction contained the following lines.
\begin{verbatim}
^$STARLINK_DIR/share/smurf/dimmconfig_bright_extended.lis
numiter=-300
flt.filt_edge_largescale=600
maptol=0.001
itermap=1
noi.box_size=-15 
flagfast=600 
flagslow=200 
ast.skip=5 
flt.zero_snr=5
flt.zero_snrlo=3
noi.box_type=1 
flt.ring_box1=0.5
flt.filt_order=4
com.sig_limit=5
ast.zero_snr=3
ast.zero_snrlo=2
ast.filt_diff=600
ast.zero_lowhits = 0.1
ast.zero_union=0
\end{verbatim}

The dimmconfig file for the DR2 external-mask reduction contained nearly identical lines,
with the final five lines above being replaced with the following lines.
\begin{verbatim}
ast.zero_mask=1
ast.zero_snr=0
\end{verbatim}

For mosaicking, we combined the observations using a {\it median} combination scheme for
the automask mosaic, first clipping each observation to the same zone as considered
for the automask via the \texttt{ast.zero\_lowhits} parameter (i.e., excluding the
noisy edge pixels).  A mean combination scheme was used for the external-mask mosaic.  Masks
were created uniformly across regions for DR2.  We used all pixels in the automask mosaic
lying above a signal-to-noise threshold of 3 which were in zones of 20 or more contiguous
pixels (determined using CUPID's {\it clumpfind}).

\subsection{Summary of Differences Between DR1 and DR2}
Many of the key differences in DR1 and DR2 have already been extensively discussed in 
\citet{Mairs15}, particularly the change in the parameters
\texttt{ast.zero\_snr} and \texttt{ast.zero\_snrlo},
which effectively allow more pixels to be recognized for having real 
astronomical signal in the automask reduction in DR2.  An important parameter not discussed
in \citet{Mairs15} is the removal of the parameter 
\texttt{flt.filt\_edge\_largescale\_last} in DR2.  When included, this parameter
allowed for a stronger filtering of the map outside of the automask or external mask
area
in the final iteration.  Excluding it allowed more real large and faint structures to be
present in the final reduced map, with the downside of also increasing large-scale noise
features.  Switching the mosaicking method to use a median combination 
for DR2 helped to reduce
the presence of these large-scale noise features in the final automask mosaic\footnote{We
therefore emphasize that the DR2 automask settings should not be applied for reductions
where only a few observations were taken.  In this case, large-scale noise features
are likely to propagate through to the final automask mosaic and hence also be included
in the mask used for the second round of reductions.}.
Neither the \texttt{flt.filt\_edge\_largescale\_last} parameter nor the median mosaic
method had been tested at the time of the publication of \citet{Mairs15}.

\subsection{CO Subtraction}
The 850~$\mu$m observing band contains the $^{12}$CO(3--2) emission line
\citep[e.g.,][]{Johnstone03}, which in
some instances can contribute significantly to the total emission observed.
A full discussion on CO emission and best practices for removing it from the
850~$\mu$m continuum data is given in \citet{Drabek12}, and an updated version
is given in \citet{Parsons18}.
Here, we provide a summary of the process used by the GBS for reference.

In short, the procedure involves using the $^{12}$CO(3--2) integrated intensity map
to estimate the contribution to emission observed by SCUBA-2.  This emission is subtracted
directly from the raw-data time stream so that it will be subject to the same filtering,
etc, as the 850~$\mu$m observations are.

We convert the CO integrated intensity map 
into the effective continuum emission based on the weather conditions present for
each 850~$\mu$m observation.  We multiply by the following factor, $C$, updated from those
originally presented in \citet{Drabek12} to account for the SCUBA-2 beamsize measurements
presented in \citet{Dempsey13}.\\
\begin{math}
0 \le \tau < 0.03, C=2.93e-3 \\
0.03 \le \tau < 0.07, C=3.14e-3 \\
0.07 \le \tau < 0.10, C=3.24e-3 \\
0.10 \le \tau < 0.16, C=3.45e-3 \\
0.16 \le \tau < 0.20, C=3.55e-3 \\
\end{math}
where $C$ is in units of (mJy~arcsec$^{-2}$)(K~km~s$^{-1}$)$^{-1}$ and $\tau$ is the optical
depth of the atmosphere measured at 250~GHz.  Note that these 
scale factors were recently updated in \citet{Parsons18}, but as the difference is
$<<$5\%, we did not re-run CO subtraction with the updated values. 

The scaled CO integrated intensity map is then aligned with the SCUBA-2 external mask, and
subtracted using the \texttt{fakemap} parameter in {\it makemap}.
For DR3 only, we additionally eliminated noisy pixels in the CO integrated intensity map.
To do this, we slightly smoothed the CO integrated intensity map (using KAPPA's {\it gausmooth}
command with a smoothing scale of 2 pixels), and zeroed out pixels with a signal-to-noise
ratio of less than 5.  Testing by the data reduction team showed that this procedure is
able to reduce the over-subtraction of CO when the HARP CO map has very noisy edges.

\section{Difference Maps}
\label{app_diff_quant}
\subsection{Visual Comparison}
As noted in Section~4, by subtracting the original mosaics with no artificial sources
added from the reduced maps where the artificial Gaussians had been added into the
time stream, we are able to determine the precise contribution of the artificial
sources to the final map.  This allows us to test the effects of filtering alone, without
including the influence of noise.

\begin{figure}[htb]
\plotone{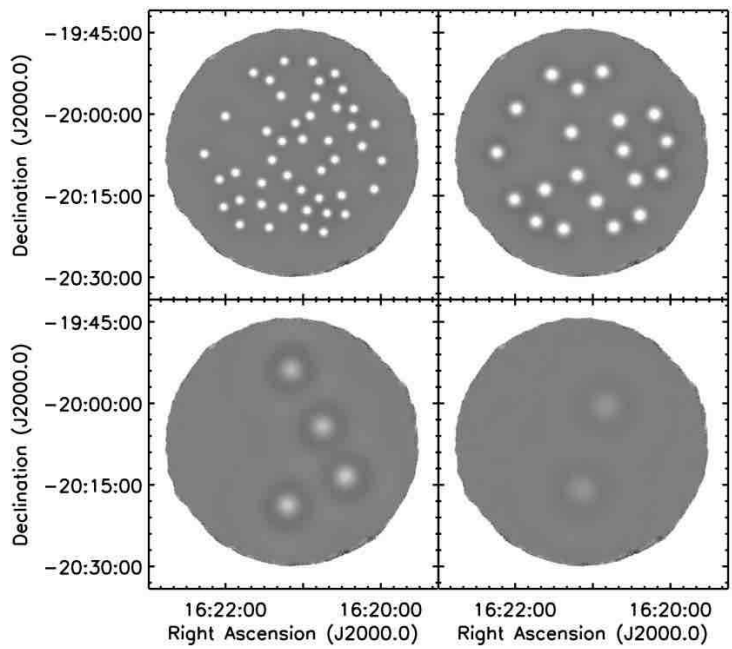}
\caption{The difference between the DR2 external-mask reductions with artificial
	Gaussians added prior to processing and the original reduction with no artificial 
	Gaussians added.  This figure shows the same artificial Gaussian fields as 
	Figures~\ref{fig_test_image} and \ref{fig_sample_reducs}, using the same
	greyscale range and other plotting conventions.}
\label{fig_diff_maps}
\end{figure}

Figure~\ref{fig_diff_maps} shows four examples of these difference maps,
examining the same artificial Gaussian cases as in the previous 
Figures~\ref{fig_test_image} and \ref{fig_sample_reducs}.
As can be seen from comparing Figure~\ref{fig_diff_maps} and Figure~\ref{fig_test_image},
the compact artificial Gaussians in the reduced images appear similar to
their initial models.  Wide artificial 
Gaussians (particularly the bottom-right panel
example), however, 
are substantially fainter after passing through the reduction pipeline.

\subsection{Quantitative Measures}
\begin{figure}[htb]
\begin{tabular}{cc}
\includegraphics[width=3in]{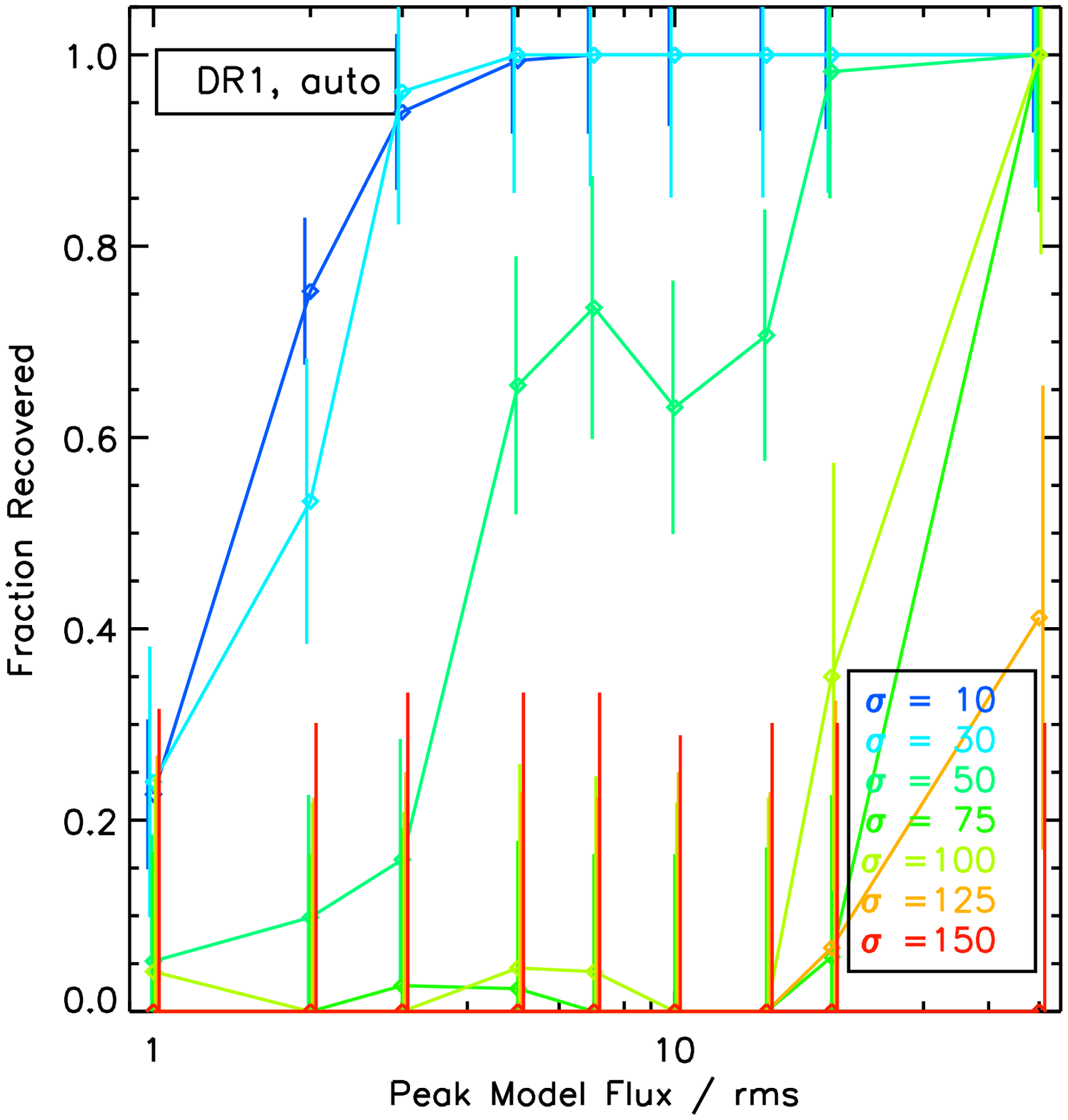} &
\includegraphics[width=3in]{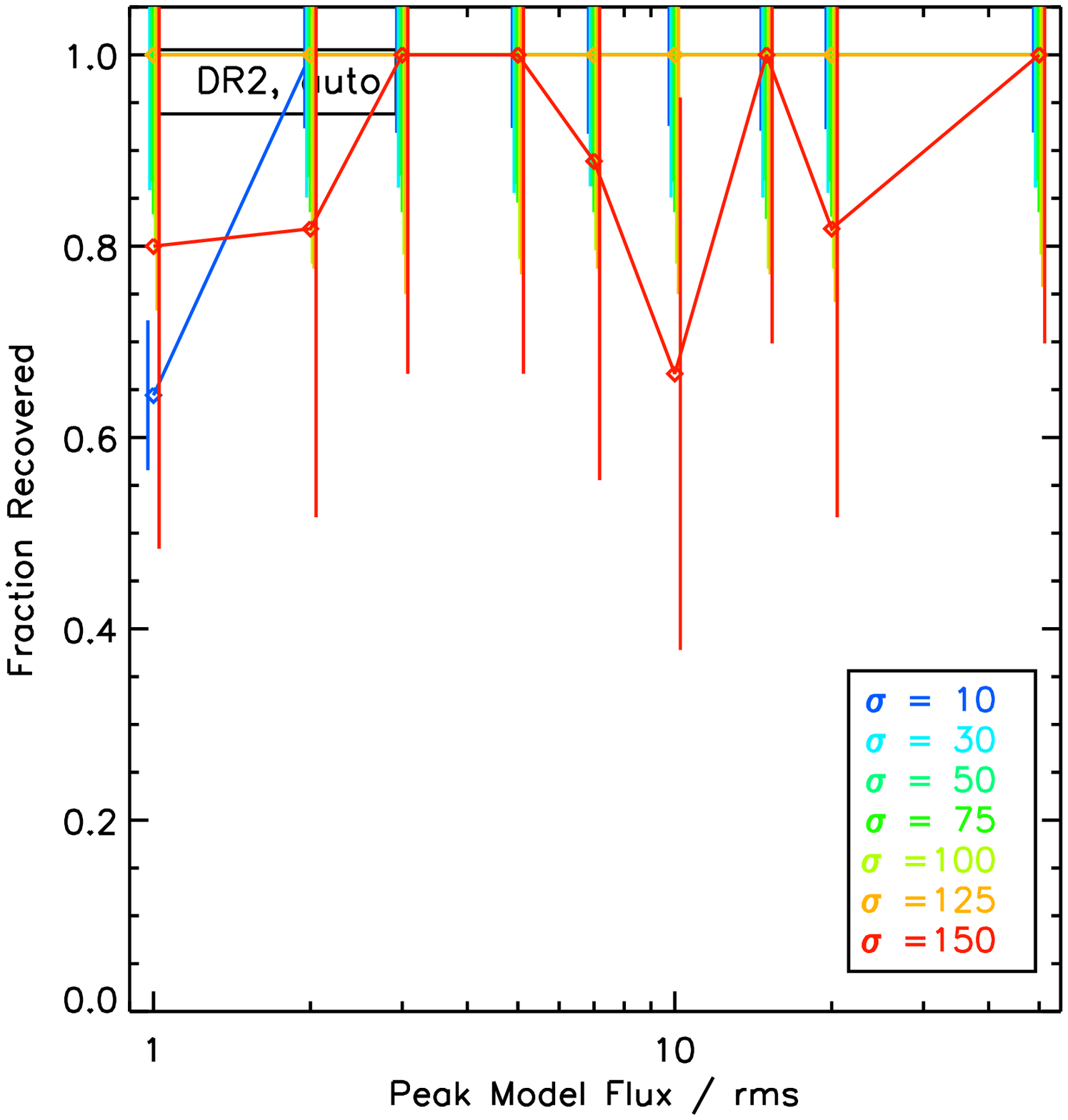} \\
\includegraphics[width=3in]{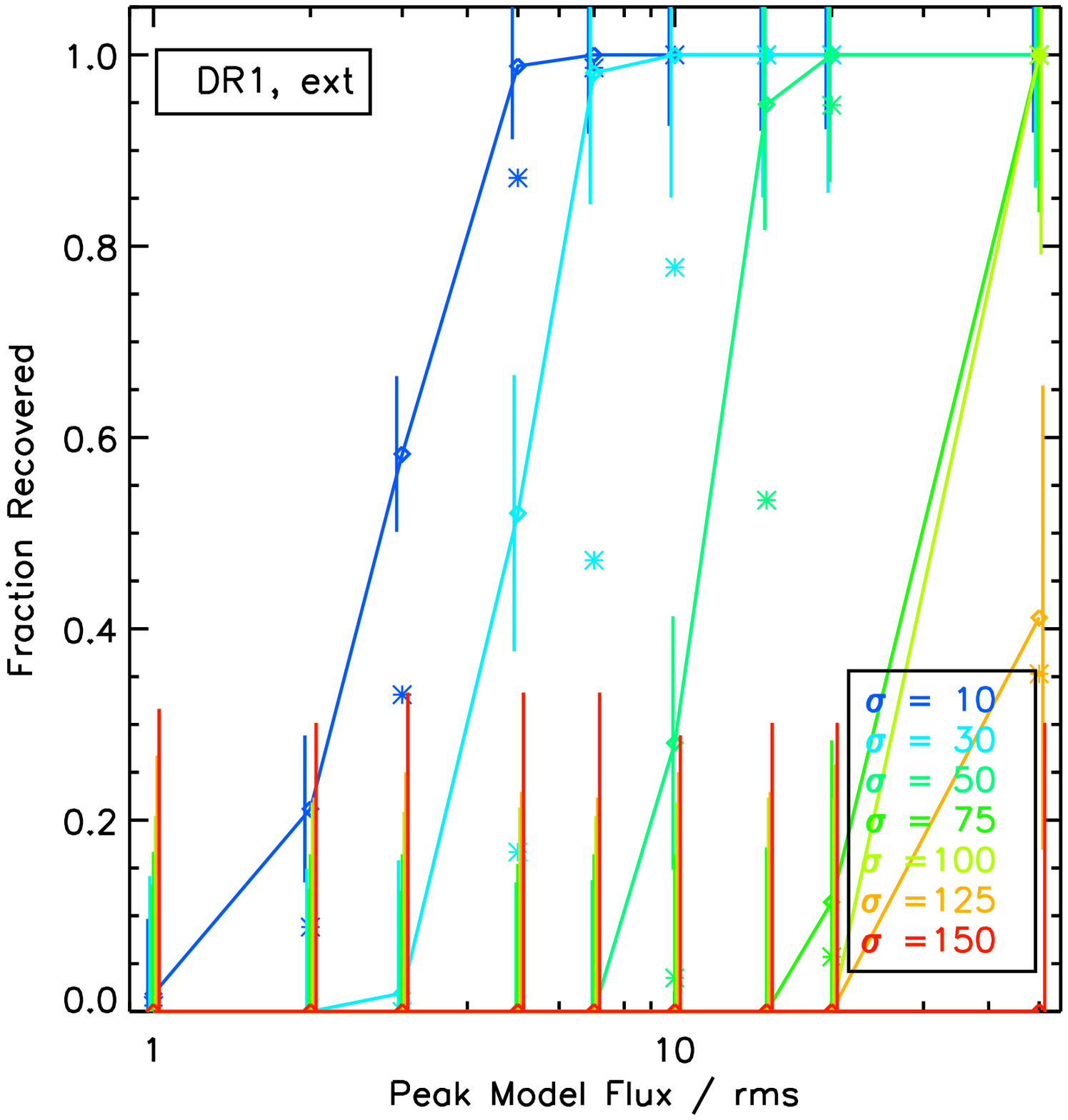} &
\includegraphics[width=3in]{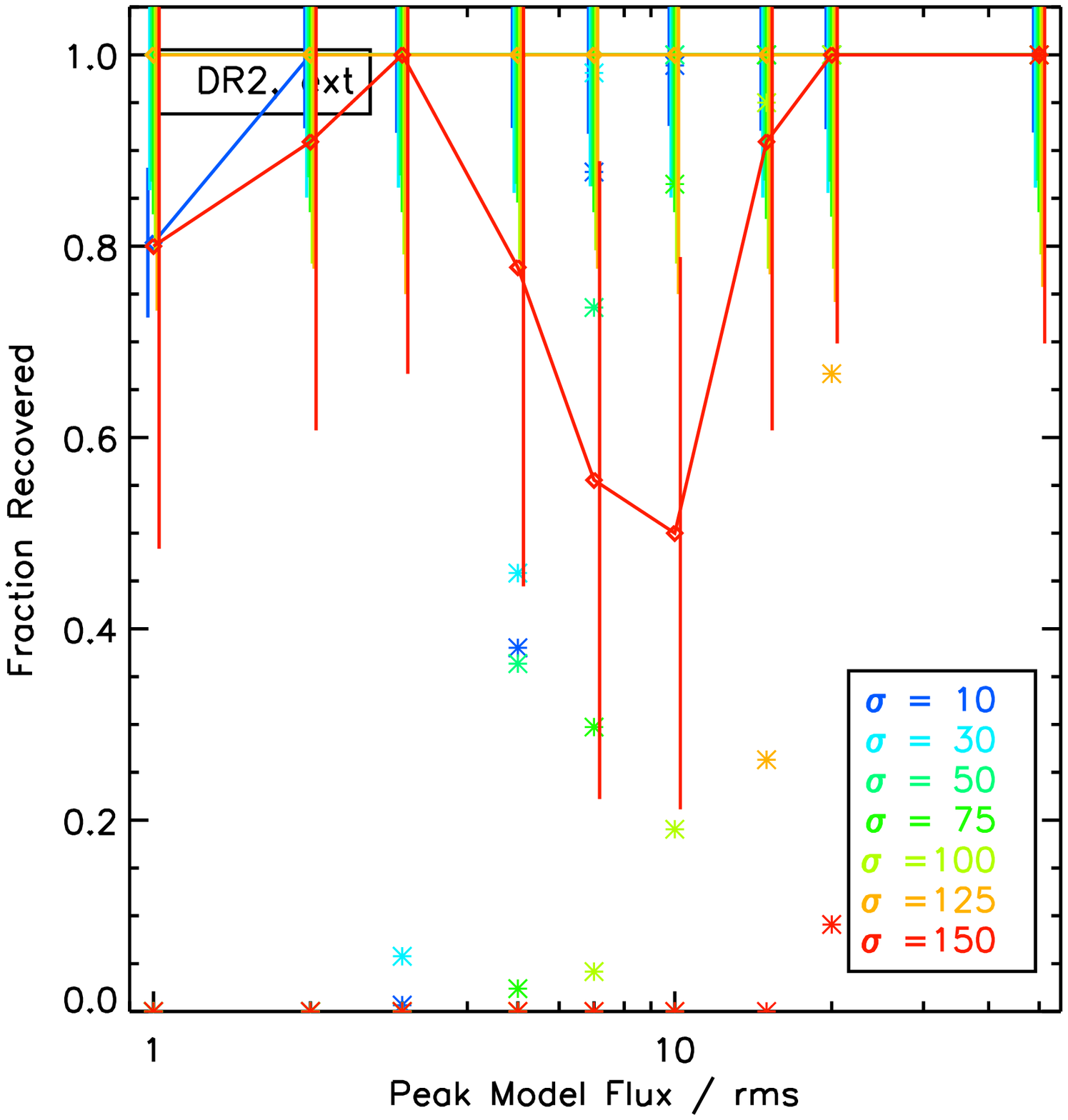} \\
\end{tabular}
\caption{The fraction of artificial Gaussian sources recovered in the background-subtracted
	maps illustrated in Figure~\ref{fig_diff_maps}, as a function of input Gaussian peak
	brightness.  The four panels show the reductions for the automask (top panels) and 
	external-mask (bottom panels) reductions using the GBS DR1 procedure (left panels) 
	and DR2 procedure (right panels).  Different colours denote artificial Gaussians
	of different initial widths, with Gaussian $\sigma$ ranging from 10\arcsec\ to
	150\arcsec.  The vertical bars show counting errors estimated using the
	square root of the total number of Gaussians inserted.}
\label{fig_diff_compl}
\end{figure}

In Figure~\ref{fig_diff_compl}, we show the fraction of artificial Gaussians recovered
within each of the difference maps.  While this measurement is never possible in
real observations, it is helpful to examine the circumstances under which artificial
Gaussian sources pass through the data-reduction pipeline.
Figure~\ref{fig_diff_compl} shows the fraction of artificial Gaussians
that are recovered as a function of Gaussian input peak flux (horizontal axis) 
and split by 
Gaussian input size (different colours).  Across all reduction methods, it is clear that
brighter and more compact Gaussians are the easiest to recover, as expected.  The difference
between DR1 and DR2 is also stark, where larger and fainter structures 
are much more likely
to be lost following the DR1 procedure.
This finding confirms our decision to switch to the DR2 procedure.
We note that DR1 includes a harsher filtering level during the final iteration,
which is undoubtedly responsible for the major loss of larger-scale structures in the
automask reduction compared with DR2.  Such filtering would then propagate through to the 
external-mask reduction through the use of a more compact mask.

\begin{figure}[htb]
\begin{tabular}{cc}
\includegraphics[width=3in]{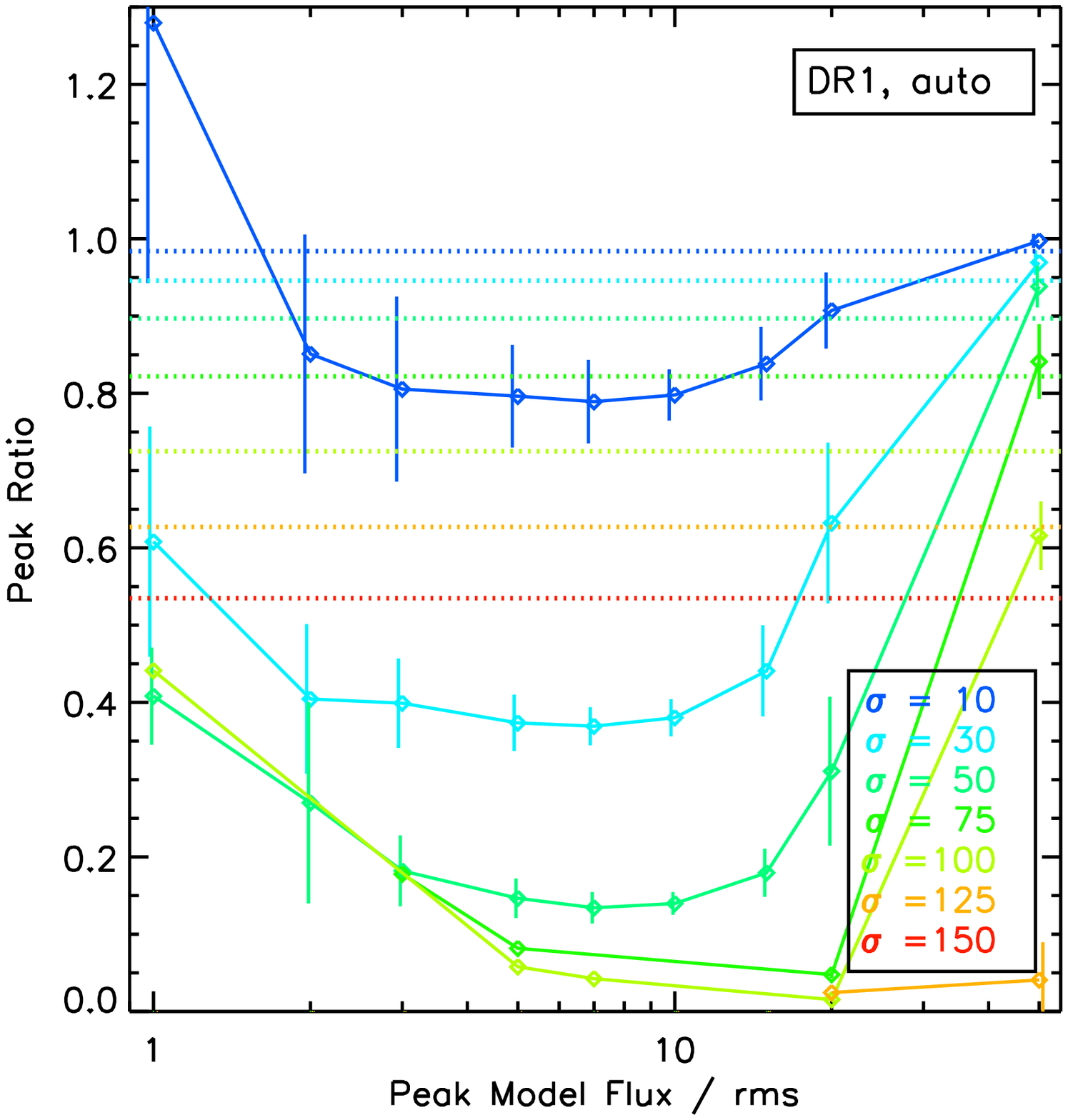} &
\includegraphics[width=3in]{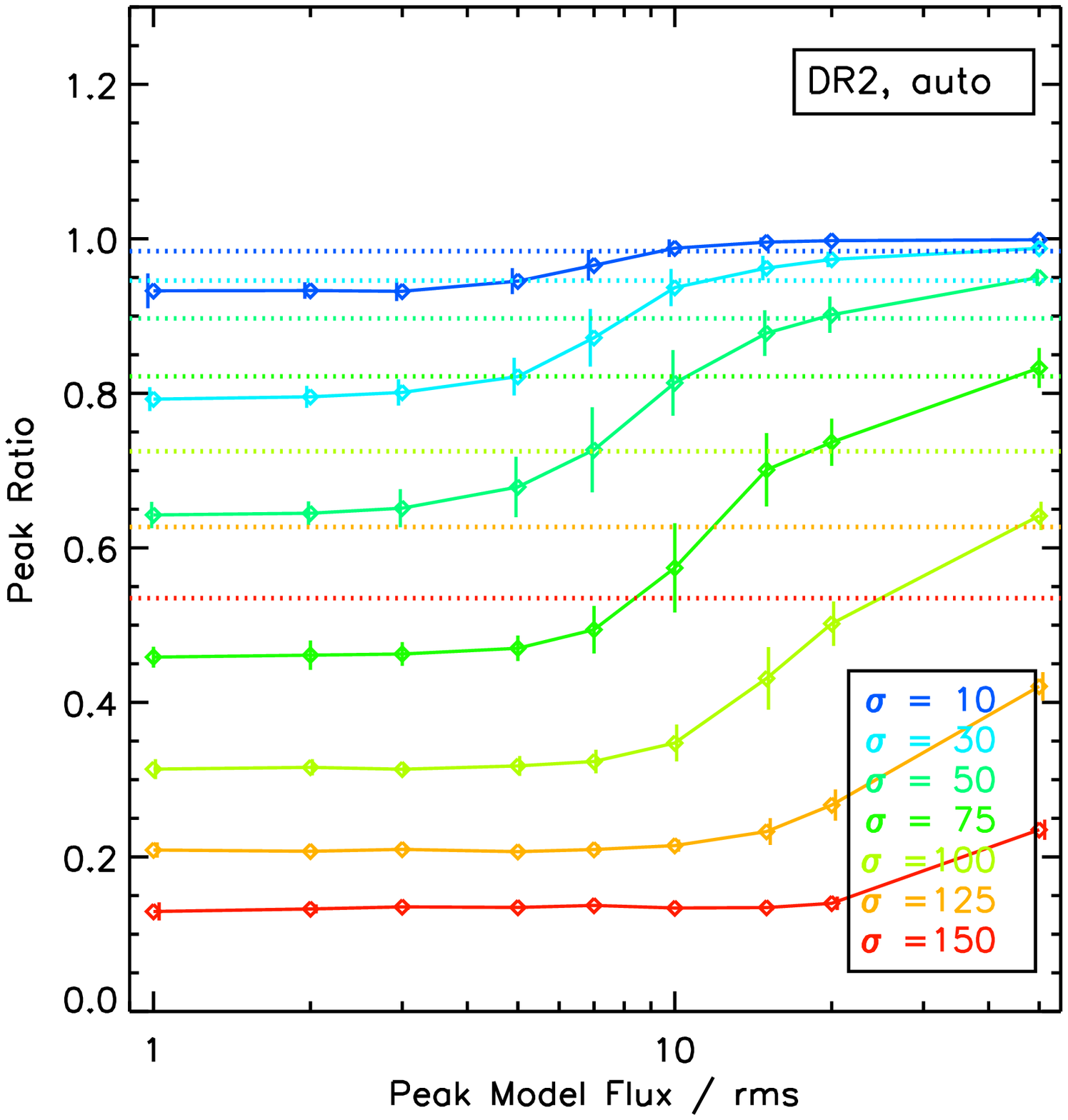} \\
\includegraphics[width=3in]{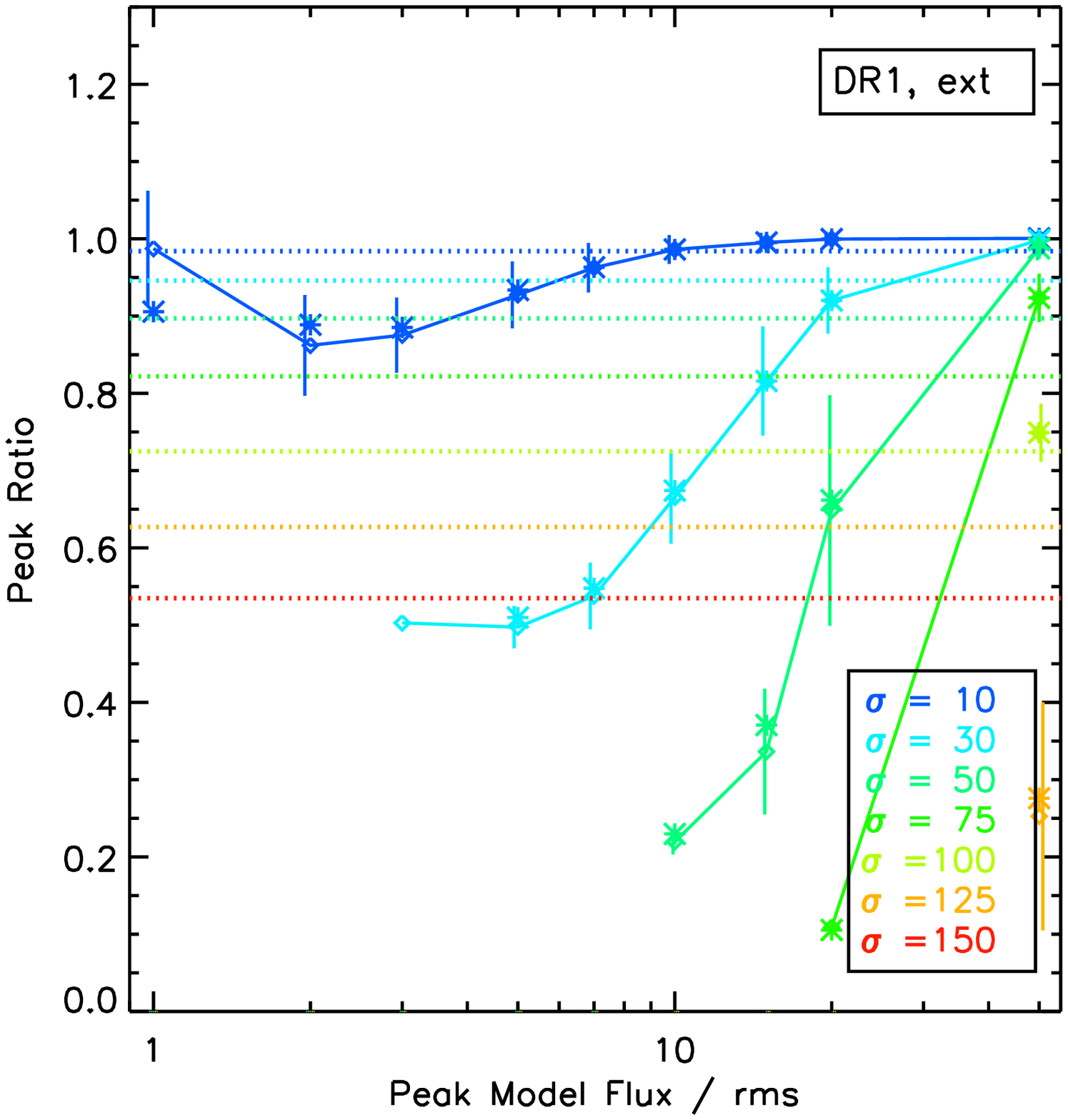} &
\includegraphics[width=3in]{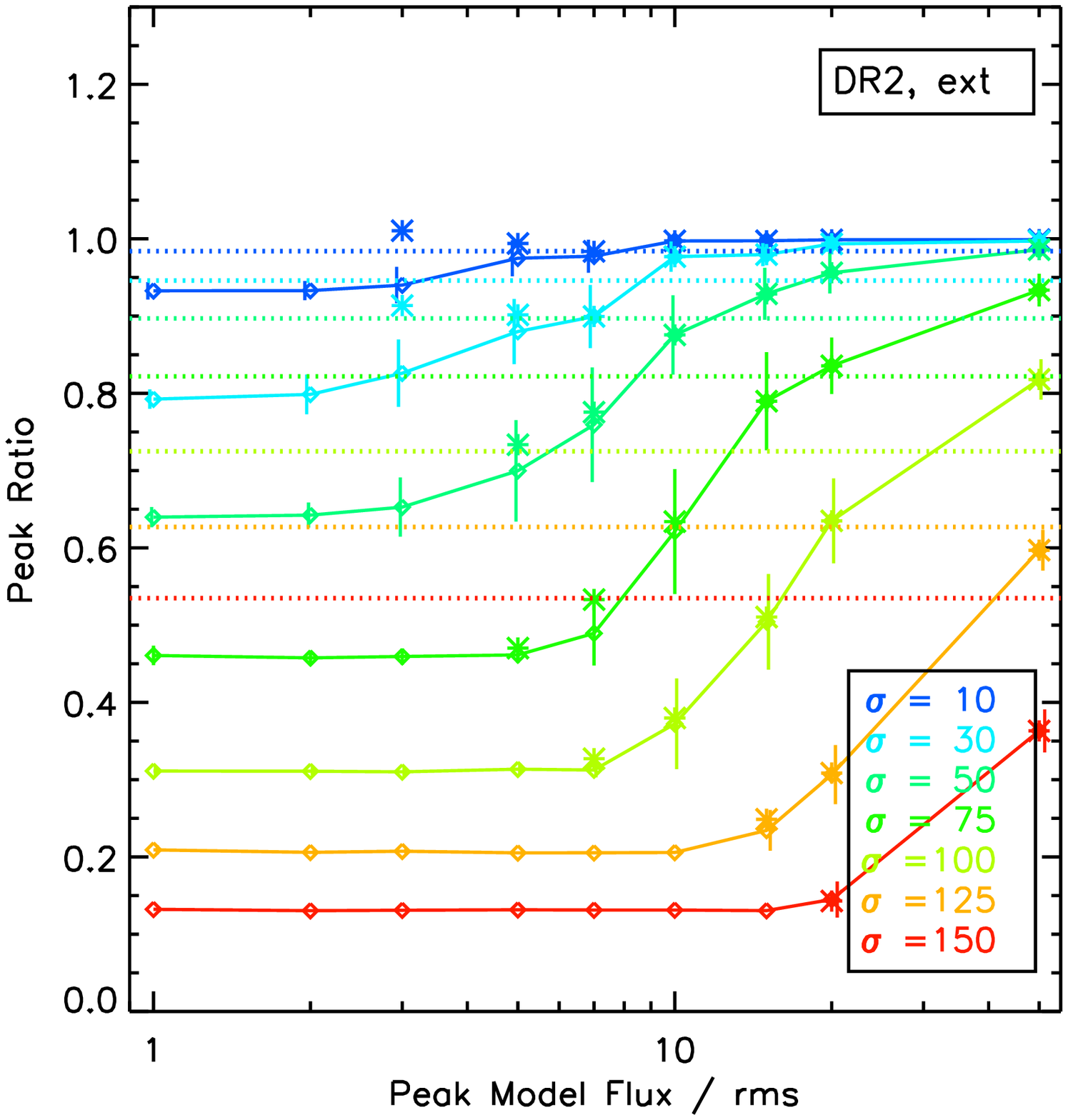} \\
\end{tabular}
\caption{Peak brightnesses measured for the recovered artificial Gaussian sources, as a fraction
        of the input values for the background-subtracted
	maps illustrated in Figure~\ref{fig_diff_maps}, as a function of input Gaussian peak
	brightness.  See Figure~\ref{fig_diff_compl} for the plotting conventions.  Here,
	vertical lines indicate the standard deviation in the values measured for each
	set of Gaussians.  The dotted horizontal lines indicate the expected peak flux
	ratio for sources filtered at a 600\arcsec\ scale.}
\label{fig_diff_peak}
\end{figure}

\begin{figure}[htb]
\begin{tabular}{cc}
\includegraphics[width=3in]{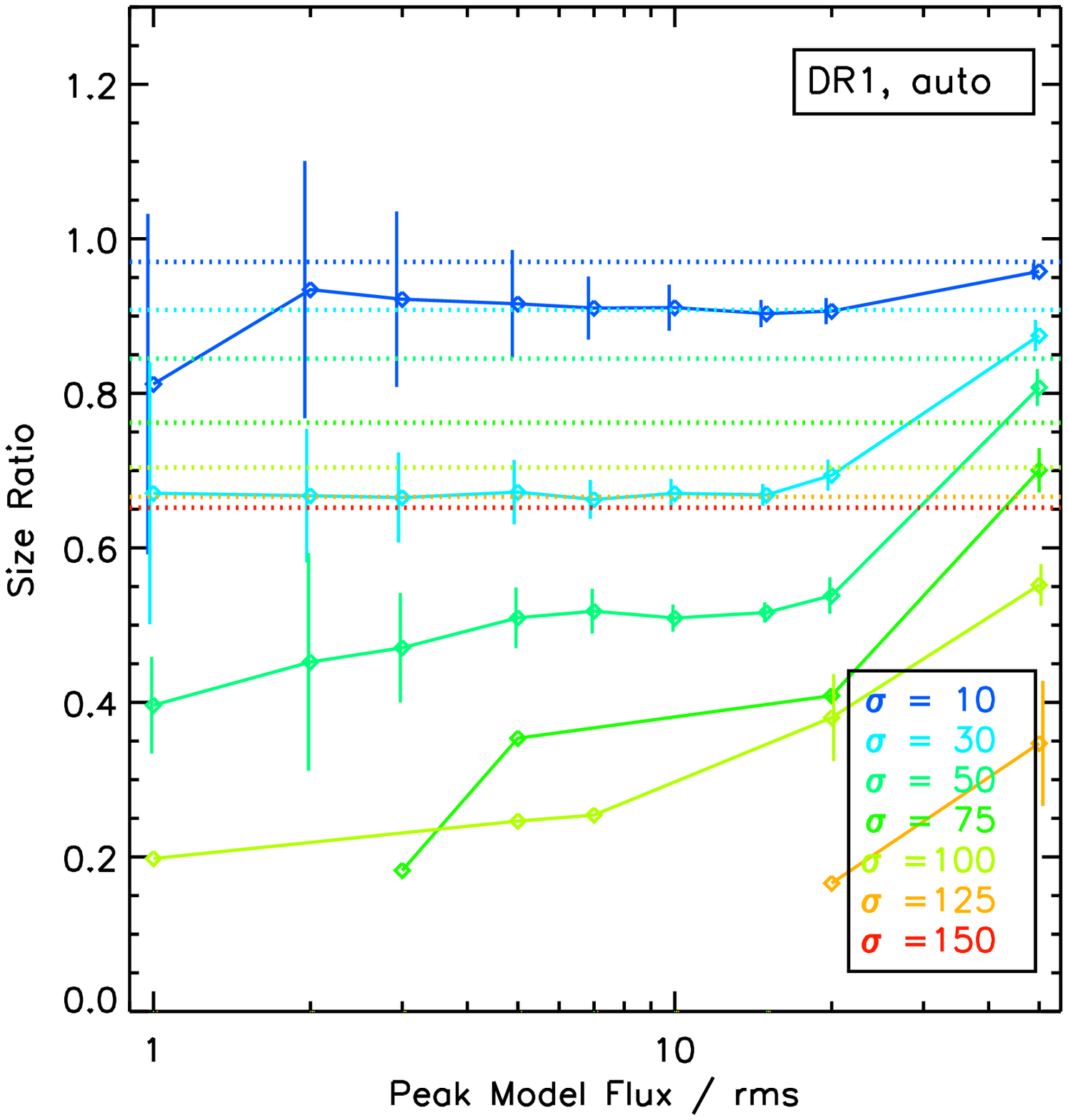} &
\includegraphics[width=3in]{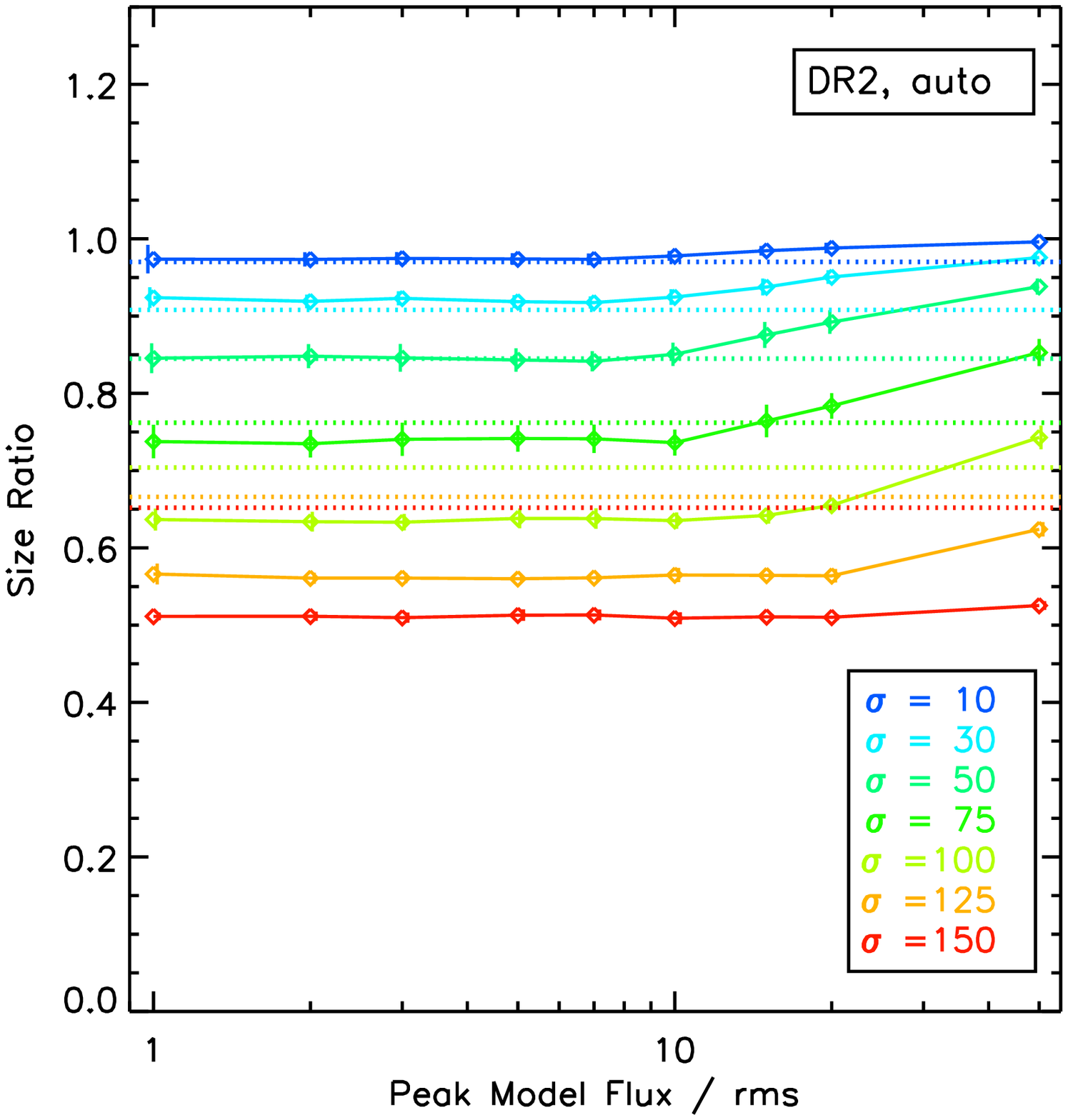} \\
\includegraphics[width=3in]{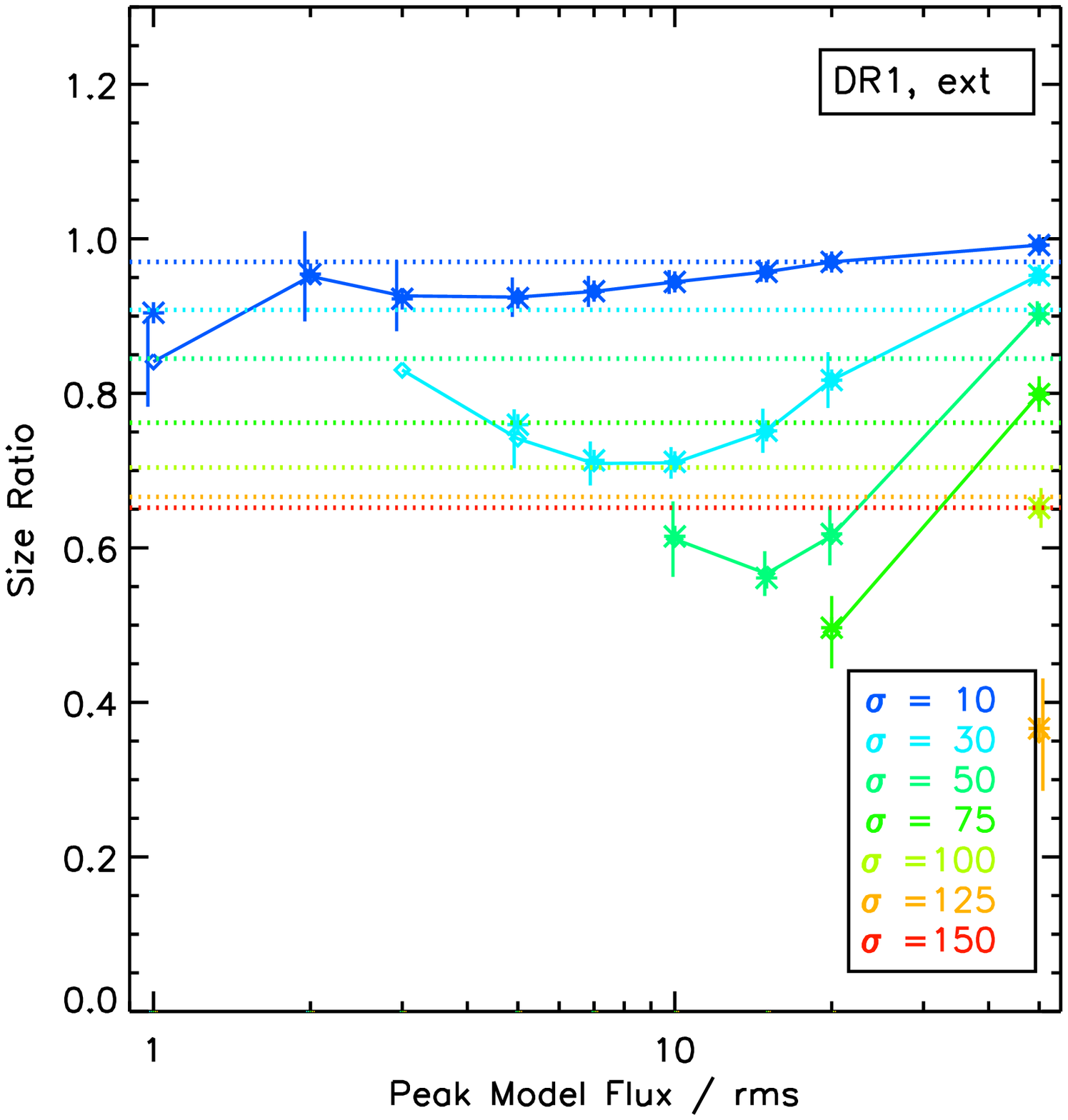} &
\includegraphics[width=3in]{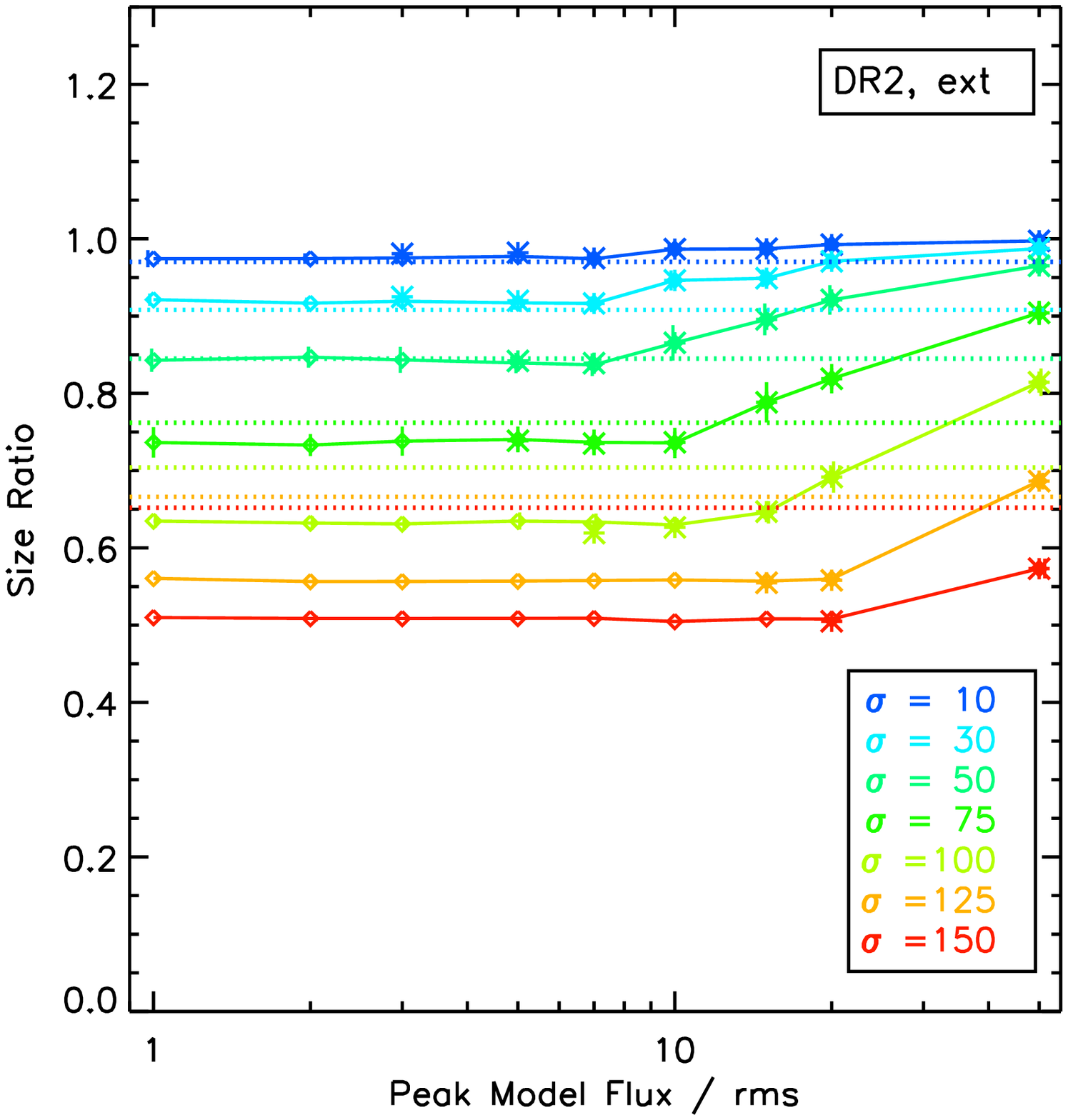} \\
\end{tabular}
\caption{Widths measured for the recovered artificial Gaussian sources, as a fraction
        of the input values for the artificial Gaussian sources recovered in the 
	background-subtracted
	maps illustrated in Figure~\ref{fig_diff_maps}, as a function of input Gaussian peak
	brightness.  
	See Figure~\ref{fig_diff_compl} for the plotting conventions.  Here,
        vertical lines indicate the standard deviation in the values measured for each
        set of Gaussians.  
	The dotted horizontal lines 
	indicate the expected size ratio for sources filtered at a 600\arcsec\ scale.} 
\label{fig_diff_width}
\end{figure}

A comparison between the automask and external-mask reductions shows at best marginal
improvements in the fraction of sources recovered.  This trend is understandable, as 
structures not recovered in the automask reduction will by definition not be included
in the mask used for the external-mask reduction.  Instead, we expect improvements in the
external-mask reduction to come primarily in the form of more accurate recovery of source 
properties (i.e., peak flux, size, and total flux).  Figures~\ref{fig_diff_peak} and 
\ref{fig_diff_width} examine this point in more detail.

Figure~\ref{fig_diff_peak} shows the ratio of the measured peak flux to the initial input 
peak flux for each artificial Gaussian that was found in the difference maps. 
As in Figure~\ref{fig_diff_compl}, a comparison between DR1 and DR2 shows that DR2 
provides significantly more-accurate peak-flux measurements across the entire grid of
artificial Gaussian parameters.  A comparison of the external-mask reductions and the
automask reductions similarly shows that the external-mask reductions improve peak-flux
recovery, particularly for the largest Gaussians.  Despite the overall better performance
of DR2, however, we note that the largest emission structures ($\sigma=150$\arcsec) 
are still poorly recovered, with measured peak fluxes of less than 15\% of their true 
value for moderately bright sources.  
Nevertheless, the GBS is mainly focused on dense cores which have typical sizes
of $\sigma \sim 10$\arcsec\footnote{For GBS cloud distances of 100~pc to 500~pc,
$\sigma = 10$\arcsec\ corresponds to a physical diameter of 0.01~pc to 0.06~pc.}, 
which are generally well recovered.
For a compact Gaussian with a typical flux cutoff of five times the local noise, we 
recover peak fluxes to better than 95\% of their input value.

Figure~\ref{fig_diff_width} similarly shows the Gaussian sizes recovered for each of the 
reductions, plotting the ratio of the recovered Gaussian size to the input size for
all sources that were recovered.  As with the previous figures, DR2 shows a clear
improvement over DR1 in returning accurate source sizes, while the difference between
automask and external-mask reductions is subtler, and is primarily apparent for the largest
and brightest artificial sources.  As in Figure~\ref{fig_diff_peak}, source properties
are poorly recovered for the largest Gaussians, regardless of their peak brightness,
but sources with properties similar to dense cores are well recovered.

\begin{figure}[htb]
\begin{tabular}{cc}
\includegraphics[width=3in]{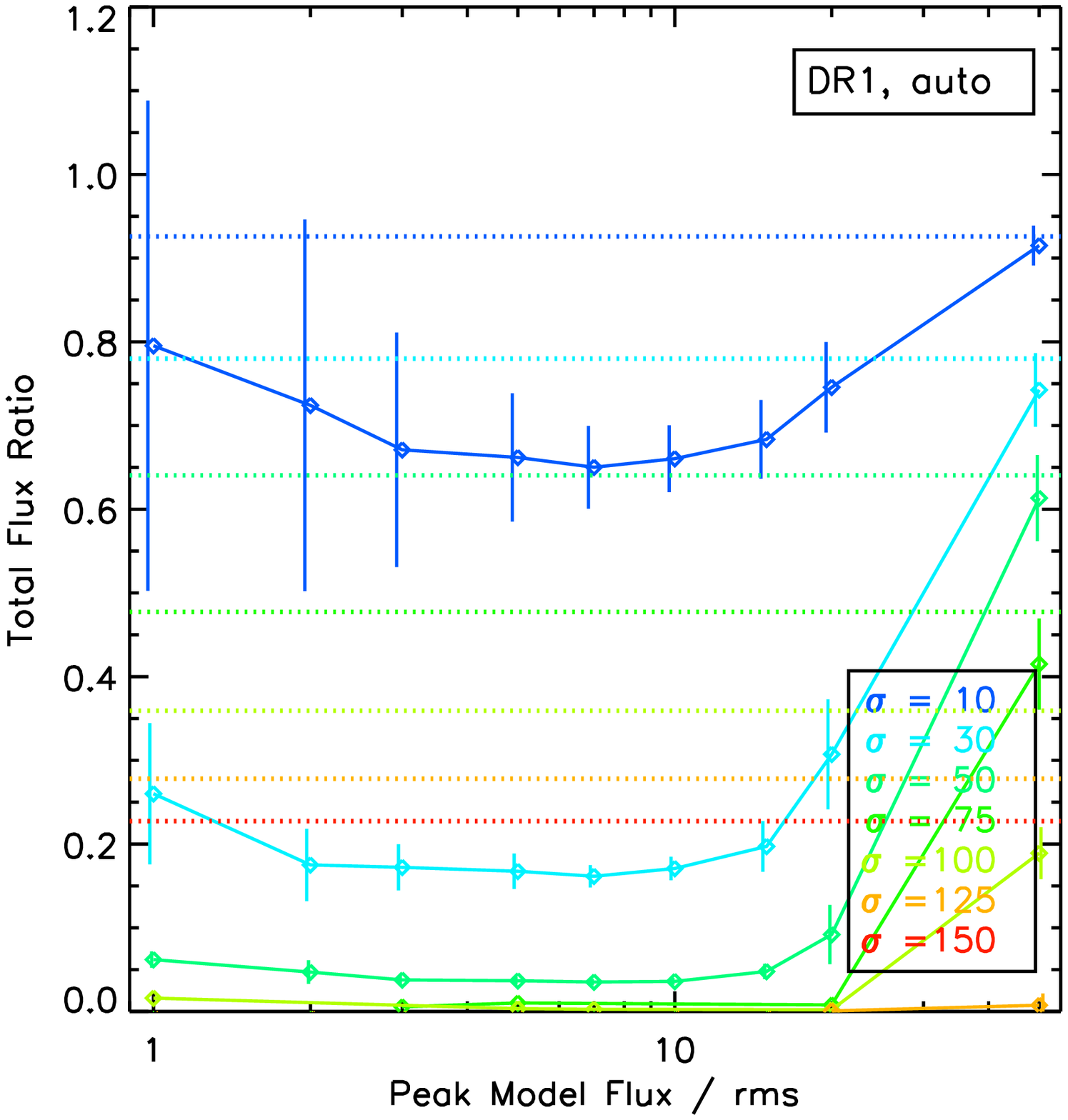} &
\includegraphics[width=3in]{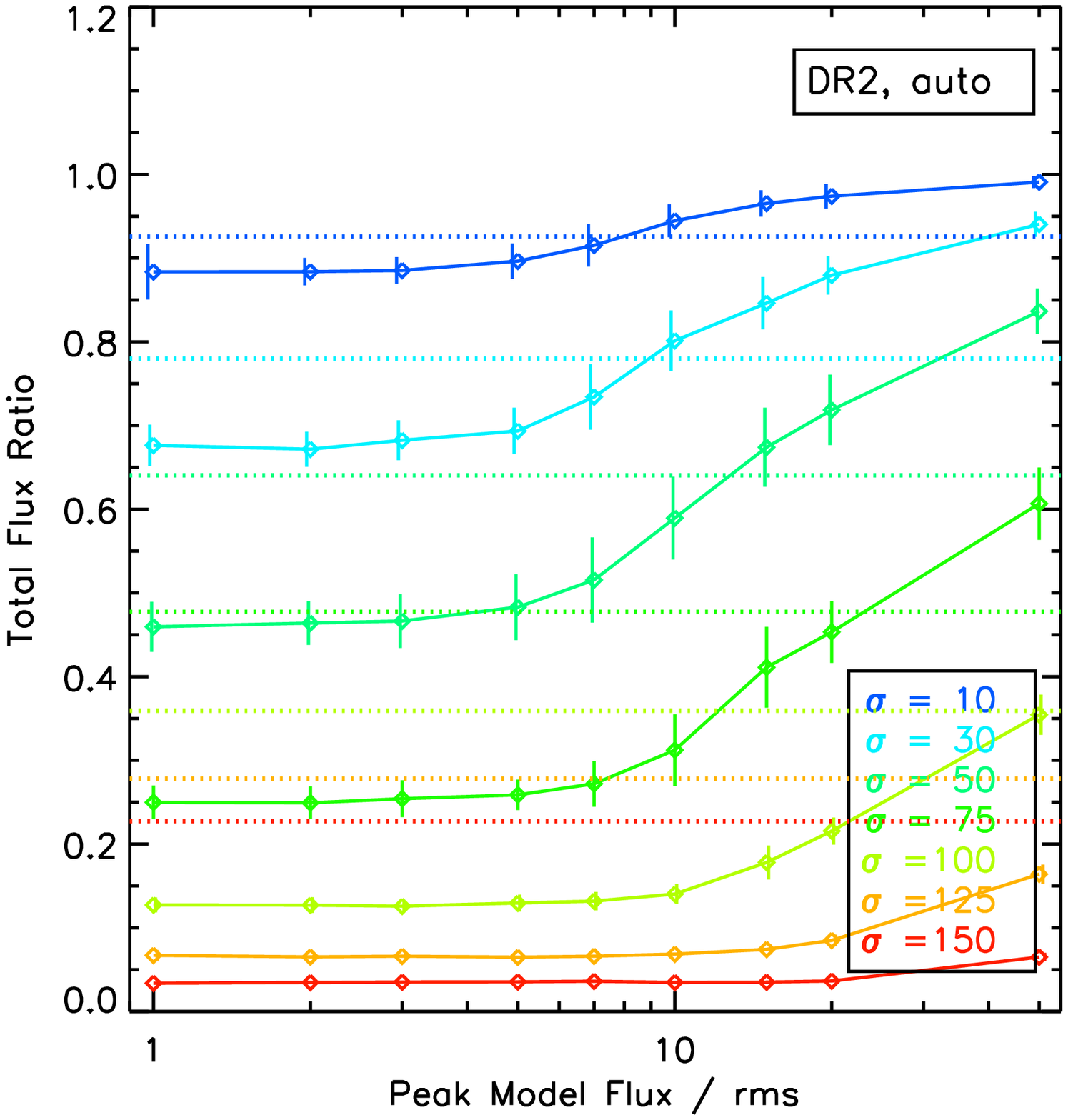} \\
\includegraphics[width=3in]{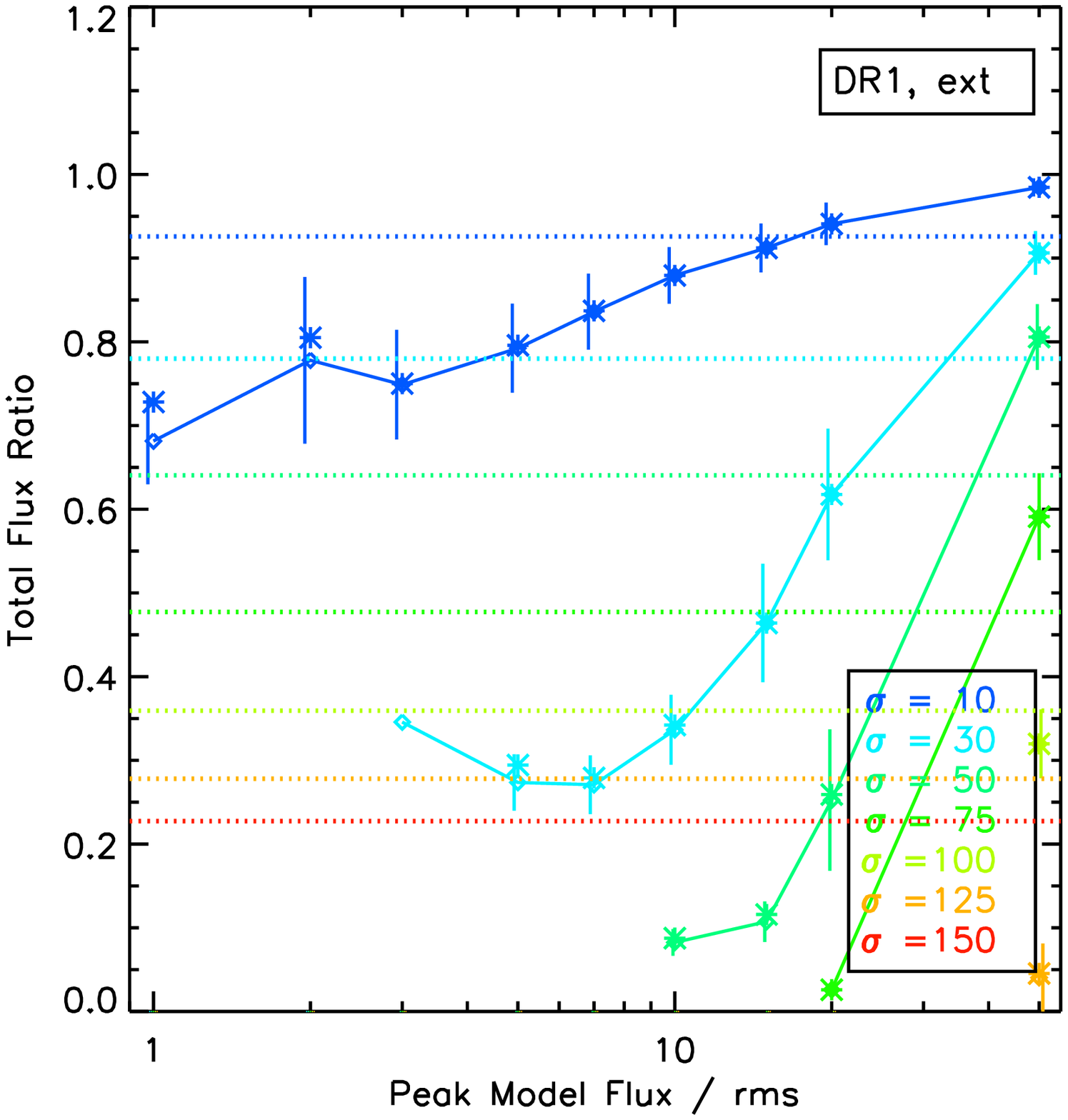} &
\includegraphics[width=3in]{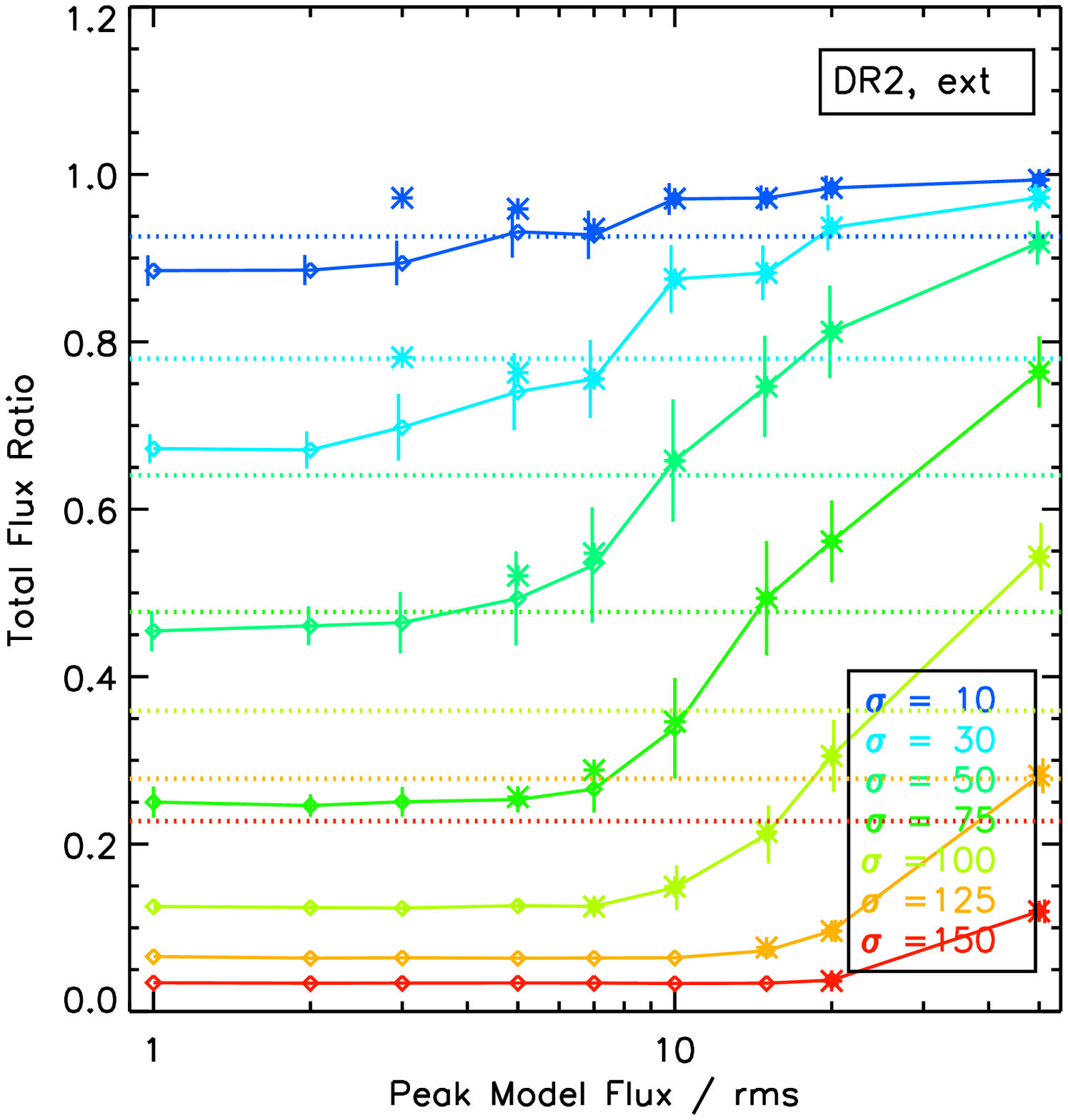} \\
\end{tabular}
\caption{Total fluxes measured for the recovered artificial Gaussian sources, as a fraction
        of the input values for the artificial Gaussian sources recovered in the 
	background-subtracted
	maps illustrated in Figure~\ref{fig_diff_maps}, as a function of input Gaussian peak
	brightness.  
	See Figure~\ref{fig_diff_compl} for the plotting conventions.  Here,
        vertical lines indicate the standard deviation in the values measured for each
        set of Gaussians.  
	The dotted horizontal lines 
	indicate the expected total flux ratio for sources filtered at a 600\arcsec\ scale.}
\label{fig_diff_totflux}
\end{figure}

Finally, Figure~\ref{fig_diff_totflux} similarly shows the ratio of the total flux recovered 
for each of the reductions compared to its input value. 
The DR2 reductions again show a clear improvement over DR1, while the external-mask
reduction improves the total flux recovered for the largest and faintest sources, although
these are still poorly recovered.  The total flux is, however, well recovered for compact
sources.

\section{Large Tables}
Here, we include the large tables discussed earlier in the paper.  In Table~\ref{tab_pong_rms},
we present a summary of the noise properties of each field observed for the survey.

\startlongtable
\begin{deluxetable}{cccccclllc}
\tablecolumns{10}
\tablewidth{0pc}
\tabletypesize{\scriptsize}
\tablecaption{Approximate Noise Per Observed Area\label{tab_pong_rms}}
\tablehead{
\colhead{Region\tablenotemark{a}} &
\colhead{Mosaic\tablenotemark{a}} &
\colhead{Region\tablenotemark{a}}&
\colhead{Field\tablenotemark{a}} &
\colhead{R.A.\tablenotemark{a}} & 
\colhead{Decl\tablenotemark{a}} &
\colhead{N$_\mathrm{obs}$\tablenotemark{b}} &
\colhead{rms$_\mathrm{850}$\tablenotemark{b}} &
\colhead{rms$_\mathrm{450}$\tablenotemark{b}} &
\colhead{Notes\tablenotemark{a}} \\
\colhead{Name} & \colhead{Name} & \colhead{Code} & \colhead{} & 
\colhead{(J2000)} & \colhead{(J2000)} &
\colhead{} &
\multicolumn{2}{c}{(mJy~arcsec$^{-2}$)} &
\colhead{}
}

\startdata
            Perseus &        PerseusWest & MJLSG38 &          L1448-S &   03:25:21.4 &  30:15:46.9 & 7 & 0.054 &  1.63 & \\
            Perseus &        PerseusWest & MJLSG38 &          L1448-N &   03:25:25.1 &  30:41:57.0 & 4 & 0.046 &  0.61 & \\
            Perseus &        PerseusWest & MJLSG38 &          L1455-S &   03:28:00.2 &  30:09:26.1 & 4 & 0.052 &  0.88 & \\
            Perseus &        PerseusWest & MJLSG38 &        NGC1333-S &   03:28:40.1 &  30:53:49.1 & 6 & 0.054 &  1.81 & \\
            Perseus &        PerseusWest & MJLSG38 &        NGC1333-N &  03:29:06.9 &  31:22:44.7 & 5 & 0.049 &  1.28 & * \\
            Perseus &        PerseusWest & MJLSG38 &          L1455-N &   03:29:44.6 &  30:27:12.6 & 6 & 0.057 &  1.79 & \\
            Perseus &        PerseusWest & MJLSG38 &             B1-S &   03:31:32.9 &  30:46:05.2 & 6 & 0.047 &  1.02 & \\
            Perseus &        PerseusWest & MJLSG38 &               B1 &  03:33:12.0 &  31:07:18.0 & 8 & 0.045 &  0.63 & * \\
            Perseus &        PerseusWest & MJLSG38 &             B1-E &   03:36:29.2 &  31:12:58.1 & 7 & 0.045 &  1.35 & \\
            Perseus &       PerseusIC348 & MJLSG38 &          IC348-W &   03:39:49.8 &  31:54:24.4 & 6 & 0.047 &  1.01 & \\
            Perseus &       PerseusIC348 & MJLSG38 &          IC348-C &   03:42:10.5 &  31:51:47.5 & 6 & 0.053 &  1.41 & \\
            Perseus &       PerseusIC348 & MJLSG38 &          IC348-E &   03:44:23.6 &  32:02:03.1 & 4 & 0.051 &  0.78 & \\
            Perseus &       PerseusIC348 & MJLSG38 &               B5 &   03:47:37.4 &  32:52:36.5 & 6 & 0.047 &  1.18 & \\
      Taurus/Auriga &        AurigaNorth & MJLSG37 &           AUR\_NW &   04:10:08.3 &  40:07:55.2 & 6 & 0.049 &  1.03 & \\
      Taurus/Auriga &        AurigaNorth & MJLSG37 &    AUR\_CENTRAL-N &   04:10:50.3 &  38:09:23.3 & 5 & 0.059 &  2.33 & \\
      Taurus/Auriga &        TaurusL1495 & MJLSG37 &     L1495-1800-2 &   04:14:11.1 &  28:14:11.0 & 6 & 0.051 &  1.53 & *,e \\
      Taurus/Auriga &        TaurusL1495 & MJLSG37 &        L1495\_NW &   04:16:40.5 &  28:37:05.0 & 2 & 0.122 &  5.47 & *,f \\
      Taurus/Auriga &        TaurusL1495 & MJLSG37 &        L1495\_SW &   04:16:40.5 &  28:05:25.0 & 2 & 0.127 &  6.44 & *,f \\
      Taurus/Auriga &        TaurusL1495 & MJLSG37 &     L1592-1800-1  &   04:17:54.4 &  27:47:55.3 & 6 & 0.047 &  1.13 & \\
      Taurus/Auriga &        TaurusL1495 & MJLSG37 &     L1495-1800-1 &   04:17:54.5 &  28:18:45.3 & 6 & 0.053 &  1.07 & *,e \\
      Taurus/Auriga &        TaurusL1495 & MJLSG37 &     L1592-1800-2  &   04:18:48.9 &  27:19:32.4 & 6 & 0.048 &  1.11 & \\
      Taurus/Auriga &        TaurusL1495 & MJLSG37 &     L1592-1800-3  &   04:20:49.1 &  27:04:07.4 & 6 & 0.052 &  1.55 & \\
      Taurus/Auriga &      AurigaCentral & MJLSG37 &    AUR\_CENTRAL-W &   04:20:47.3 &  37:29:31.8 & 6 & 0.049 &  0.88 & \\
      Taurus/Auriga &      AurigaCentral & MJLSG37 &    AUR\_CENTRAL-E &   04:25:10.4 &  37:10:09.2 & 6 & 0.049 &  0.89 & \\
      Taurus/Auriga &        TaurusSouth & MJLSG37 &     TAURUSSOUTH5  &   04:23:21.7 &  25:03:35.9 & 6 & 0.049 &  1.11 & \\
      Taurus/Auriga &        TaurusSouth & MJLSG37 &     TAURUSSOUTH4 &   04:26:57.6 &  24:35:07.5 & 6 & 0.050 &  0.82 & \\
      Taurus/Auriga &        TaurusSouth & MJLSG37 &     TAURUSSOUTH3 &   04:29:38.4 &  24:34:57.9 & 6 & 0.051 &  1.26 & \\
      Taurus/Auriga &        TaurusSouth & MJLSG37 &     TAURUSSOUTH2 &   04:32:20.9 &  24:23:50.3 & 6 & 0.061 &  1.80 & \\
      Taurus/Auriga &        TaurusSouth & MJLSG37 &     TAURUSSOUTH1 &   04:35:17.3 &  24:07:43.6 & 6 & 0.052 &  1.21 & \\
      Taurus/Auriga &      AurigaLkHa101 & MJLSG37 &       LKHA-101-S &   04:30:16.3 &  35:17:50.9 & 4 & 0.051 &  0.73 & \\
      Taurus/Auriga &      AurigaLkHa101 & MJLSG37 &       LKHA-101-N &   04:30:42.5 &  35:48:12.8 & 4 & 0.055 &  1.09 & \\
      Taurus/Auriga &          TaurusTMC & MJLSG37 &           TMC1-N &   04:38:54.6 &  26:23:47.8 & 6 & 0.047 &  1.05 & \\
      Taurus/Auriga &          TaurusTMC & MJLSG37 &          TMC1-SW &   04:39:02.0 &  25:52:10.4 & 6 & 0.056 &  1.68 & \\
      Taurus/Auriga &          TaurusTMC & MJLSG37 &          TMC1-NE &   04:41:00.3 &  26:09:25.2 & 7 & 0.059 &  2.22 & \\
      Taurus/Auriga &          TaurusTMC & MJLSG37 &           TMC1-S &   04:41:07.7 &  25:37:47.8 & 6 & 0.058 &  2.14 & \\
            Orion~A &            Orion~A & MJLSG31 &      OMC1\_TILE17 &   05:33:09.6 & -05:37:52.0 & 7 & 0.046 &  0.88 & \\
            Orion~A &            Orion~A & MJLSG31 &       OMC1\_TILE1 &  05:34:20.7 & -05:09:53.4 & 4 & 0.056 &  1.02 & * \\
            Orion~A &            Orion~A & MJLSG31 &       OMC1\_TILE2 &   05:34:57.7 & -05:40:10.4 & 4 & 0.053 &  0.75 & \\
            Orion~A &            Orion~A & MJLSG31 &  OMC1\_ISF\_CENTRE & 05:35:14.2 & -05:22:21.5 & 1 & 0.179 &  6.72 & *,c \\
            Orion~A &            Orion~A & MJLSG31 &      OMC1\_TILE56 &   05:35:45.8 & -06:07:04.8 & 4 & 0.053 &  1.05 & \\
            Orion~A &            Orion~A & MJLSG31 &       OMC1\_TILE4 &   05:35:51.2 & -04:46:25.5 & 5 & 0.051 &  0.76 & \\
            Orion~A &            Orion~A & MJLSG31 &       OMC1\_TILE7 &   05:36:13.2 & -06:31:42.8 & 6 & 0.044 &  0.66 & \\
            Orion~A &            Orion~A & MJLSG31 &       OMC1\_TILE9 &   05:38:16.4 & -06:39:54.6 & 7 & 0.045 &  1.27 & \\
            Orion~A &            Orion~A & MJLSG31 &       OMC1\_TILE3 &   05:36:24.1 & -05:17:00.8 & 4 & 0.052 &  0.67 & \\
            Orion~A &            Orion~A & MJLSG31 &       OMC1\_TILE8 &   05:36:46.2 & -07:02:18.1 & 6 & 0.049 &  1.22 & \\
            Orion~A &            Orion~A & MJLSG31 &      OMC1\_TILE10 &  05:38:49.5 & -07:10:29.9 & 6 & 0.048 &  1.34 & * \\
            Orion~A &            Orion~A & MJLSG31 &      OMC1\_TILE11 &  05:40:07.2 & -07:33:28.8 & 6 & 0.043 &  0.88 & * \\
            Orion~A &            Orion~A & MJLSG31 &      OMC1\_TILE16 &   05:40:58.3 & -09:04:00.2 & 6 & 0.046 &  1.07 & \\
            Orion~A &            Orion~A & MJLSG31 &      OMC1\_TILE12 &   05:40:58.4 & -08:00:40.2 & 7 & 0.046 &  1.18 & \\
            Orion~A &            Orion~A & MJLSG31 &      OMC1\_TILE14 &   05:40:58.4 & -08:32:20.2 & 6 & 0.047 &  1.13 & \\
            Orion~A &            Orion~A & MJLSG31 &      OMC1\_TILE13 &   05:42:49.2 & -08:16:30.2 & 7 & 0.045 &  1.11 & \\
            Orion~A &            Orion~A & MJLSG31 &      OMC1\_TILE15 &   05:42:49.3 & -08:48:10.2 & 6 & 0.048 &  1.01 & \\
            Orion~B &       OrionB\_N2023 & MJLSG41 &    ORIONBS\_450\_W &   05:40:33.9 & -01:48:50.9 & 4 & 0.052 &  0.92 & \\
            Orion~B &       OrionB\_N2023 & MJLSG41 &    ORIONBS\_450\_S &   05:41:17.3 & -02:18:36.3 & 4 & 0.051 &  0.75 & g \\
            Orion~B &       OrionB\_N2023 & MJLSG41 &    ORIONBS\_850\_S &   05:41:55.4 & -01:24:35.4 & 7 & 0.043 &  1.17 & \\
            Orion~B &       OrionB\_N2023 & MJLSG41 &    ORIONBS\_450\_E &   05:42:38.8 & -01:54:20.8 & 6 & 0.049 &  1.07 & \\
            Orion~B &       OrionB\_N2023 & MJLSG41 &    ORIONBS\_850\_N &   05:43:39.4 & -01:09:35.4 & 6 & 0.047 &  0.99 & \\
            Orion~B &       OrionB\_N2068 & MJLSG41 &    ORIONBN\_450\_W &   05:45:56.6 &  00:24:38.9 & 6 & 0.055 &  1.70 & \\
            Orion~B &       OrionB\_N2068 & MJLSG41 &    ORIONBN\_450\_S &   05:46:18.6 & -00:06:32.3 & 6 & 0.050 &  1.14 & \\
            Orion~B &       OrionB\_N2068 & MJLSG41 &    ORIONBN\_850\_N &   05:47:33.6 &  00:45:00.1 & 6 & 0.047 &  0.94 & \\
            Orion~B &       OrionB\_N2068 & MJLSG41 &    ORIONBN\_450\_E &   05:47:55.6 &  00:13:49.0 & 6 & 0.050 &  1.03 & \\
            Orion~B &       OrionB\_L1622 & MJLSG41 & ORIONBN\_850\_solo &   05:54:33.0 &  01:49:04.7 & 6 & 0.053 &  1.95 & \\
              Lupus &              Lupus & MJLSG34 &        LUPUSI-SW &   15:39:33.7 & -34:41:30.0 & 7 & 0.063 &  5.24 & d \\
              Lupus &              Lupus & MJLSG34 &        LUPUSI-NW &   15:42:45.3 & -34:04:30.3 & 6 & 0.053 &  1.70 & \\
              Lupus &              Lupus & MJLSG34 &         LUPUSI-E &   15:45:22.8 & -34:21:31.7 & 6 & 0.060 &  2.88 & \\
 Ophiuchus/Scorpius &           OphScoN6 & MJLSG32 &           OPHN-6 &   16:21:09.5 & -20:07:01.4 & 7 & 0.045 &  1.64 & \\
 Ophiuchus/Scorpius &         OphScoMain & MJLSG32 &          L1688-3 &   16:25:09.3 & -24:23:47.7 & 5 & 0.048 &  1.20 & \\
 Ophiuchus/Scorpius &         OphScoMain & MJLSG32 &          L1688-1 &   16:27:03.5 & -24:41:57.5 & 4 & 0.053 &  1.10 & \\
 Ophiuchus/Scorpius &         OphScoMain & MJLSG32 &          L1688-2 &   16:27:15.7 & -24:10:24.7 & 4 & 0.058 &  1.31 & \\
 Ophiuchus/Scorpius &         OphScoMain & MJLSG32 &          L1688-4 &   16:29:09.9 & -24:28:34.5 & 4 & 0.057 &  1.41 & \\
 Ophiuchus/Scorpius &         OphScoMain & MJLSG32 &          L1689-2 &   16:32:04.5 & -24:58:26.4 & 6 & 0.057 &  2.57 & \\
 Ophiuchus/Scorpius &         OphScoMain & MJLSG32 &          L1689-1 &   16:32:27.4 & -24:28:53.7 & 6 & 0.047 &  1.70 & \\
 Ophiuchus/Scorpius &         OphScoMain & MJLSG32 &          L1709-1 &   16:32:19.1 & -23:56:40.9 & 7 & 0.050 &  1.73 & \\
 Ophiuchus/Scorpius &         OphScoMain & MJLSG32 &          L1689-3 &   16:34:34.8 & -24:36:39.6 & 6 & 0.050 &  1.87 & \\
 Ophiuchus/Scorpius &         OphScoMain & MJLSG32 &          L1712-1 &   16:39:02.2 & -24:14:40.8 & 6 & 0.053 &  1.67 & \\
 Ophiuchus/Scorpius &           OphScoN2 & MJLSG32 &           OPHN-2 &   16:47:40.0 & -12:05:00.0 & 6 & 0.048 &  1.49 & \\
 Ophiuchus/Scorpius &           OphScoN3 & MJLSG32 &           OPHN-3 &   16:50:53.6 & -15:21:46.0 & 6 & 0.051 &  1.88 & \\
               Pipe &            PipeB59 & MJLSG39 &         PIPE-B59 &   17:11:33.4 & -27:26:35.2 & 6 & 0.045 &  0.69 & \\
               Pipe &             PipeE1 & MJLSG39 &          PIPE-E1 &   17:34:06.9 & -25:39:27.4 & 6 & 0.049 &  1.34 & \\
     Serpens/Aquila &      SerpensMWC297 & MJLSG33 &     SERPENS-MWC297 & 18:28:13.8 & -03:43:58.3 & 6 & 0.053 &  1.72 & \\
     Serpens/Aquila &        SerpensMain & MJLSG33 &       SERPENSNH3 &   18:29:11.2 &  00:28:37.7 & 6 & 0.045 &  0.74 & \\
     Serpens/Aquila &        SerpensMain & MJLSG33 &     SERPENSMAIN1 &   18:29:59.7 &  01:14:21.9 & 4 & 0.060 &  0.99 & \\
     Serpens/Aquila &             Aquila & MJLSG33 &      SERPENSS-NW &   18:29:30.6 & -01:47:30.3 & 5 & 0.051 &  0.79 & \\
     Serpens/Aquila &             Aquila & MJLSG33 &      SERPENSS-SW &   18:30:09.8 & -02:17:37.3 & 5 & 0.051 &  0.79 & \\
     Serpens/Aquila &             Aquila & MJLSG33 &      SERPENSS-NE &   18:31:34.6 & -01:54:05.3 & 4 & 0.053 &  0.66 & \\
     Serpens/Aquila &             Aquila & MJLSG33 &      SERPENSS-SE &   18:32:13.8 & -02:24:12.3 & 7 & 0.045 &  1.16 & \\
     Serpens/Aquila &           SerpensE & MJLSG33 &       SERPENS-E3 &   18:36:27.4 & -01:17:45.4 & 6 & 0.046 &  1.32 & \\
     Serpens/Aquila &           SerpensE & MJLSG33 &       SERPENS-E1 &   18:37:48.8 & -01:42:00.9 & 6 & 0.050 &  1.69 & \\
     Serpens/Aquila &           SerpensE & MJLSG33 &       SERPENS-E2 &   18:38:32.1 & -01:12:15.5 & 6 & 0.049 &  1.82 & \\
     Serpens/Aquila &           SerpensN & MJLSG33 &        SERPENS-N &   18:39:05.5 &  00:27:26.6 & 8 & 0.042 &  0.70 & \\
   Corona~Australis &                CrA & MJLSG35 &            CRA-1 &   19:01:35.0 & -36:55:56.6 & 6 & 0.053 &  2.04 & \\
   Corona~Australis &                CrA & MJLSG35 &            CRA-2 &   19:03:34.1 & -37:13:58.6 & 5 & 0.060 &  2.37 & \\
   Corona~Australis &                CrA & MJLSG35 &            CRA-E &   19:10:23.8 & -37:07:53.6 & 6 & 0.051 &  1.16 & \\
            Cepheus &       CepheusSouth & MJLSG40 &          L1157-W &   20:37:24.0 &  67:57:31.9 & 6 & 0.049 &  2.11 & \\
            Cepheus &       CepheusSouth & MJLSG40 &          L1157-E &   20:44:10.8 &  67:50:05.9 & 7 & 0.052 &  2.36 & \\
            Cepheus &       CepheusSouth & MJLSG40 &          L1172-N &   21:01:37.3 &  68:14:21.3 & 4 & 0.064 &  0.98 & \\
            Cepheus &       CepheusSouth & MJLSG40 &          L1172-S &   21:02:33.5 &  67:44:48.6 & 6 & 0.051 &  2.12 & \\
            Cepheus &       CepheusL1228 & MJLSG40 &            L1228 &   20:57:42.9 &  77:38:19.9 & 6 & 0.046 &  0.80 & \\
            Cepheus &       CepheusL1251 & MJLSG40 &          L1251-W &   22:29:41.4 &  75:14:53.0 & 6 & 0.049 &  0.88 & \\
            Cepheus &       CepheusL1251 & MJLSG40 &          L1251-E &   22:37:32.7 &  75:14:53.0 & 6 & 0.052 &  1.82 & \\
             IC5146 &             IC5146 & MJLSG36 &         IC5146-W &   21:45:35.3 &  47:37:05.1 & 6 & 0.048 &  1.40 & \\
             IC5146 &             IC5146 & MJLSG36 &         IC5146-E &   21:48:30.3 &  47:31:52.5 & 6 & 0.045 &  1.03 & \\
             IC5146 &             IC5146 & MJLSG36 &        IC5146-H2 &   21:53:42.6 &  47:15:24.2 & 6 & 0.050 &  1.39 & \\
\enddata
\tablenotetext{a}{Region name, mosaic name, region code, field name, and centre position.  
	The region name corresponds to the name of the molecular cloud, while the
	mosaic name corresponds to the GBS name given for each mosaic of (usually
	contiguous) SCUBA-2 observations.  There are several mosaics for some
	molecular clouds.  The region (or observing) code
	and field name are listed in the JCMT archive as Proposal ID and Target Name 
	({\tt http://www.cadc-ccda.hia-iha.nrc-cnrc.gc.ca/en/jcmt/}).  Observations taken
	during science verification (SV) were observed under the code MJLSG22.  Here, they
	are combined with observations taken under regular observing, using the standard
	code for that cloud.  An asterisk in the `Notes' column denotes observations that 
	include SV data.}
\tablenotetext{b}{Number of integrations, and the estimated rms noise for the mosaic of 
	all observations of this field.  Note that the number of integrations may include
	partially completed observations.  See the text for details.  For an effective
	beamsize of 14\farcs6 and 9\farcs8 at 850~$\mu$m and 450~$\mu$m respectively
	\citep[see][]{Dempsey13}, 1~mJy~arcsec$^{-2}$ corresponds to 73~mJy~beam$^{-1}$ and
	49~mJy~beam$^{-1}$, respectively.}
\tablenotetext{c}{Overlaps with OMC1 Tiles 1 through 4; field name modified from target name 
	in archive of OrionBN-KL for clarity.}
\tablenotetext{d}{For historical interest, we note that observations of this field appear
	to be the final SCUBA-2 data obtained at JCMT before ownership of the telescope
	was transferred to the East Asian Observatory.}
\tablenotetext{e}{SV centres offset by $\sim$ 10 arcmin.}
\tablenotetext{f}{Only observed during SV; full survey depth not reached over entire field,
	however, there is significant overlap with standard survey fields.}
\tablenotetext{g}{One of the observations is mis-labeled with code MJLSG31 in the archive.}
\end{deluxetable}

Table~\ref{tab_recov_props} meanwhile provides a summary of the recovered source properties
for the two external mask reductions tested.

\startlongtable
\begin{deluxetable}{crr|rrrrrr|rrrrrr}
\tablecolumns{15}
\tablewidth{0pc}
\tabletypesize{\scriptsize}
\tablecaption{Properties Recovered for External Mask Reductions\label{tab_recov_props}}
\tablehead{
\colhead{DR} &
\colhead{$\sigma$\tablenotemark{a}} &
\colhead{Peak\tablenotemark{a}} & 
\multicolumn{2}{c}{Peak$_\mathrm{rec}$\tablenotemark{b}}&
\multicolumn{2}{c}{Sigma$_\mathrm{rec}$\tablenotemark{b}} & 
\multicolumn{2}{c}{Tot$_\mathrm{rec}$\tablenotemark{b}} &
\multicolumn{2}{c}{Peak$_\mathrm{rec,mask}$\tablenotemark{c}}&
\multicolumn{2}{c}{Sigma$_\mathrm{rec,mask}$\tablenotemark{c}} &
\multicolumn{2}{c}{Tot$_\mathrm{rec,mask}$\tablenotemark{c}} \\ 
\colhead{Method} & 
\colhead{(\arcsec)} & 
\colhead{($N_\mathrm{rms}$)} & 
\colhead{Mean} & 
\colhead{Dev} & 
\colhead{Mean} & 
\colhead{Dev} & 
\colhead{Mean} & 
\colhead{Dev} & 
\colhead{Mean} & 
\colhead{Dev} & 
\colhead{Mean} & 
\colhead{Dev} & 
\colhead{Mean} & 
\colhead{Dev}  
}

\startdata
 DR1 &  10 &  1 &   1.44  &   0.28  &   0.76  &   0.15  &   0.82  &   0.29  &   1.26  &   0.22  &   0.83  &   0.07  &   0.87  &   0.31 \\
 DR1 &  10 &  2 &   0.92  &   0.18  &   0.95  &   0.20  &   0.82  &   0.30  &   0.99  &   0.18  &   0.91  &   0.17  &   0.80  &   0.29 \\
 DR1 &  10 &  3 &   0.90  &   0.15  &   0.94  &   0.12  &   0.78  &   0.16  &   0.96  &   0.14  &   0.91  &   0.12  &   0.78  &   0.17 \\
 DR1 &  10 &  5 &   0.95  &   0.09  &   0.94  &   0.07  &   0.82  &   0.11  &   0.96  &   0.08  &   0.93  &   0.06  &   0.83  &   0.10 \\
 DR1 &  10 &  7 &   0.97  &   0.07  &   0.94  &   0.04  &   0.86  &   0.08  &   0.97  &   0.07  &   0.94  &   0.04  &   0.86  &   0.08 \\
 DR1 &  10 & 10 &   0.99  &   0.04  &   0.95  &   0.03  &   0.90  &   0.06  &   0.99  &   0.04  &   0.95  &   0.03  &   0.90  &   0.06 \\
 DR1 &  10 & 15 &   0.99  &   0.03  &   0.96  &   0.02  &   0.93  &   0.05  &   0.99  &   0.03  &   0.96  &   0.02  &   0.93  &   0.05 \\
 DR1 &  10 & 20 &   1.00  &   0.02  &   0.98  &   0.02  &   0.95  &   0.04  &   1.00  &   0.02  &   0.98  &   0.02  &   0.95  &   0.04 \\
 DR1 &  10 & 50 &   1.00  &   0.01  &   1.00  &   0.01  &   0.99  &   0.01  &   1.00  &   0.01  &   1.00  &   0.01  &   0.99  &   0.01 \\
 DR1 &  30 &  1 &   1.14  &    -1  &   0.34  &    -1  &   0.13  &    -1  &  -1  &  -1  &  -1  &  -1  &  -1  &  -1 \\
 DR1 &  30 &  2 &   0.64  &   0.09  &   0.65  &   0.05  &   0.28  &   0.06  &   0.71  &  -1  &   0.69  &  -1  &   0.34  &  -1 \\
 DR1 &  30 &  3 &   0.48  &   0.07  &   0.70  &   0.06  &   0.23  &   0.04  &   0.52  &   0.04  &   0.71  &   0.04  &   0.26  &   0.04 \\
 DR1 &  30 &  5 &   0.48  &   0.06  &   0.71  &   0.05  &   0.24  &   0.04  &   0.51  &   0.04  &   0.70  &   0.05  &   0.25  &   0.04 \\
 DR1 &  30 &  7 &   0.54  &   0.06  &   0.71  &   0.03  &   0.27  &   0.03  &   0.56  &   0.06  &   0.71  &   0.03  &   0.28  &   0.04 \\
 DR1 &  30 & 10 &   0.67  &   0.07  &   0.73  &   0.02  &   0.36  &   0.05  &   0.69  &   0.06  &   0.73  &   0.02  &   0.37  &   0.05 \\
 DR1 &  30 & 15 &   0.82  &   0.07  &   0.76  &   0.03  &   0.48  &   0.07  &   0.82  &   0.07  &   0.76  &   0.03  &   0.48  &   0.07 \\
 DR1 &  30 & 20 &   0.92  &   0.04  &   0.83  &   0.04  &   0.64  &   0.08  &   0.92  &   0.04  &   0.83  &   0.04  &   0.64  &   0.08 \\
 DR1 &  30 & 50 &   1.00  &   0.01  &   0.96  &   0.01  &   0.93  &   0.03  &   1.00  &   0.01  &   0.96  &   0.01  &   0.93  &   0.03 \\
 DR1 &  50 &  1 &   1.00  &    -1  &   0.19  &    -1  &   0.04  &    -1  &  -1  &  -1  &  -1  &  -1  &  -1  &  -1 \\
 DR1 &  50 &  2 &  -1  &  -1  &  -1  &  -1  &  -1  &  -1  &  -1  &  -1  &  -1  &  -1  &  -1  &  -1 \\
 DR1 &  50 &  3 &  -1  &  -1  &  -1  &  -1  &  -1  &  -1  &  -1  &  -1  &  -1  &  -1  &  -1  &  -1 \\
 DR1 &  50 &  5 &   0.23  &   0.02  &   0.51  &   0.05  &   0.06  &   0.01  &   0.20  &  -1  &   0.55  &  -1  &   0.06  &  -1 \\
 DR1 &  50 &  7 &   0.19  &   0.02  &   0.55  &   0.03  &   0.06  &   0.01  &   0.19  &   0.02  &   0.57  &   0.03  &   0.06  &   0.00 \\
 DR1 &  50 & 10 &   0.21  &   0.03  &   0.56  &   0.02  &   0.06  &   0.01  &   0.21  &   0.02  &   0.56  &   0.02  &   0.07  &   0.01 \\
 DR1 &  50 & 15 &   0.33  &   0.09  &   0.56  &   0.01  &   0.11  &   0.03  &   0.38  &   0.09  &   0.56  &   0.01  &   0.12  &   0.03 \\
 DR1 &  50 & 20 &   0.65  &   0.16  &   0.62  &   0.04  &   0.26  &   0.09  &   0.66  &   0.15  &   0.62  &   0.04  &   0.26  &   0.09 \\
 DR1 &  50 & 50 &   0.99  &   0.02  &   0.91  &   0.02  &   0.83  &   0.05  &   0.99  &   0.02  &   0.91  &   0.02  &   0.83  &   0.05 \\
 DR1 &  75 &  1 &  -1  &  -1  &  -1  &  -1  &  -1  &  -1  &  -1  &  -1  &  -1  &  -1  &  -1  &  -1 \\
 DR1 &  75 &  2 &  -1  &  -1  &  -1  &  -1  &  -1  &  -1  &  -1  &  -1  &  -1  &  -1  &  -1  &  -1 \\
 DR1 &  75 &  3 &  -1  &  -1  &  -1  &  -1  &  -1  &  -1  &  -1  &  -1  &  -1  &  -1  &  -1  &  -1 \\
 DR1 &  75 &  5 &  -1  &  -1  &  -1  &  -1  &  -1  &  -1  &  -1  &  -1  &  -1  &  -1  &  -1  &  -1 \\
 DR1 &  75 &  7 &  -1  &  -1  &  -1  &  -1  &  -1  &  -1  &  -1  &  -1  &  -1  &  -1  &  -1  &  -1 \\
 DR1 &  75 & 10 &  -1  &  -1  &  -1  &  -1  &  -1  &  -1  &  -1  &  -1  &  -1  &  -1  &  -1  &  -1 \\
 DR1 &  75 & 15 &  -1  &  -1  &  -1  &  -1  &  -1  &  -1  &  -1  &  -1  &  -1  &  -1  &  -1  &  -1 \\
 DR1 &  75 & 20 &   0.09  &   0.03  &   0.46  &   0.02  &   0.02  &   0.01  &   0.09  &   0.02  &   0.47  &   0.04  &   0.02  &   0.00 \\
 DR1 &  75 & 50 &   0.92  &   0.04  &   0.81  &   0.03  &   0.60  &   0.06  &   0.92  &   0.04  &   0.81  &   0.03  &   0.60  &   0.06 \\
 DR1 & 100 &  1 &  -1  &  -1  &  -1  &  -1  &  -1  &  -1  &  -1  &  -1  &  -1  &  -1  &  -1  &  -1 \\
 DR1 & 100 &  2 &  -1  &  -1  &  -1  &  -1  &  -1  &  -1  &  -1  &  -1  &  -1  &  -1  &  -1  &  -1 \\
 DR1 & 100 &  3 &  -1  &  -1  &  -1  &  -1  &  -1  &  -1  &  -1  &  -1  &  -1  &  -1  &  -1  &  -1 \\
 DR1 & 100 &  5 &  -1  &  -1  &  -1  &  -1  &  -1  &  -1  &  -1  &  -1  &  -1  &  -1  &  -1  &  -1 \\
 DR1 & 100 &  7 &  -1  &  -1  &  -1  &  -1  &  -1  &  -1  &  -1  &  -1  &  -1  &  -1  &  -1  &  -1 \\
 DR1 & 100 & 10 &  -1  &  -1  &  -1  &  -1  &  -1  &  -1  &  -1  &  -1  &  -1  &  -1  &  -1  &  -1 \\
 DR1 & 100 & 15 &  -1  &  -1  &  -1  &  -1  &  -1  &  -1  &  -1  &  -1  &  -1  &  -1  &  -1  &  -1 \\
 DR1 & 100 & 20 &  -1  &  -1  &  -1  &  -1  &  -1  &  -1  &  -1  &  -1  &  -1  &  -1  &  -1  &  -1 \\
 DR1 & 100 & 50 &   0.75  &   0.04  &   0.66  &   0.03  &   0.32  &   0.04  &   0.75  &   0.04  &   0.66  &   0.03  &   0.32  &   0.04 \\
 DR1 & 125 &  1 &  -1  &  -1  &  -1  &  -1  &  -1  &  -1  &  -1  &  -1  &  -1  &  -1  &  -1  &  -1 \\
 DR1 & 125 &  2 &  -1  &  -1  &  -1  &  -1  &  -1  &  -1  &  -1  &  -1  &  -1  &  -1  &  -1  &  -1 \\
 DR1 & 125 &  3 &  -1  &  -1  &  -1  &  -1  &  -1  &  -1  &  -1  &  -1  &  -1  &  -1  &  -1  &  -1 \\
 DR1 & 125 &  5 &  -1  &  -1  &  -1  &  -1  &  -1  &  -1  &  -1  &  -1  &  -1  &  -1  &  -1  &  -1 \\
 DR1 & 125 &  7 &  -1  &  -1  &  -1  &  -1  &  -1  &  -1  &  -1  &  -1  &  -1  &  -1  &  -1  &  -1 \\
 DR1 & 125 & 10 &  -1  &  -1  &  -1  &  -1  &  -1  &  -1  &  -1  &  -1  &  -1  &  -1  &  -1  &  -1 \\
 DR1 & 125 & 15 &  -1  &  -1  &  -1  &  -1  &  -1  &  -1  &  -1  &  -1  &  -1  &  -1  &  -1  &  -1 \\
 DR1 & 125 & 20 &  -1  &  -1  &  -1  &  -1  &  -1  &  -1  &  -1  &  -1  &  -1  &  -1  &  -1  &  -1 \\
 DR1 & 125 & 50 &   0.19  &   0.15  &   0.34  &   0.06  &   0.03  &   0.03  &   0.25  &   0.14  &   0.35  &   0.07  &   0.04  &   0.04 \\
 DR1 & 150 &  1 &  -1  &  -1  &  -1  &  -1  &  -1  &  -1  &  -1  &  -1  &  -1  &  -1  &  -1  &  -1 \\
 DR1 & 150 &  2 &  -1  &  -1  &  -1  &  -1  &  -1  &  -1  &  -1  &  -1  &  -1  &  -1  &  -1  &  -1 \\
 DR1 & 150 &  3 &  -1  &  -1  &  -1  &  -1  &  -1  &  -1  &  -1  &  -1  &  -1  &  -1  &  -1  &  -1 \\
 DR1 & 150 &  5 &  -1  &  -1  &  -1  &  -1  &  -1  &  -1  &  -1  &  -1  &  -1  &  -1  &  -1  &  -1 \\
 DR1 & 150 &  7 &  -1  &  -1  &  -1  &  -1  &  -1  &  -1  &  -1  &  -1  &  -1  &  -1  &  -1  &  -1 \\
 DR1 & 150 & 10 &  -1  &  -1  &  -1  &  -1  &  -1  &  -1  &  -1  &  -1  &  -1  &  -1  &  -1  &  -1 \\
 DR1 & 150 & 15 &  -1  &  -1  &  -1  &  -1  &  -1  &  -1  &  -1  &  -1  &  -1  &  -1  &  -1  &  -1 \\
 DR1 & 150 & 20 &  -1  &  -1  &  -1  &  -1  &  -1  &  -1  &  -1  &  -1  &  -1  &  -1  &  -1  &  -1 \\
 DR1 & 150 & 50 &  -1  &  -1  &  -1  &  -1  &  -1  &  -1  &  -1  &  -1  &  -1  &  -1  &  -1  &  -1 \\
\hline
 DR2 &  10 &  1 &   1.59  &   0.30  &   0.96  &   0.25  &   1.44  &   0.65  &  -1  &  -1  &  -1  &  -1  &  -1  &  -1 \\
 DR2 &  10 &  2 &   1.02  &   0.16  &   1.06  &   0.21  &   1.14  &   0.43  &  -1  &  -1  &  -1  &  -1  &  -1  &  -1 \\
 DR2 &  10 &  3 &   0.97  &   0.12  &   1.03  &   0.15  &   1.04  &   0.35  &   1.17  &  -1  &   1.12  &  -1  &   1.42  &  -1 \\
 DR2 &  10 &  5 &   0.99  &   0.09  &   1.00  &   0.08  &   0.99  &   0.16  &   1.03  &   0.08  &   1.00  &   0.06  &   1.04  &   0.12 \\
 DR2 &  10 &  7 &   0.98  &   0.07  &   0.98  &   0.05  &   0.95  &   0.11  &   0.99  &   0.06  &   0.98  &   0.05  &   0.95  &   0.10 \\
 DR2 &  10 & 10 &   1.00  &   0.04  &   0.99  &   0.04  &   0.99  &   0.08  &   1.00  &   0.04  &   0.99  &   0.04  &   0.99  &   0.08 \\
 DR2 &  10 & 15 &   1.00  &   0.03  &   0.99  &   0.02  &   0.98  &   0.05  &   1.00  &   0.03  &   0.99  &   0.02  &   0.98  &   0.05 \\
 DR2 &  10 & 20 &   1.00  &   0.02  &   1.00  &   0.02  &   0.99  &   0.04  &   1.00  &   0.02  &   1.00  &   0.02  &   0.99  &   0.04 \\
 DR2 &  10 & 50 &   1.00  &   0.01  &   1.00  &   0.01  &   1.00  &   0.01  &   1.00  &   0.01  &   1.00  &   0.01  &   1.00  &   0.01 \\
 DR2 &  30 &  1 &   1.43  &   0.31  &   1.21  &   0.30  &   2.19  &   1.14  &  -1  &  -1  &  -1  &  -1  &  -1  &  -1 \\
 DR2 &  30 &  2 &   0.97  &   0.22  &   1.14  &   0.16  &   1.26  &   0.42  &  -1  &  -1  &  -1  &  -1  &  -1  &  -1 \\
 DR2 &  30 &  3 &   0.91  &   0.13  &   1.00  &   0.16  &   0.93  &   0.33  &   1.08  &   0.10  &   1.07  &   0.09  &   1.22  &   0.12 \\
 DR2 &  30 &  5 &   0.87  &   0.09  &   0.95  &   0.07  &   0.78  &   0.16  &   0.91  &   0.07  &   0.95  &   0.08  &   0.82  &   0.19 \\
 DR2 &  30 &  7 &   0.89  &   0.08  &   0.93  &   0.04  &   0.78  &   0.11  &   0.89  &   0.08  &   0.93  &   0.04  &   0.78  &   0.11 \\
 DR2 &  30 & 10 &   0.98  &   0.05  &   0.97  &   0.03  &   0.92  &   0.09  &   0.98  &   0.05  &   0.97  &   0.03  &   0.92  &   0.09 \\
 DR2 &  30 & 15 &   0.98  &   0.03  &   0.96  &   0.02  &   0.90  &   0.07  &   0.98  &   0.03  &   0.96  &   0.02  &   0.90  &   0.07 \\
 DR2 &  30 & 20 &   1.00  &   0.03  &   0.98  &   0.02  &   0.96  &   0.06  &   1.00  &   0.03  &   0.98  &   0.02  &   0.96  &   0.06 \\
 DR2 &  30 & 50 &   1.00  &   0.01  &   0.99  &   0.01  &   0.98  &   0.03  &   1.00  &   0.01  &   0.99  &   0.01  &   0.98  &   0.03 \\
 DR2 &  50 &  1 &   1.22  &   0.14  &   1.03  &   0.19  &   1.30  &   0.57  &  -1  &  -1  &  -1  &  -1  &  -1  &  -1 \\
 DR2 &  50 &  2 &   0.87  &   0.18  &   0.99  &   0.23  &   0.87  &   0.41  &  -1  &  -1  &  -1  &  -1  &  -1  &  -1 \\
 DR2 &  50 &  3 &   0.72  &   0.17  &   0.96  &   0.21  &   0.68  &   0.32  &  -1  &  -1  &  -1  &  -1  &  -1  &  -1 \\
 DR2 &  50 &  5 &   0.72  &   0.14  &   0.91  &   0.12  &   0.60  &   0.18  &   0.79  &   0.13  &   0.93  &   0.07  &   0.68  &   0.14 \\
 DR2 &  50 &  7 &   0.77  &   0.13  &   0.88  &   0.07  &   0.59  &   0.15  &   0.79  &   0.11  &   0.88  &   0.06  &   0.62  &   0.14 \\
 DR2 &  50 & 10 &   0.88  &   0.08  &   0.88  &   0.05  &   0.69  &   0.12  &   0.88  &   0.08  &   0.88  &   0.05  &   0.69  &   0.12 \\
 DR2 &  50 & 15 &   0.92  &   0.05  &   0.91  &   0.03  &   0.76  &   0.09  &   0.92  &   0.05  &   0.91  &   0.03  &   0.76  &   0.09 \\
 DR2 &  50 & 20 &   0.95  &   0.05  &   0.93  &   0.03  &   0.83  &   0.08  &   0.95  &   0.05  &   0.93  &   0.03  &   0.83  &   0.08 \\
 DR2 &  50 & 50 &   0.99  &   0.02  &   0.97  &   0.01  &   0.93  &   0.04  &   0.99  &   0.02  &   0.97  &   0.01  &   0.93  &   0.04 \\
 DR2 &  75 &  1 &   1.49  &   0.03  &   0.83  &   0.07  &   0.98  &   0.17  &  -1  &  -1  &  -1  &  -1  &  -1  &  -1 \\
 DR2 &  75 &  2 &   0.74  &   0.13  &   0.81  &   0.08  &   0.48  &   0.13  &  -1  &  -1  &  -1  &  -1  &  -1  &  -1 \\
 DR2 &  75 &  3 &   0.64  &   0.09  &   0.82  &   0.07  &   0.44  &   0.09  &  -1  &  -1  &  -1  &  -1  &  -1  &  -1 \\
 DR2 &  75 &  5 &   0.48  &   0.09  &   0.78  &   0.11  &   0.30  &   0.10  &   0.54  &  -1  &   0.82  &  -1  &   0.36  &  -1 \\
 DR2 &  75 &  7 &   0.49  &   0.10  &   0.78  &   0.07  &   0.30  &   0.08  &   0.59  &   0.06  &   0.80  &   0.05  &   0.37  &   0.06 \\
 DR2 &  75 & 10 &   0.63  &   0.12  &   0.76  &   0.05  &   0.36  &   0.10  &   0.65  &   0.11  &   0.76  &   0.05  &   0.38  &   0.10 \\
 DR2 &  75 & 15 &   0.80  &   0.09  &   0.81  &   0.04  &   0.52  &   0.09  &   0.80  &   0.09  &   0.81  &   0.04  &   0.52  &   0.09 \\
 DR2 &  75 & 20 &   0.83  &   0.05  &   0.83  &   0.02  &   0.58  &   0.06  &   0.83  &   0.05  &   0.83  &   0.02  &   0.58  &   0.06 \\
 DR2 &  75 & 50 &   0.93  &   0.03  &   0.92  &   0.02  &   0.78  &   0.05  &   0.93  &   0.03  &   0.92  &   0.02  &   0.78  &   0.05 \\
 DR2 & 100 &  1 &   1.23  &    -1  &   0.53  &    -1  &   0.33  &    -1  &  -1  &  -1  &  -1  &  -1  &  -1  &  -1 \\
 DR2 & 100 &  2 &   0.68  &   0.07  &   0.71  &   0.11  &   0.34  &   0.08  &  -1  &  -1  &  -1  &  -1  &  -1  &  -1 \\
 DR2 & 100 &  3 &   0.55  &   0.08  &   0.67  &   0.11  &   0.25  &   0.08  &  -1  &  -1  &  -1  &  -1  &  -1  &  -1 \\
 DR2 & 100 &  5 &   0.39  &   0.06  &   0.71  &   0.08  &   0.19  &   0.04  &  -1  &  -1  &  -1  &  -1  &  -1  &  -1 \\
 DR2 & 100 &  7 &   0.35  &   0.06  &   0.67  &   0.05  &   0.16  &   0.04  &   0.47  &  -1  &   0.64  &  -1  &   0.19  &  -1 \\
 DR2 & 100 & 10 &   0.40  &   0.09  &   0.66  &   0.03  &   0.17  &   0.04  &   0.43  &   0.03  &   0.65  &   0.04  &   0.19  &   0.03 \\
 DR2 & 100 & 15 &   0.53  &   0.08  &   0.67  &   0.03  &   0.24  &   0.05  &   0.53  &   0.08  &   0.67  &   0.03  &   0.24  &   0.05 \\
 DR2 & 100 & 20 &   0.65  &   0.08  &   0.70  &   0.02  &   0.32  &   0.05  &   0.65  &   0.08  &   0.70  &   0.02  &   0.32  &   0.05 \\
 DR2 & 100 & 50 &   0.82  &   0.03  &   0.83  &   0.02  &   0.57  &   0.04  &   0.82  &   0.03  &   0.83  &   0.02  &   0.57  &   0.04 \\
 DR2 & 125 &  1 &   1.21  &   0.08  &   0.51  &   0.00  &   0.31  &   0.02  &  -1  &  -1  &  -1  &  -1  &  -1  &  -1 \\
 DR2 & 125 &  2 &   0.62  &   0.04  &   0.55  &   0.04  &   0.18  &   0.02  &  -1  &  -1  &  -1  &  -1  &  -1  &  -1 \\
 DR2 & 125 &  3 &   0.45  &   0.06  &   0.58  &   0.06  &   0.14  &   0.02  &  -1  &  -1  &  -1  &  -1  &  -1  &  -1 \\
 DR2 & 125 &  5 &   0.33  &   0.06  &   0.54  &   0.07  &   0.10  &   0.03  &  -1  &  -1  &  -1  &  -1  &  -1  &  -1 \\
 DR2 & 125 &  7 &   0.27  &   0.03  &   0.61  &   0.07  &   0.10  &   0.02  &  -1  &  -1  &  -1  &  -1  &  -1  &  -1 \\
 DR2 & 125 & 10 &   0.22  &   0.03  &   0.63  &   0.07  &   0.09  &   0.02  &  -1  &  -1  &  -1  &  -1  &  -1  &  -1 \\
 DR2 & 125 & 15 &   0.25  &   0.06  &   0.58  &   0.05  &   0.08  &   0.02  &   0.28  &   0.04  &   0.58  &   0.05  &   0.09  &   0.02 \\
 DR2 & 125 & 20 &   0.31  &   0.06  &   0.59  &   0.03  &   0.11  &   0.02  &   0.31  &   0.06  &   0.58  &   0.03  &   0.11  &   0.02 \\
 DR2 & 125 & 50 &   0.60  &   0.03  &   0.70  &   0.01  &   0.29  &   0.03  &   0.60  &   0.03  &   0.70  &   0.01  &   0.29  &   0.03 \\
 DR2 & 150 &  1 &  -1  &  -1  &  -1  &  -1  &  -1  &  -1  &  -1  &  -1  &  -1  &  -1  &  -1  &  -1 \\
 DR2 & 150 &  2 &  -1  &  -1  &  -1  &  -1  &  -1  &  -1  &  -1  &  -1  &  -1  &  -1  &  -1  &  -1 \\
 DR2 & 150 &  3 &   0.41  &   0.04  &   0.43  &   0.04  &   0.07  &   0.01  &  -1  &  -1  &  -1  &  -1  &  -1  &  -1 \\
 DR2 & 150 &  5 &  -1  &  -1  &  -1  &  -1  &  -1  &  -1  &  -1  &  -1  &  -1  &  -1  &  -1  &  -1 \\
 DR2 & 150 &  7 &   0.22  &   0.04  &   0.47  &   0.04  &   0.05  &   0.01  &  -1  &  -1  &  -1  &  -1  &  -1  &  -1 \\
 DR2 & 150 & 10 &   0.16  &   0.01  &   0.61  &   0.06  &   0.06  &   0.01  &  -1  &  -1  &  -1  &  -1  &  -1  &  -1 \\
 DR2 & 150 & 15 &   0.14  &   0.03  &   0.51  &   0.03  &   0.04  &   0.01  &  -1  &  -1  &  -1  &  -1  &  -1  &  -1 \\
 DR2 & 150 & 20 &   0.14  &   0.05  &   0.54  &   0.05  &   0.04  &   0.01  &   0.17  &  -1  &   0.51  &  -1  &   0.04  &  -1 \\
 DR2 & 150 & 50 &   0.36  &   0.03  &   0.59  &   0.01  &   0.12  &   0.02  &   0.36  &   0.03  &   0.59  &   0.01  &   0.12  &   0.02 \\
\enddata
\tablenotetext{a}{Properties of inserted artificial Gaussians.  The Gaussian width, sigma,
	is given in arcsec, while the peak flux is given in units of the rms noise of 
	the map.}
\tablenotetext{b}{Typical recovered properties of artificial Gaussians, as a fraction of
	their true input value.  For the recovered peak flux, width sigma, and total
	flux, we report
	both the mean recovered value and the standard deviation.  A value of -1 denotes
	cases where no artificial Gaussians were recovered, as well as 
	cases where only one artificial Gaussian was recovered and the standard
	deviation is therefore not measureable.}
\tablenotetext{c}{Typical recovered properties of artificial Gaussians, for the 
	subset of sources that lie within the external mask.}
\end{deluxetable}

\acknowledgements{
The authors wish to recognize and acknowledge the very significant cultural role 
and reverence that the summit of Maunakea has always had within the indigenous 
Hawaiian community.  We are most fortunate to have the opportunity to conduct 
observations from this mountain.
The JCMT has historically been operated by the Joint Astronomy Centre on behalf of the 
Science and Technology Facilities Council of the United Kingdom, the National Research 
Council of Canada, and the Netherlands Organisation for Scientific Research. Additional 
funds for the construction of SCUBA-2 were provided by the Canada Foundation for 
Innovation. The identification number for the programme under which the SCUBA-2 data 
used in this paper is MJLSG*.
The authors thank the anonymous referee for their constructive feedback which improved
this paper.
The authors thank the JCMT staff for their support of
the GBS team in data collection and reduction efforts.
The Starlink software \citep{Currie14} is supported by 
the East Asian Observatory.  These data were reduced using a development version from
2015 December 15, version 01683d2c3d.
This research used the services of the Canadian Advanced Network for
Astronomy Research (CANFAR), which in turn is supported by CANARIE,
Compute Canada, University of Victoria, the National Research Council of
Canada, and the Canadian Space Agency.
This research used the facilities of the Canadian Astronomy Data Centre operated by the 
National Research Council of Canada with the support of the Canadian Space Agency.
Figures in this paper were creating using the NASA IDL astronomy library
\citep{idlastro} and the Coyote IDL library ({\tt http://www.idlcoyote.com/index.html}).
JCM acknowledges support from the European Research Council under the European
Community's Horizon 2020 framework program (2014-2020) via the ERC Consolidator
grant `From Cloud to Star Formation (CSF)' (project number 648505).
KP acknowledges support from the Ministry of Science and Technology (Taiwan) 
(Grant No. 106-2119-M-007 -021 -MY3) and the Science and Technology Facilities 
Council (Grant No. ST/M000877/1).
ADC is supported by the STFC consolidated grant ST/N000706/1.
}

\facility{JCMT (SCUBA-2)}
\software{Starlink \citep{Currie14}, CUPID \citep{Berry13}, KAPPA \citep{Currie13}, 
SMURF \citep{Jenness13,smurf},  
mpfit \citep{Markwardt09}, IDL astronomy library, and Coyote Graphics}

\bibliographystyle{apj}
\bibliography{orionbib}{}

\end{document}